%% file: FASER_ALPsMP_JHEP.tex
\documentclass[amsmath,amssymb,aps,prd,11pt,tightenlines,superscriptaddress,nofootinbib,preprintnumbers,notitlepage]{revtex4-2}

\input{header}

\makeatletter
\renewcommand{\p@subsection}{}
\renewcommand{\p@subsubsection}{}
\makeatother

\usepackage{makecell}
\begin{document}

\preprint{CERN-EP-2024-262}
\title{{\Large  Shining Light on the Dark Sector: Search for Axion-like Particles and Other New Physics in Photonic Final States with FASER \\  \vspace{1cm}
FASER Collaboration
}}

\begin{figure*}[h]
\vspace*{-0.6in}
\begin{flushleft}
\includegraphics[width=0.19\textwidth]{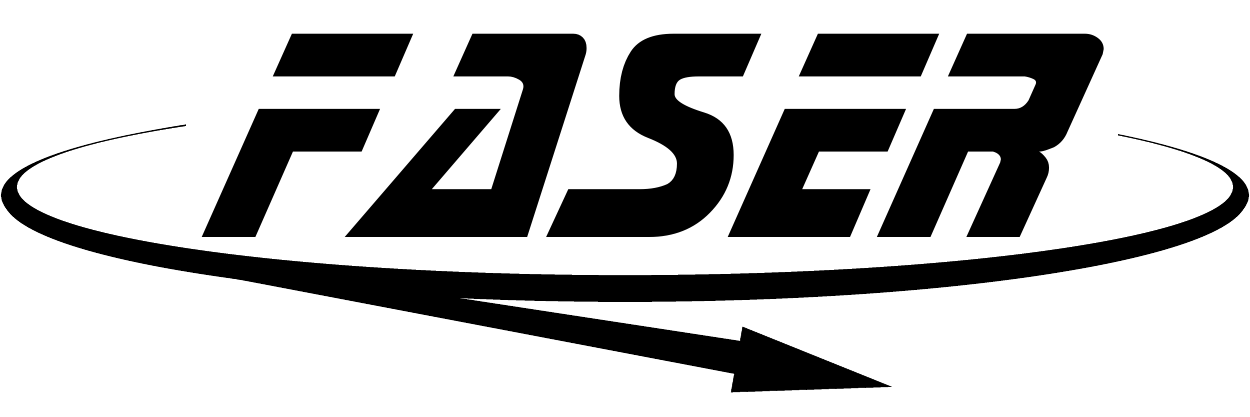}
\end{flushleft}
\end{figure*}

\input{authorlist240903}

\begin{abstract}
\bigskip
The first FASER search for a light, long-lived particle decaying into a pair of photons is reported. The search uses LHC proton-proton collision data at $\sqrt{s}=13.6~\tev$ collected in 2022 and 2023, corresponding to an integrated luminosity of $57.7~\ifb$. A model with axion-like particles (ALPs) dominantly coupled to weak gauge bosons is the primary target.  
Signal events are characterised by high-energy deposits in the electromagnetic calorimeter and no signal in the veto scintillators. One event is observed, compared to a background expectation of $0.44 \pm 0.39$ events, which is entirely dominated by neutrino interactions. World-leading constraints on ALPs are obtained for masses up to $300~\mev$ and couplings to the Standard Model W gauge boson, $g_{aWW}$, around $10^{-4}$ GeV$^{-1}$, testing a previously unexplored region of parameter space. Other new particle models that lead to the same experimental signature, including ALPs coupled to gluons or photons, U(1)$_B$ gauge bosons, up-philic scalars, and a Type-I two-Higgs doublet model, are also considered for interpretation, and new constraints on previously viable parameter space are presented in this paper.

\end{abstract}

\maketitle

\begin{center}
\copyright~2024 CERN for the benefit of the FASER Collaboration. Reproduction of this article or parts of it is allowed as specified in the CC-BY-4.0 license.  
\end{center}
\clearpage
\tableofcontents

\section{Introduction \label{sec:Introduction}}
The ForwArd Search ExpeRiment (FASER)~\cite{Feng:2017uoz, FASER:2018ceo, FASER:2018bac} is an experiment at the CERN Large Hadron Collider (LHC) designed to search for light, weakly-interacting particles produced at the ATLAS interaction point (IP1).
These particles include the neutrinos of the Standard Model (SM) and also new particles associated with beyond-the-SM (BSM) phenomena.

The FASER detector~\cite{FASER:2022hcn} is positioned on the beam collision axis, or line of sight (LOS), 480~m from IP1. The detector is in an ideal location to study light particles that are produced in proton-proton collisions and are so feebly interacting that they can travel through a hundred meters of concrete and rock, offering a relatively large acceptance and efficiency for long-lived particle decays. 
High-energy neutrinos and muons are the only Standard Model particles that can pass through the rock and reach the detector. While muons are deflected by the LHC magnets, not all of them are affected. Neutrinos from colliders were first directly observed using the electronic components of FASER using 2022 data~\cite{FASER:2023zcr}. Neutrinos can also be detected through their interactions with the neutrino-specific passive detector component, FASER$\nu$~\cite{FASER:2020gpr}, which has been used to observe the first electron neutrinos at colliders and also measure neutrino cross sections in the previously unprobed TeV energy range~\cite{FASER:2024hoe}. 

Long-lived particles are predicted by a wide range of BSM models that can accommodate a suitable dark matter (DM) candidate, absent in the SM. These models often involve new particles in a hidden sector, feebly interacting with standard matter. Depending on their mass and couplings to SM particles, they can be long-lived and potentially decay within the FASER detector volume. FASER's reach for BSM long-lived particles has been studied in Ref.~\cite{FASER:2018eoc}. The first search for new physics published by FASER focused on models characterized by the presence of dark photons using the 2022 dataset~\cite{FASER:2023tle}. It exploited the signature of an electron-positron pair appearing within the detector and reconstructed as charged particle tracks associated with a high-energy deposit in the electromagnetic (EM) calorimeter of the FASER detector.

In this paper, a search for light, neutral new particles  decaying to a pair of photons is reported. Events are required to have a high-energy deposit in the EM calorimeter, corresponding to the signature of energetic photon pairs, in addition to no upstream activity. Incoming charged particles are vetoed through FASER's scintillator systems.

Axion-like particles (ALPs) form a broad category of pseudoscalar particles that encompasses axions. Axions are particularly significant due to their potential to solve the strong CP problem ~\cite{Peccei:1977hh,Wilczek:1977pj,Weinberg:1977ma}. Furthermore, axions and ALPs within extended, UV-complete models can offer a variety of phenomenological benefits across a large range of masses~\cite{Rubakov:1997vp,Berezhiani:2000gh,Hook:2014cda,Fukuda:2015ana,Gherghetta:2016fhp,Dimopoulos:2016lvn}. They can be viable dark matter candidates or mediate the interactions between SM and DM particles, and several classes of models exist differing by the kind of coupling between the ALP and SM particles.  Models with ALPs dominantly coupling to weak gauge bosons and decaying into a pair of photons are the primary target of this paper, probing a  range of masses, $m_a$, between $50$ and $500~\mev$ and couplings to the SM particles, $g_{aWW}$, between $10^{-5}$ and $10^{-3}$~GeV$^{-1}$. Models where ALPs interact either exclusively with gluons or photons \cite{Beacham:2019nyx, Aloni:2018vki, Bauer:2018uxu, Feng:2018pew} are also considered for interpretation. Furthermore, events characterised by the presence of a pair of highly-energetic EM deposits can be a signature for other new physics models. Results are therefore also interpreted in the gauged U(1)B model \cite{Tulin:2014tya, Ilten:2018crw, Foguel:2022ppx}, with a new light gauge boson coupled to baryon number, in the up-philic model \cite{Kling:2021fwx, Batell:2018fqo}, predicting light scalars predominantly coupled to up-quarks, and in Type-I two-Higgs doublet models (2HDM), where light scalars with relatively long lifetime can also be accommodated \cite{Kling:2022uzy,Branco:2011iw,Chen:2019pkq}.

The paper is structured as follows: \cref{sec:SignalModels} reports the details of the new physics models used as primary target or considered for interpretation.  The FASER detector is briefly presented in \cref{sec:FASER}. \cref{sec:SamplesAndReco} summarises the event reconstruction and details about the data and simulation samples used. \cref{sec:EventSelection}--\cref{sec:Uncertainties} describe the core of the analysis, from event selection, to the estimation of the backgrounds and the evaluation of systematic uncertainties.~\cref{sec:Results} reports the results and their statistical interpretation as constraints on the parameter space of the ALPs and other hidden sector models.  The appendices contain additional details of the analysis, comparisons to existing bounds, an event display, and a reinterpretation of the results of this paper in the dark photon model, where electron-positron pairs are identified as EM deposits with no tracking requirements.
\section{Signal Models}
\label{sec:SignalModels}

\begin{figure}[tbp]
    \centering
\subfigure[ALP-W production]{\includegraphics[width=0.32\linewidth]{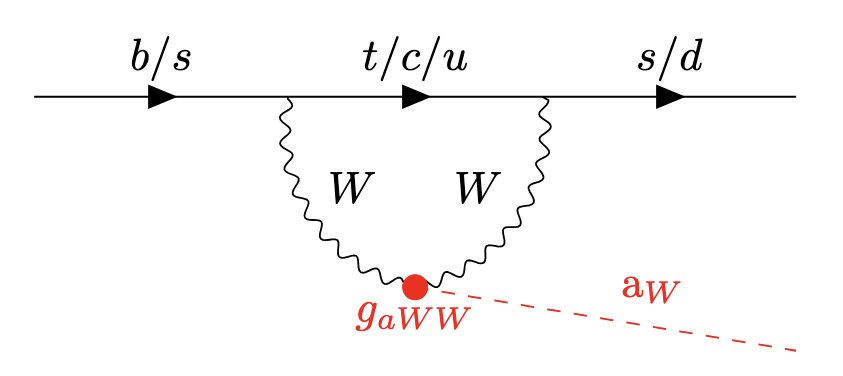}}
    \subfigure[ALP-W decay]{\includegraphics[width=0.32\linewidth]{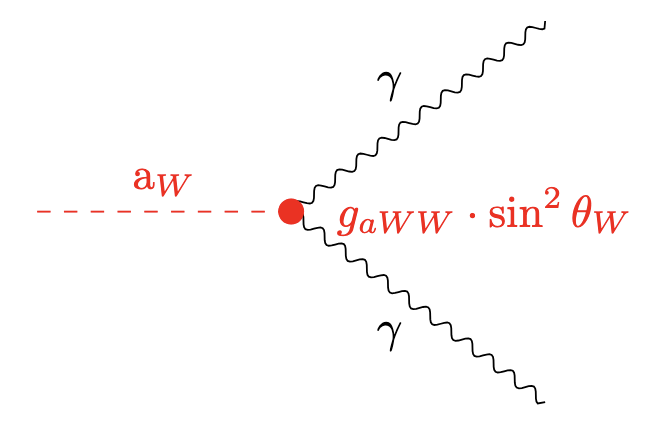}}
    \caption{Primary Feynman diagrams of production and decay of the ALP-W model.}
    \label{fig:ALP-Wdiagrams}
\end{figure}

FASER can probe several types of new physics models characterised by the presence of long-lived particles produced through a variety of processes. Depending on their mass and couplings to SM particles, several interesting decay modes can be searched for at FASER, helping constrain new regions in the parameter space.
In models where the ALP couples to the SU(2)$_L$ gauge bosons before electroweak symmetry breaking (EWSB)~\cite{izaguirre2016new, Gori:2020xvq, Kling_2020}, referred to as the electroweak-philic ALP model (ALP-W), the phenomenology is particularly favourable for FASER. In this case, after EWSB, the ALP ($a_W$) couples to both photons and weak gauge bosons, as detailed in the following. 
The corresponding Lagrangian is~\cite{izaguirre2016new, Gori:2020xvq, Kling_2020}
\begin{align} \label{eq:alpWLagrangian}
        \mathcal{L} \supset -\frac{1}{2}m_{a}^2a_W^2 - \frac{1}{4}g_{aWW}a_WW^{a,\mu\nu}\tilde{W}^{a}_{\mu\nu} \ ,
\end{align}
where $m_{a}$ is the ALP mass, $g_{aWW}$ is the ALP-W coupling parameter, and $W^{\mu\nu}$ is the SU(2)$_L$ field strength tensor. At the LHC, these particles can be produced in decays of $b$- or $s$-flavoured hadrons produced at IP1, primarily via flavour-changing neutral current (FCNC) decays. 
The dominant production processes, at almost equal rates, include the decays of $B^0$ and $B^\pm$ mesons into ALPs and various possible strange hadrons. The production of $B_s$ mesons is suppressed primarily because the strange quark is heavier than the up/down quarks and, consequently, it contributes significantly less to the ALP-W production rate. In the low $m_{a}$ range, where kinematically allowed, kaons can also decay into a pion and an ALP. Once produced, the dominant ALP-W decay mode is into two photons, $a \to \gamma \gamma$. The decay $a \to \gamma ee$ through an off-shell photon has a branching fraction at the percent level and is negligible for this study. An example production mechanism and decay within the ALP-W model is shown in \cref{fig:ALP-Wdiagrams}. 

\begin{figure}[t!]
    \centering
    \subfigure[Acceptance \label{fig:ALPWAcceptance}]{\includegraphics[width=0.29\textwidth]{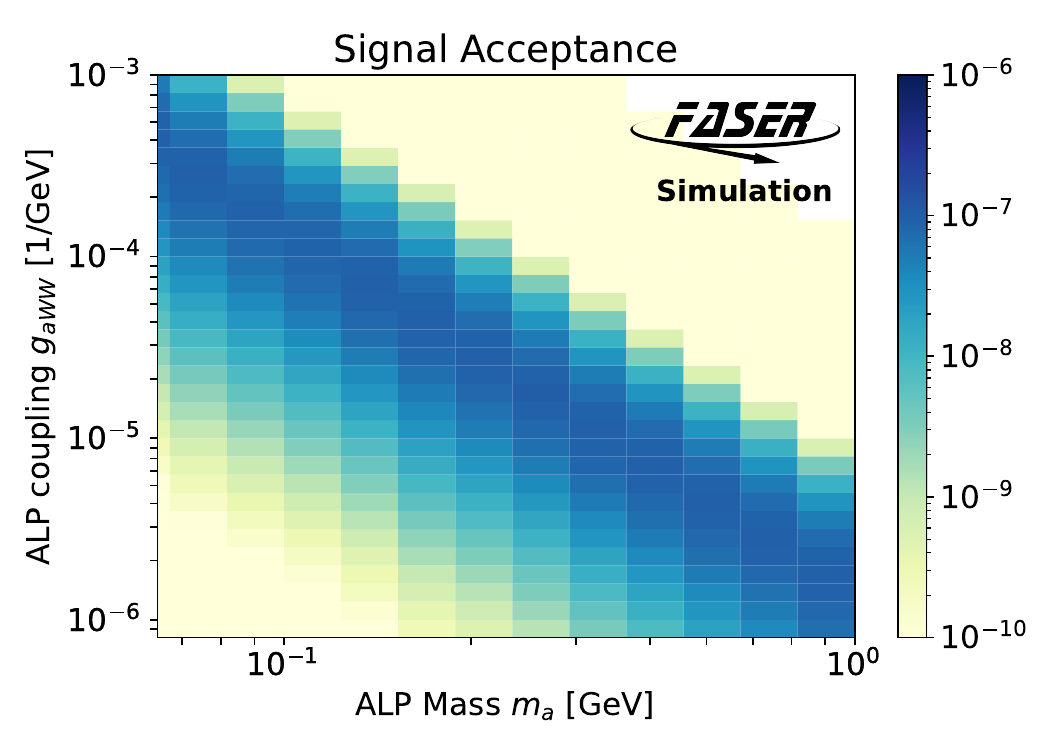}}
    \subfigure[ Fraction of events with E$>$1.5~TeV \label{fig:ALPWEfficiency}]{\includegraphics[width=0.29\textwidth]{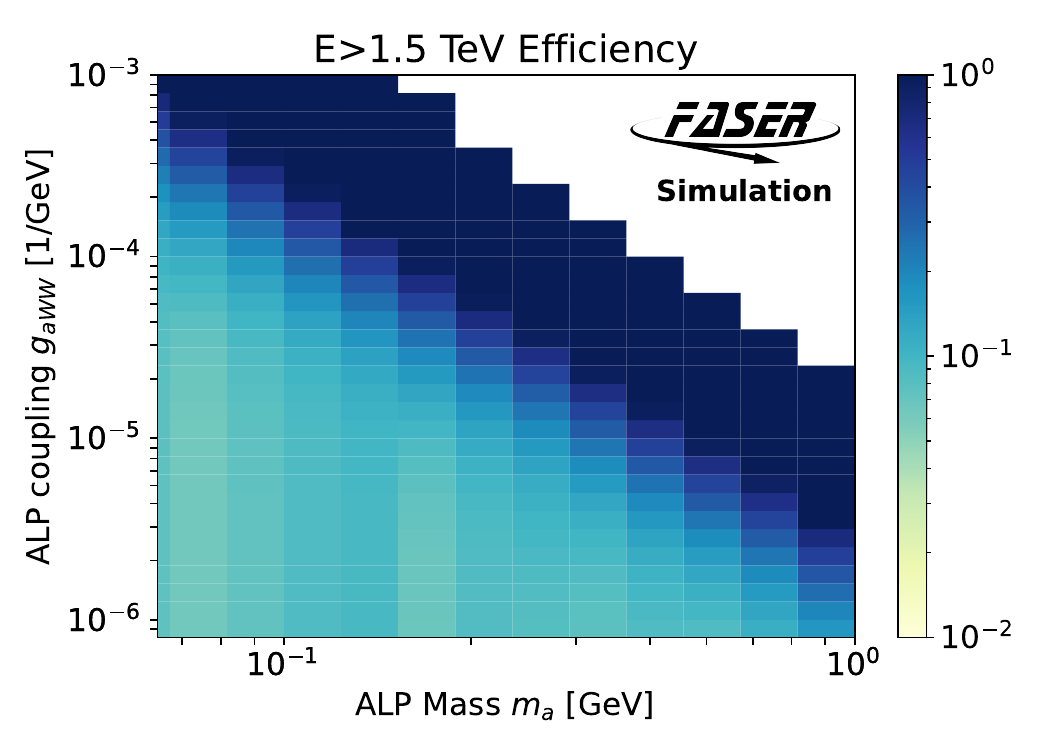}} 
    \subfigure[Signal events \label{fig:ALPWSignalEvents}]{\includegraphics[width=0.29\textwidth]{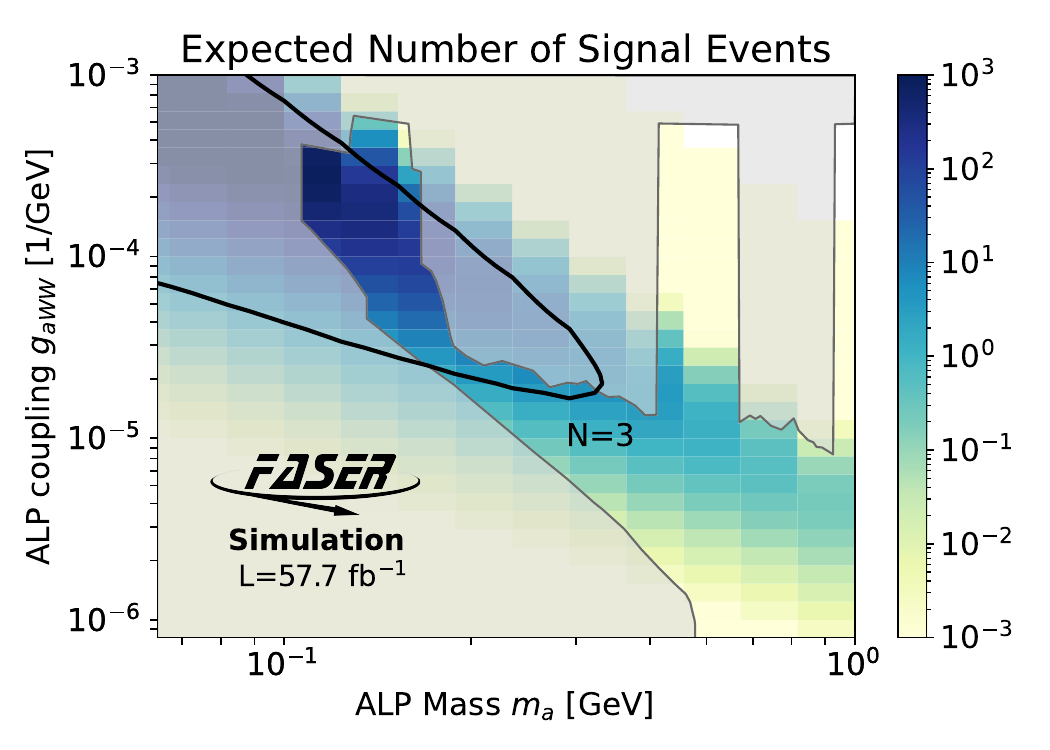}}
    \caption{(a) The acceptance for events from the ALP-W model at truth level to decay inside FASER. (b) The fraction of ALP events in FASER with ALP energies above 1.5 TeV. 
    (c) Expected number of ALP-W signal events in FASER, assuming $57.7~\ifb$ and accounting for the typical signal selection efficiency on top of a 1.5 TeV energy requirement. Shaded areas highlight previously excluded parameter space.
    \label{fig:AcceptancesAndEffiencies}}
\end{figure}
The ALP-W is the primary target of the study presented here, with the largest projected sensitivity expected within FASER among the many models initially investigated in Ref.~\cite{FASER:2018eoc}. The expected acceptance, the fraction of ALPs with energy $>$1.5 TeV and the signal yield  as a function of $m_{a}$ and $g_{aWW}$ are shown in \cref{fig:AcceptancesAndEffiencies}. Predictions are obtained using Monte Carlo (MC) samples generated with the \texttt{FORESEE}~\cite{Kling:2021fwx} package before the detector response simulation, referred to as truth level, and assuming an integrated luminosity of $57.7~\ifb$.    
In the parameter space of interest, typical signal acceptances in the FASER detector volume are of the order of $10^{-6}$ to $10^{-7}$ (\cref{fig:ALPWAcceptance}). FASER covers around $10^{-8}$ of the solid angle of the ATLAS IP and the probability to decay in the FASER decay-volume is $< 0.3\%$, this highlights the beneficial location of FASER for such BSM searches. 
Since forward hadrons inherit a sizable fraction of the beam energy, the ALPs produced in their decay and reaching FASER can have multi-TeV momenta~\cite{FASER:2018eoc}. In the ALP-W case, the fraction of events with energy above 1.5~TeV (as shown in \cref{fig:ALPWEfficiency}) corresponds to the selection criterion for the signal region used in the analysis. For high coupling values, this fraction is well above 50\%. Across a broad portion of the parameter space, such as around masses of 100~MeV and coupling values as low as $10^{-5}~\text{GeV}^{-1}$, the fraction remains at 10\% or higher. At a luminosity of 57.7 fb$^{-1}$, signal yields evaluated at truth level as a function of $m_a$ and coupling $g_{aWW}$ are shown in \cref{fig:ALPWSignalEvents}, where the effects of the $a_W$ momentum selection and the typical signal selection efficiency are included. The latter takes into account as well the efficiency of the other selections applied in the analysis and is around 80\%. As evident in this figure, with the current dataset, FASER has the sensitivity to see hundreds of $a_W$ events in currently unconstrained regions of parameter space. Assuming a background-free analysis, also shown is the contour highlighting the region with more than three expected signal events, which indicates that previously unexplored parameter space with masses $m_a \sim$ 60 MeV $-$ 400 MeV and couplings $g_{aWW} \sim 10^{-5} - 10^{-3}~\text{GeV}^{-1}$ are expected to be probed by FASER. \\

Additional models with similar expected detector signatures have also been considered for interpretation and are described in the following. 
\begin{itemize}

\item \textbf{ALP-photon} A benchmark model of axion-like particles suggested by the Physics-Beyond-Colliders group \cite{Beacham:2019nyx} is an effective Lagrangian with an ALP only coupling to photons, referred to as the ALP-photon model, defined by its coupling to photons $g_{a\gamma\gamma}$ and its mass \cite{Feng:2018pew}. The ALP-W model introduced in \cref{eq:alpWLagrangian} can be seen as a UV-completion of the ALP-photon model. The dominant production mode for ALP-photons to enter FASER is through the so-called \textit{Primakoff} process \cite{Primakoff:1951iae, Tsai:1986tx, Beacham:2019nyx}, where a photon interacts in matter and converts to an ALP. This can happen in material between IP1 and FASER, most importantly the TAN \cite{TAN}, which is an absorber for neutral particles. In the mass and coupling ranges allowing the ALP-photon to decay within FASER, ALPs decay almost solely into two photons. 
\item \textbf{ALP-gluon} Models where axion-like particles can interact exclusively with gluons through coupling $g_{agg}$ are also considered. These models are especially intriguing because axions were originally proposed as a solution to the strong CP problem, with their mass being directly related to the strength of their interaction with gluons. Interactions between ALPs and ordinary matter 
affect quarks properties (at one loop), induce changes in quark flavours (at two loops), and extend to leptons and photons (at three loops). The model considered in this paper, referred to as the ALP-gluon model, was studied in previous works \cite{Aloni:2018vki}, and FASER's reach was first evaluated in Ref.~\cite{FASER:2018eoc}. The dominant production mode arises from mixing with neutral scalars such as $\pi^0$, $\eta$, and $\eta'$. Flavour-changing couplings to quarks can produce ALP-gluons via $B$-meson decays ($B \rightarrow X_s a$), but these are loop-suppressed and thus not considered in this study. At low masses, ALP-gluons decay only to two photons, with hadronic decays such as to $3\pi$ and $\pi^+ \pi^- \gamma$ occurring above approximately $m_a = 0.4$~GeV. Estimates of the decay widths for hadronic decays for the relevant mass range are taken from Ref.~\cite{Aloni:2018vki}. 

\item \textbf{Up-philic scalar} Long-lived scalar or pseudoscalar particles such as ALPs offer interesting model-building prospects, since they can avoid many symmetry-based constraints restricting the properties of vector-like new particles. 
A light scalar predominantly coupling to up-type quarks \cite{Batell:2018fqo} (also referred to as the up-philic scalar, $S$) can be introduced avoiding strict constraints from $B$-meson decays and other flavour-related experimental data measurements, while still allowing for high production rates in proton-proton collisions. The model hereby considered can be entirely defined through the scalar mass, $m_S$, and its  coupling $g_u$ to SM up quarks. The primary production modes for this up-philic scalar are rare decays of $\eta$ and $\eta'$, specifically $\eta, \eta' \to \pi^0 S$, with suppressed kaon and $B$-meson decays \cite{Kling:2021fwx}. The scalar decays mainly into hadronic final states, predominantly a neutral pion pair, if kinematically allowed. For lower scalar masses, the lifetime of $S$ is significantly extended due to loop-induced decays into photons.
\item \textbf{U(1)$_B$ vector boson} A common approach to add potential dark sector mediators is by introducing an additional U(1) symmetry, with a new vector gauge boson. A well-known class of such models are dark photons, which only interact with the SM through kinetic mixings with the SM fermions. This can be extended to a direct gauge coupling between the new boson and SM fields. Introducing a dark sector through a U(1) gauge symmetry conserving baryon number would lead to an additional vector boson coupling to the baryon number ($g_B$), here referred to as the U(1)$_B$ gauge boson \cite{Tulin:2014tya, Ilten:2018crw, Foguel:2022ppx}. The existence of these new bosons could explain the accidental baryon number conservation of the SM. The main production of U(1)$_B$ gauge bosons at the LHC is through dark bremsstrahlung or light meson decays. The U(1)$_B$ vector boson can decay into an electron-positron pair at small masses, below 0.2 GeV. Above that, its main decay channel is into a neutral pion and a photon, leaving a photonic final state in FASER. At higher masses, a decay into a charged pion pair and a neutral pion is also possible. 
\item \textbf{Type-I 2HDM} Type-I ``fermiophobic" two Higgs doublet models (2HDM) predict a second Higgs doublet \(\Phi_2\) coupling to fermions and leading to a new CP-even scalar Higgs, \(H\). For large values of tan$\beta$ (the ratio of the vacuum expectation values of the two doublets) and small values of \(\cos(\beta - \alpha)\) (parametrizing the coupling of \(H\) to gauge bosons, with the mass-squared matrix of the scalars diagonalised by the angle $\alpha$), the scalar \(H\) is very weakly coupled to fermions and gauge bosons \cite{Kling:2022uzy,Branco:2011iw,Chen:2019pkq}. Hence it can be long-lived and potentially detectable in FASER. Following Ref.~\cite{Kling:2022uzy}, we further choose \(\cos(\beta - \alpha)\) = 1/\(\tan(\beta)\) to suppress strong bounds from $h \to HH$ decays and $m_A = m_{H^+} = 600$~GeV to decouple the pseudo-scalar and charged Higgs boson. The scalar \(H\) is mainly produced through rare B-meson decays, \(B \rightarrow X_s H\), with small contributions from \(B \rightarrow X_s HH\). The single scalar production rate scales with 1/\(\tan(\beta)^2\), while a double scalar production is independent of \(\tan(\beta)\). The additional scalar primarily decays into two photons, \(H \rightarrow \gamma\gamma\).
\end{itemize}

Examples of decay diagrams for new particles as predicted by the above-listed additional models considered in this paper are shown in \cref{fig:decay2}. This showcases the variety of coupling and model specifications FASER can probe exploiting photon-based signatures. An overview of the production and decay mechanisms for all models discussed are given in~\cref{app:SigProdAndDecay}. The expected acceptance, efficiency, and signal yield for each model are similar or lower than those evaluated for the ALP-W model and depend on the mass, coupling of the new predicted particle to ordinary matter and the production mode. Additionally, the photon-signature-based analysis presented in the following can be used to infer interpretations in models with electron-signatures, such as given in \cref{app:darkPhoton}.

\begin{figure}[h]
\centering
 \subfigure[ALP-photon Decay]{ \includegraphics[width=0.24\linewidth]{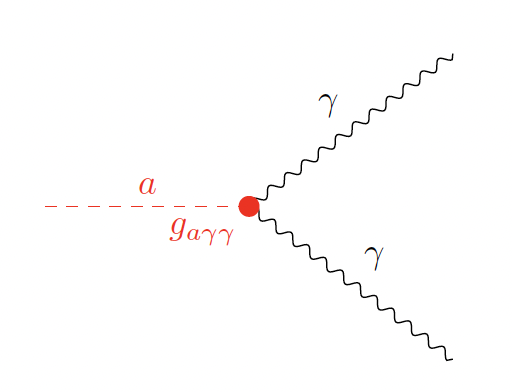}} 
 \subfigure[ALP-gluon Decay \label{fig:ALPg-decay}]{\includegraphics[width=0.24\linewidth]{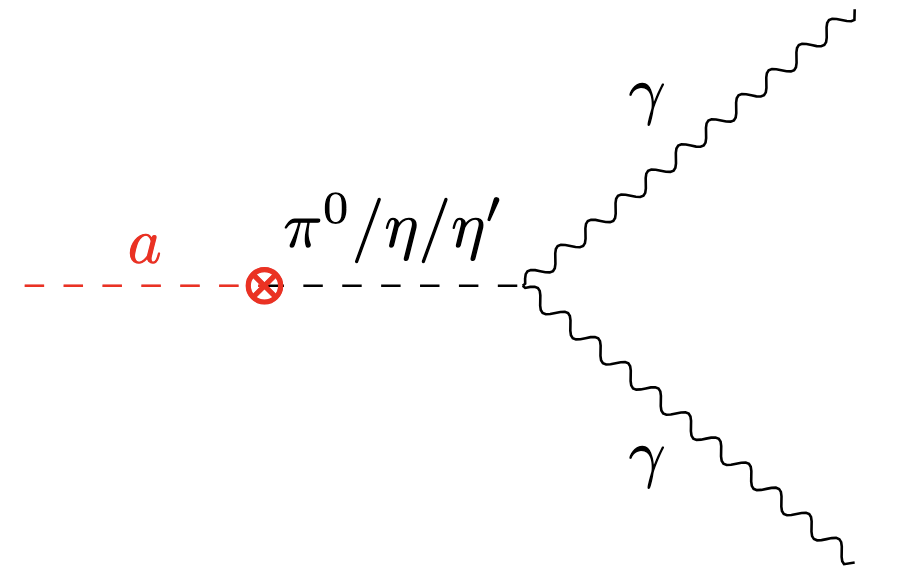}}
\subfigure[U(1)$_B$ Decay]{\includegraphics[width=0.24\linewidth]{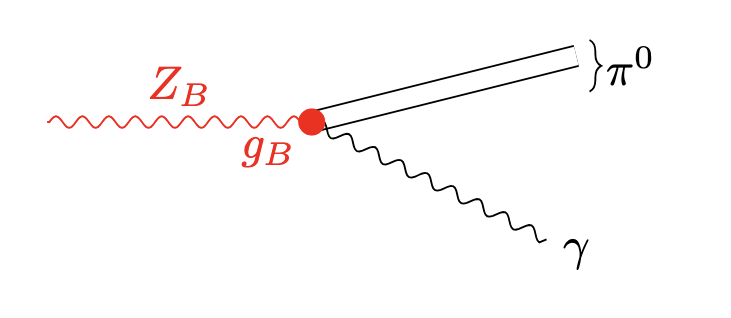}}\\
\subfigure[Up-philic Decay ($m_S < 2 \times m_{\pi^0}$) \label{fig:UPless}]{\includegraphics[width=0.24\linewidth]{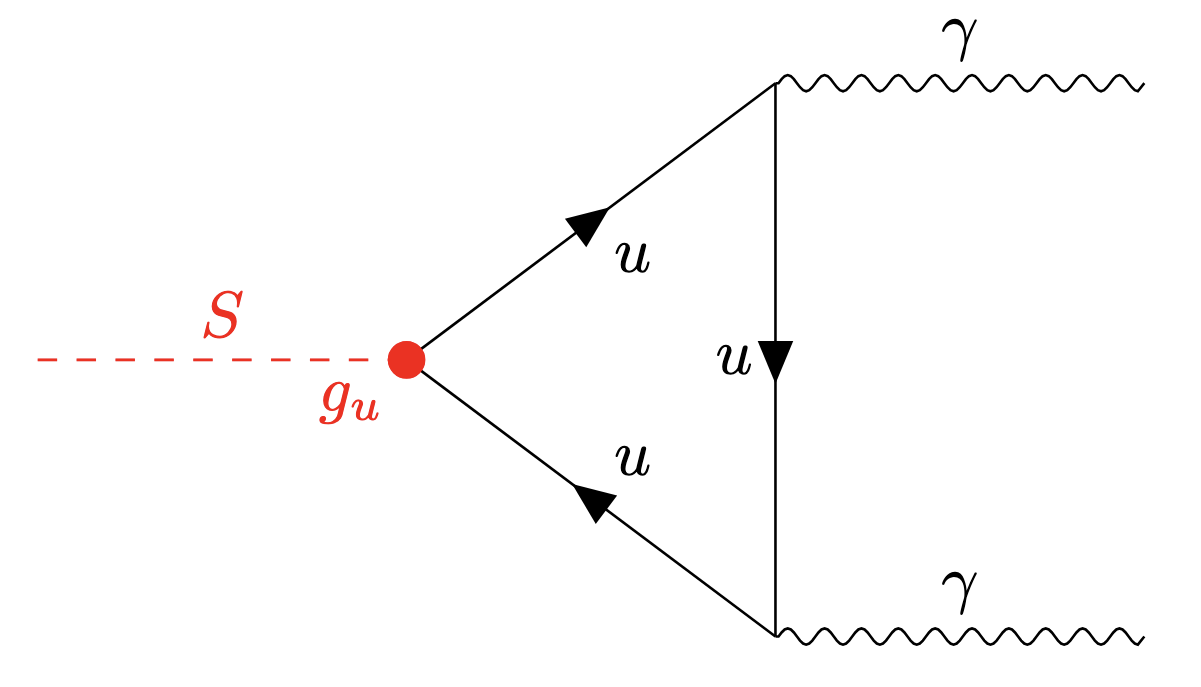}}
\subfigure[Up-Philic Decay ($m_S > 2 \times m_{\pi^0}$) \label{fig:UPmore}]{\includegraphics[width=0.24\linewidth]{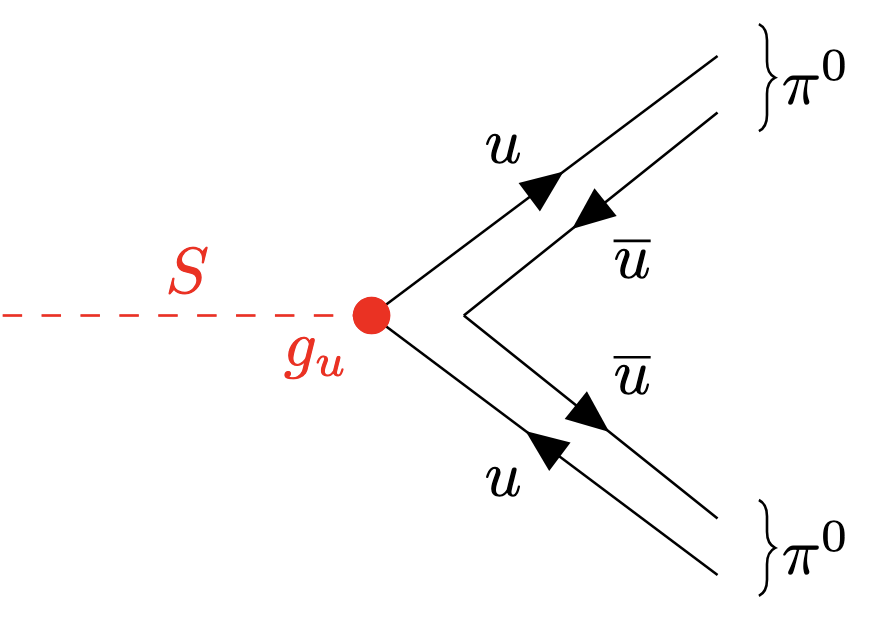}}
\subfigure[2HDM Decay \label{fig:2HDM-decay}]{\includegraphics[width=0.24\linewidth]{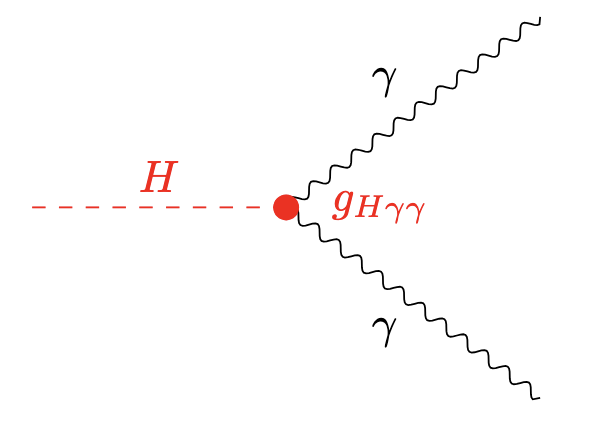}}
\caption{Examples of decay diagrams for new particles as predicted by (left to right, top to bottom): ALP-photon (a), ALP-gluon (b), U(1)$_B$ (c), up-philic (d and e) and 2HDM (f) models. In the case of up-philic models, decays for scalar mass ranges above or below twice the mass of the $\pi^0$ are shown.} \label{fig:decay2}
\end{figure}

\section{The FASER Detector \label{sec:FASER}}

The FASER experiment, described in detail in Ref.~\cite{FASER:2022hcn}, is located in the TI12 connection tunnel about 5~m away from the LHC. It is aligned with the LOS of the IP1 collision axis.\footnote{FASER uses a right-handed coordinate system. The origin of the coordinate system in the transverse plane is the centre of the detector axis.  In the longitudinal direction, the origin is at the front of the first tracking station of the spectrometer, which is 477.759~m from IP1. The $x$-axis points horizontally towards the center of the LHC, the $y$-axis vertically towards the Earth's surface, and the $z$-axis along the central detector axis, away from IP1. The radius from the centre of the detector is calculated as $r=\sqrt{(x^2+y^2) \label{footnote1}}$, with the azimuthal angle $\phi$ around the $z$-axis. Pseudorapidity is $\eta = -\ln \tan(\theta/2)$, where $\theta$ is the polar angle from the z axis } However, due to the crossing angle in IP1 \footnote{In 2022, the half-crossing angle was fixed at -160 $\mu$rad, while in 2023 it varied slightly between -165 $\mu$rad and -135 $\mu$rad during physics fills, with most luminosity recorded around -160 $\mu$rad.}, the LOS is offset vertically by up to 6.5~cm with respect to the centre of the detector. With a 10 cm-radius active transverse size, it covers an angular acceptance corresponding to a pseudorapidity $\eta>9.2$ around the LOS with respect to IP1. FASER also includes a sub-detector, FASER$\nu$~\cite{FASER:2020gpr}, designed to detect neutrinos produced in the LHC collisions and to study their properties.

 \begin{figure}[tpb]
 \centering
 \includegraphics[width=0.95\textwidth]{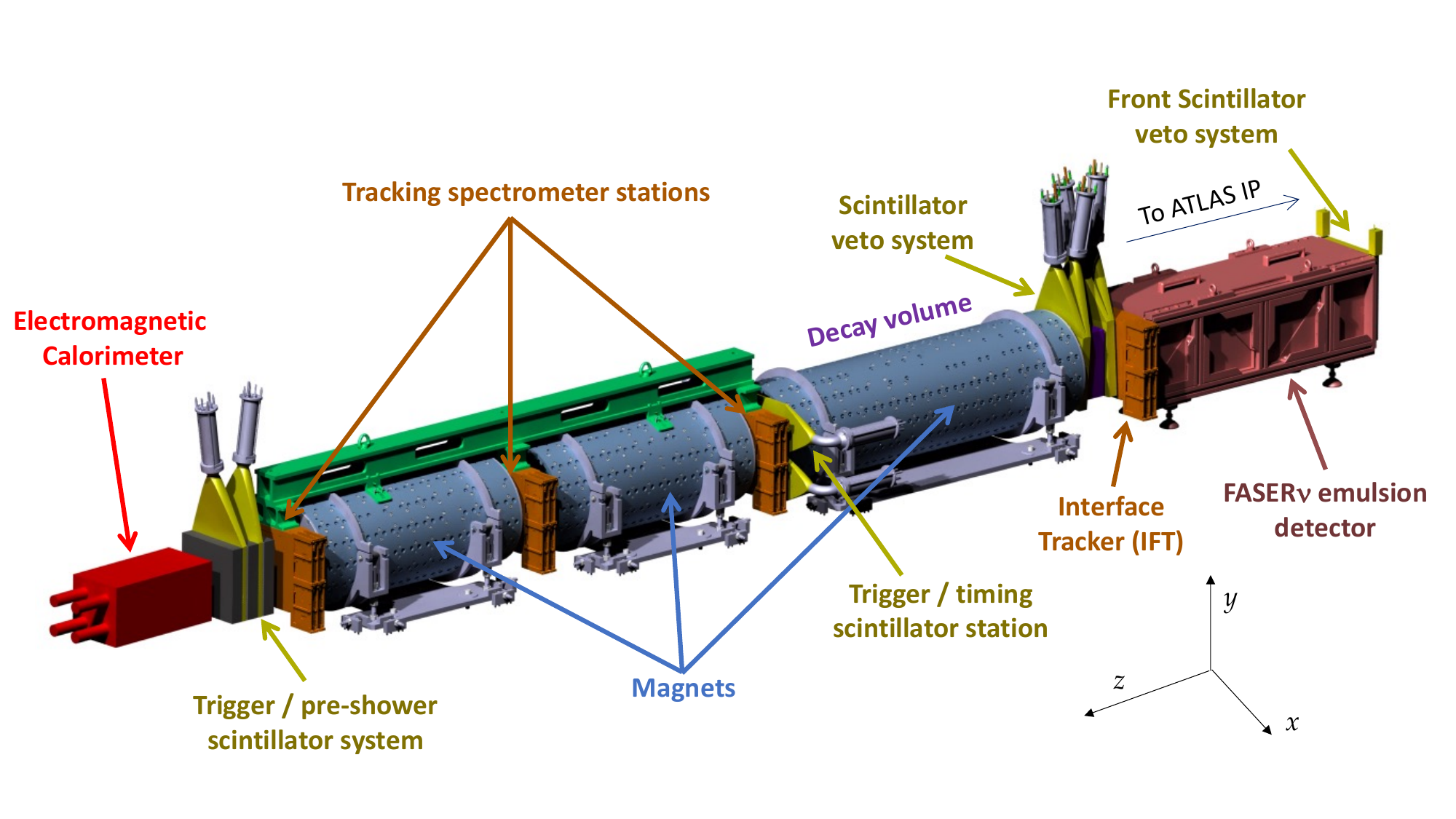}
 \caption{A sketch of the FASER detector, showing the different detector systems. The detector coordinate system is also shown~\cite{FASER:2022hcn}.}
 \label{fig:Faser_Dec}
 \end{figure}

The detector (as illustrated in ~\cref{fig:Faser_Dec}) consists of a front scintillator veto system, the FASER$\nu$ emulsion detector, the interface tracker,  the FASER scintillator veto station, the decay volume, the timing scintillator station, the FASER tracking spectrometer,  the preshower (PS) scintillator system, and the EM calorimeter system. The detector includes three 0.57~T dipole magnets, one surrounding the decay volume and the other two embedded in the tracking spectrometer. The key components of interest for this analysis are the scintillator (including the preshower) systems, and the EM calorimeter. Furthermore, the calorimeter trigger is used for the data set considered.
 
The scintillator system includes four stations, each featuring multiple scintillator layers as follows:

\begin{itemize}
[itemsep=0.03cm, topsep=0.15cm, leftmargin=1.5em]
\item At the front of the detector is the vetoNu station, comprising of two scintillator layers positioned in front of the FASER$\nu$ tungsten/emulsion detector. While the FASER$\nu$ detector is not utilized in this analysis, the eight interaction lengths of tungsten play a role in suppressing potential backgrounds. 
\item In front of the decay volume is the veto scintillator station composed of three scintillator layers, with a 10~cm-thick lead absorber placed between the two downstream scintillators and the upstream one. 
\item For triggering and timing measurements and as an additional veto layer, the timing scintillator station is placed after the decay volume and in front of the tracking spectrometer. This station consists of two scintillator layers separately covering the top and bottom half of the detector (with a small overlap region in between). The scintillators are 1~cm thick to minimize the material in the detector volume. The timing scintillators cover a bigger transverse area with respect to the LOS, facilitating identification of incoming large-angle muons that miss the veto scintillators and could lead to a high-energy deposit in the calorimeter.
\item  A preshower detector, consisting of two scintillator layers, is placed after the tracking spectrometer and in front of the calorimeter. Both layers are preceded by a 3 mm-thick layer of tungsten radiator to create a simple preshower detector. The purpose of the preshower in FASER is to differentiate between an incoming EM shower formed by a high-energy photon or electron and a neutrino interacting in the calorimeter material. To reduce backsplash from the calorimeter and preshower radiator into the last tracking station, a 5 cm-thick graphite block is placed in front of each layer of tungsten and between the final scintillator layer and the calorimeter.
\end{itemize}

All scintillators (except the preshower) are used to veto events with incoming charged particles, which are mostly muons.

Since the signal of interest does not rely on the reconstruction of tracks through the tracking spectrometer (see ~\cref{sec:EventSelection}), the analysis is sensitive to ALPs decaying within an extended decay volume comprising the region between the veto scintillators and the preshower scintillator station, which consists of the decay volume, spectrometer magnets, tracker stations and the timing station (referred to as the sensitive detector volume in the following and is a total length of 4.15~m with a 10~cm radius).

The EM energy of particles traversing the detector volume is measured by the EM calorimeter, located at the furthest end of the detector. The calorimeter is composed of four spare modules from the outer electromagnetic calorimeter of the LHCb detector~\cite{LHCB:2000ab}, each comprising a total depth of 25 radiation lengths. The energy resolution has been measured to be $\mathcal{O}{(1\%)}$~\cite{FASER:2022hcn} in the high-energy range most relevant for this analysis, using data collected in the test beam at the CERN SPS carried out in July 2021.

\section{Data and Simulation Samples
\label{sec:SamplesAndReco}}

This analysis uses data events collected during 2022 and 2023 in $\sqrt{s}=13.6~\tev$ collisions provided by the LHC during its ongoing Run 3 data-taking. The dataset analysed corresponds to an integrated luminosity of $57.7~\ifb$ after data quality selections were applied to remove data taken during non-standard conditions. The ATLAS luminosity measurements and calibrations described in Refs.~\cite{ATL-DAPR-PUB-2023-001, ATL-DAPR-PUB-2024-001, Avoni:2018iuv, DAPR-2021-01} are used. 

Events are triggered by signals from the scintillators or calorimeter system, with a typical total trigger rate of 1~kHz in the relevant years. This is dominated by high-energy muons entering FASER from collision decay products in IP1. The average detector deadtime was measured to be 1.3\%. The trigger and data acquisition system is described in more detail in Ref.~\cite{FASERTDAQ:2021}. In this analysis, only the calorimeter trigger is used and is fully efficient for energy deposits above 20 GeV.

Monte Carlo (MC) simulated samples are used in the analysis to evaluate signal and background yields and optimise the event selection. Additionally, the MC was used to aid in the evaluation of some of the systematic uncertainties for the analysis. The simulation of all signal MC samples is based on \textit{FORESEE}. For each signal model, samples are simulated for a 2-dimensional grid in the mass-vs-coupling plane. To overcome the computational impracticality of running the full detector simulation for fine grids with large sample statistics, a method of parameterizing the efficiency, as a function of energy, of long-lived particles decaying into specific final states to pass analysis selections was adopted. These parameterized efficiencies were obtained for all relevant decay final states using large full detector simulation samples with a flat energy distribution and the energy dependence was modeled with parameterised functions (as given in~\cref{app:parametrisedEfficiencies}). The method was validated against the full simulations for various benchmark models, showing agreement within the MC statistical uncertainty, and then used to generate a large number of simulation points across different models, facilitating a more accurate statistical analysis.

The decay of ALPs, up-philic scalars, U(1)$_B$ bosons and 2HDM scalars up until the first preshower layer was simulated within a 100~mm radius of the detector axis. The simulation of forward $B$-mesons, from which ALPs or light scalars arise, followed the latest prescription developed in Ref.~\cite{HeavyHadronPrescription}, using \verb|POWHEG| \cite{Nason_2004, Frixione_2007, Alioli_2010} with the NNPDF3.1sx+LHCb PDF set \cite{Bertone_2019, Ball_2018} to model $B$-meson production at next-to-leading order and using parton distribution functions including small-x resummation at next-to-leading logarithmic accuracy and matched with \verb|Pythia 8.3| \cite{Bierlich:2022pfr,Sjostrand:2014zea} to model the parton shower and hadronization.
Various sources of uncertainties on the predicted signal yields were evaluated. The main uncertainties taken into account are on the $B$-meson production rates and were estimated as uncertainties on the flux based on variations of the renormalization and factorization scales of the different generators. The kaon decay rate prediction is based on \verb|EPOS-LHC| ~\cite{pierog2015epos}, with \verb|SIBYLL| 2.3d~\cite{riehn2020hadronic}, \verb|QGSJET 2.04|~\cite{ostapchenko2011monte} and a dedicated forward physics tune of \verb|PYTHIA 8.3|~\cite{Fieg:2023kld}, used as alternative MC generators to estimate the corresponding flux uncertainty.

For background studies, described in detail in \cref{sec:BkgEstimation}, various MC samples are used. Neutrinos are produced upstream of FASER through light- and charm-hadron decays and can then undergo charged- and neutral-current interactions in FASER. The neutrino fluxes were obtained using the fast neutrino flux simulation presented in Ref. \cite{Kling:2021gos}, with adjustments made to match the LHC’s configuration during Run 3 \cite{neutrinofluxPaper}.
Following the recommendations detailed in Ref.~\cite{neutrinofluxPaper}, the central prediction for the neutrino flux from light hadrons is based on \verb|EPOS-LHC|, with systematic uncertainties estimated by the spread of generator predictions from \verb|SIBYLL|, \verb|QGSJET| and \verb|Pythia|. For charm hadrons, the \verb|POWHEG+Pythia| prediction is used, with uncertainties from scale variations~\cite{Frixione_2007}. 
 Neutrino interactions in FASER were simulated with \verb|GENIE|~\cite{GENIE:2021wox,Andreopoulos:2009rq,Andreopoulos:2015wxa}. For each of the nominal samples and variations, 10~ab$^{-1}$ of events were  simulated. The Bodek-Yang model \cite{Bodek_2002, Bodek_2005, Bodek:2010km} employed in \verb|GENIE| agrees with more recent cross section models \cite{Candido:2023utz,Jeong:2023hwe} to within $6\%$  over the range of energies of interest \cite{neutrinofluxPaper}. 
Two dedicated high-energy muon simulation samples were used to aid in the evaluation of neutral-hadron and large-angle muon backgrounds (see \cref{sec:BkgEstimation}), both from dedicated \texttt{FLUKA} simulations of the production and propagation of muons produced from the collision products in IP1 ~\cite{Battistoni:2015epi, Ferrari:2005zk, Bohlen:2014buj}. The two muon samples used are: a \texttt{FLUKA} sample with $150\times10^6$ muons from 200 million \textit{pp} collisions, and a dedicated \texttt{FLUKA} sample of $4\times10^5$ muons simulated at large angles and radii. The \texttt{FLUKA} setup includes a realistic LHC infrastructure model between IP1 and FASER, validated by LHC data. The first sample contains muons entering FASER from IP1, while the second consists of muons generated upstream of the vetoNu stations, with radii between 9 and 25~cm. Additional muon simulation samples were used for systematic uncertainty evaluation.

To simulate the particles, \texttt{GEANT4} \cite{GEANT4:2002zbu} is used to simulate the propagation and interactions within the FASER detector. It includes a realistic detector geometry, including passive material. An additional 8.8\% correction factor is applied to the calorimeter EM energy based on testbeam studies, aligning the calibrated MC energy with the testbeam data, following the procedure of Ref.~\cite{FASER:2023tle}.  

\section{Event Reconstruction and Event Selection\label{sec:EventSelection}}

The reconstruction of data events recorded in FASER uses the Calypso~\cite{calypso} framework, based on the open-source ATLAS Athena reconstruction software~\cite{ATL-PHYS-PUB-2009-011,athena}. The total charge deposit \footnote{A charge deposit is defined as the measured charge in the photomultiplier tube (PMT) when particles lose energy and generate light in the scintillators.} of the calorimeter and scintillator signals is extracted by summing the digitised PMT pulse values post-pedestal subtraction. Simulation samples follow the same reconstruction as data. The detector response is calibrated using muons (as minimum-ionising-particles, MIPs), with charges being converted to EM energy/ MIP-equivalent (nMIPs) for the calorimeters or scintillators, respectively.

The typical signature of a signal event in the detector is shown in \cref{fig:ALP_sketch} for ALPs, and can be summarised as: 

  \begin{itemize}
  [itemsep=0.03cm, topsep=0.15cm, leftmargin=1.5em]
  \item No signal is observed in the veto and timing scintillators, since ALPs are electrically neutral.
  \item Preshower charge deposits consistent with an EM shower arising from the decay photons.
  \item A large energy deposit in the calorimeter left by the high-energy photon pairs.
  \end{itemize}

The same topology is expected for all the signal models considered in this paper. 
To avoid any bias in the analysis, a blinding methodology was implemented on the data sample. A ``blinded'' region was defined as events with a limited
deposited charge in any of the veto scintillators and calorimeter energy surpassing 100 GeV. Event selection, background estimation, and the consideration of systematic uncertainties were all finalised prior to investigating this blinded region.

\begin{figure}[htb]
\centering
\includegraphics[width=0.95\textwidth]{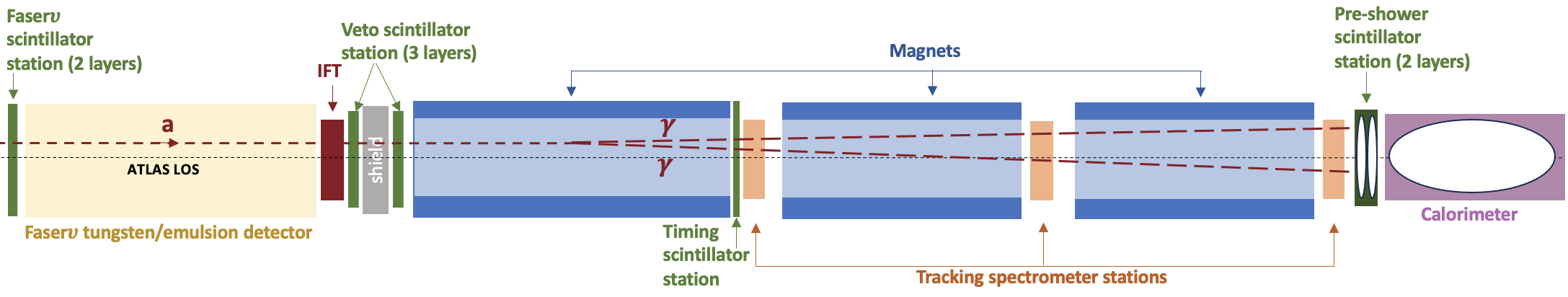}
\caption{Sketch of an ALP traversing through the FASER detector. As ALPs are electrically neutral, no signature is expected in any of the veto scintillator stations, followed by signal in the preshower and a large deposit in the calorimeter. The dotted lines show that the ALP leaves no signal in the detector and the white blobs in the preshower layers and the calorimeter depict energy deposits. The ALP is allowed to decay within the sensitive detector volume comprising the region across all three magnets, with a total length of 4.15~m with a 10~cm radius.} 
\label{fig:ALP_sketch}
\end{figure}

Following the above, the event selection requires events triggering the calorimeter and in time with the collision timing. Only events corresponding to colliding bunches are selected, with a requirement on the calorimeter timing to be $> -5$~ns and $<10$~ns to ensure consistency with collision timing. These times are with respect to the expected collision time, and are calibrated on a run-by-run basis using muon events.

Since no veto signal is expected from signal events, the charge deposited in each of the five veto scintillator stations is required to be less than half that expected from a MIP ($<40$ pC). Similarly, the charge deposited in the timing scintillator is required to be less than half a MIP too (resulting in a selection of $<20$ pC).%

The two (or more) photons produced in signal particles’ decays cannot be resolved, hence selections on the overall charge deposits in the preshower and calorimeter are applied. A positive, non-zero charge deposit is required in both the preshower layers, with the one in the second layer being greater than the charge deposit equivalent of 10 MIPs, and the preshower ratio (ratio of the charge deposited in the second preshower layer to the charge deposited in the first preshower layer, also referred to as PS ratio) being greater than 4.5. This is because the photons are expected to shower in the preshower material, releasing more energy in the second layer, hence the preshower ratio is expected to be high. Additionally, these selections will filter out events with the first preshower layer having any negative or saturated charges. Lastly, a large calorimeter energy deposit above 1.5 TeV is required, where the calorimeter energy variable considered is the summed calorimeter energy across all four modules. 

The selections applied on data and MC are summarised in~\cref{tab:sum}. The event selection is designed to ensure high acceptance of ALPs that decay anywhere in the sensitive detector volume, and is equivalently effective for light scalars and new gauge bosons as predicted by the other models considered for reinterpretation. The efficiency to select events with signal particles decaying inside the calorimeter will be suppressed by the preshower selections. Timing scintillator charge selections will also reject any event where a photon converts before the timing scintillator.

\begin{table}[tbh]
    \centering
    \begin{tabular}{|c|}    
    
    \hline   
    \multicolumn{1}{|c|}{\textbf{Trigger and Data Quality}} \\  
    \hline
    \hline
     Selecting events with calorimeter triggers\\
    \hline
    Calorimeter timing ($> -5$~ns and $<10$~ns) \\
    \hline
    \multicolumn{1}{|c|}{\textbf{Baseline Selection}} \\   
    \hline
    \hline
    Veto/VetoNu Scintillators to have no signal ($<$ 0.5 MIPs) \\
    Timing Scintillators to have no signal ($<$ 0.5 MIPs) \\
    \hline
    \multicolumn{1}{|c|}{\textbf{Signal Region}} \\  
    \hline
    \hline
    \ \ Preshower Ratio to have EM shower in the Preshower ($>$ 4.5) \ \ \\
    Second Preshower Layer to have signal ($>$ 10 MIPs) \\
    Calorimeter to have a large energy deposit ($>$ 1.5 TeV)  \\
    \hline
\end{tabular}
\caption{Selections used in the analysis.}
\label{tab:sum}
\end{table}

Cutflows showing the fraction of events that pass the above selections are shown in \cref{tab:sec:cutflowALPW2} for a representative ALP-W model point. 
Across the ($m_a, g_{aWW}$) parameter space, in regions FASER is sensitive to, the efficiency of the preshower ratio requirement is between 75 and 80\%. The selection efficiency through the Veto Scintillator signal and Timing Scintillator signal is above 99\% and 97\%, respectively throughout the ALP-W parameter space.
Across the $a_W$ mass range, the efficiency of the second preshower layer selection is above 95\%. The efficiency of the calorimeter energy selection is mostly dependent on the coupling. For low masses and high couplings the efficiency is close to 99\%, whilst at higher masses the efficiency is between 30\% and 90\%. At low masses and low couplings this falls to less than 30\%, but this region is already largely excluded as seen in \cref{fig:AcceptancesAndEffiencies}. 

\begin{table}[h]
    \centering
    \begin{tabular}{|c|c|c|}    
    \hline
    Selection  & \ \ Efficiency \ \ & \ \ Cum. Efficiency \ \ \\ 
    \hline  
    \multicolumn{3}{|c|}{$m_{a} = 140$ MeV, $g_{aWW} = 2 \times 10^{-4}$ GeV$^{-1}$}\\ 
    \hline    
    Veto/VetoNu Scintillators to have no signal ($<$ 0.5 MIPs)  &   99.6\%  &     99.6\% \\  
    Timing Scintillators to have no signal ($<$ 0.5 MIPs)    &  97.8\%  &     97.4\%  \\ 
    Preshower Ratio to have EM shower in the Preshower ($>$ 4.5) &   85.7\% &     83.5\%  \\ 
    Second Preshower Layer to have signal ($>$ 10 MIPs)  &  98.6\% &     82.3\% \\ 
    Calorimeter to have a large energy deposit ($>$ 1.5 TeV)  &  91.6\% &     75.4\% \\ 
\hline 
\end{tabular}
\caption{MC cutflow for a representative ALP-W signal point with $m_{a} = 140$ MeV and $g_{aWW} = 2 \times 10^{-4}$~GeV$^{-1}$, showing the percentage of signal events passing each selection.}

\label{tab:sec:cutflowALPW2}
\end{table}
%
\section{Background Estimation \label{sec:BkgEstimation}}

Various sources of background are considered in the analysis. The primary background results from neutrino interactions within the detector. Other physics-related backgrounds may arise from neutral hadrons entering the detector, muons that bypass the veto scintillator systems as they enter the detector at an angle, or veto inefficiencies. Additionally, non-collision backgrounds originating from cosmic rays or beam-related events are also taken into account. The following subsections provide a detailed description and quantification of the neutrino interactions as well as the evaluation and checks of the other, ultimately negligible, background sources.

\subsection{Neutrino Background}

Neutrinos are produced upstream of FASER through light- and charm-hadron decays and can then undergo charged and neutral current interactions within FASER. They will evade FASER's veto scintillator stations but interact in or near the preshower or the calorimeter, resulting in a possibly significant background contribution due to minimal upstream activity, resembling signal events. For neutrino interactions to produce $>$1.5~TeV in the calorimeter, the neutrino interaction must be a charged current $\nu_e$ interaction producing a high-energy electron (typically with ~70\% of the neutrino energy), or the large EM energy can come from a very energetic $\pi^0$ from the hadronic side of a charged current or neutral current neutrino interaction.

This background is evaluated using MC simulations (see \cref{sec:SamplesAndReco}), which are validated in regions designed specifically to target neutrinos interacting in different areas of the detector. In addition, FASER measurements of high energy neutrinos \cite{FASER:2024hoe} shows good agreement of data with the simulation predictions, albeit with large uncertainties. The MC predictions of neutrino interactions in the signal regions for the targeted dataset are summarised in \cref{tab:neut_comp}. The  prediction is shown split by neutrino flavour. Systematic uncertainties, as discussed in \cref{sec:Uncertainties}, are also reported. As they can be asymmetric, when combined they are first symmetrised by taking the maximum absolute variation from the nominal.

\begin{table}[tbh]
    \centering
    \begin{tabular}{|c|c|}
    \hline
    \multicolumn{2}{|c|}{$>$ 1.5 TeV signal region} \\ \hline
$\nu_{e}$ & 0.34 $\pm$ 0.33 (flux) $\pm$ 0.11 (exp.) $\pm$ 0.05 (stat.) \\
$\nu_{\mu}$ & 0.10 $\pm$ 0.05 (flux) $\pm$ 0.05 (exp.) $\pm$ 0.02 (stat.) \\
\textbf{Total} & \textbf{0.44 $\pm$ 0.39 (88.6\%)} \\ \hline

    \end{tabular}
\caption{Summary of the MC estimate of the neutrino background in the signal region. MC statistical uncertainties, uncertainties on the flux, as well as experimental uncertainties, further discussed in \cref{sec:Uncertainties}, are also given. The numbers are normalised to 57.7 fb$^{-1}$.}
\label{tab:neut_comp}
\end{table}

As described in \cref{sec:EventSelection}, only small charge deposits are expected in the veto, vetoNu, and timing scintillators for an ALP decaying in FASER. The neutrino background is therefore studied after the baseline selection requirements on those charge deposits described in \cref{tab:sum}. 

Imposing requirements on the preshower variables has been shown to provide selection power among neutrinos interacting in different parts of the detector.
Effective distinction between neutrinos interacting in the magnet, calorimeter, and preshower is achieved through selections on the charge deposited in the second preshower layer and the ratio of deposits of the two preshower layers. Neutrinos interacting in the magnet material deposit high charges in the second preshower layer and have a PS ratio around one when large calorimeter energy is also required, whereas neutrinos interacting in the calorimeter material deposit lower charges in the second preshower layer and have a broader PS ratio range.  Those interacting in the preshower material closely resemble signal events, posing a challenging background. Three regions are therefore defined for the validation of the background estimates on the basis of the above categorization, as illustrated in \cref{fig:neut_regions}. In the following, they are referred to as the ``magnet'', ``calorimeter'' and ``preshower'' regions, with the preshower region at high-energy representing the signal region. A fourth region, labelled as ``Other'', is in between the ``calorimeter'' and ``magnet'' regions with PS ratio between 1.5 and 4.5. In the design of the validation regions, particularly the ``magnet'' and ``calorimeter'' regions, signal contamination was taken into account. That is found to be below 30\% for the ALP-W parameter space not previously excluded. Even though this contamination can be significantly higher for the additional new physics models considered for reinterpretation, 
MC predictions are in agreement with observed data, and no background normalisations capable of absorbing a potential signal contribution are in use within this analysis. 
\begin{figure}
\centering  
\includegraphics[width=0.75\textwidth]{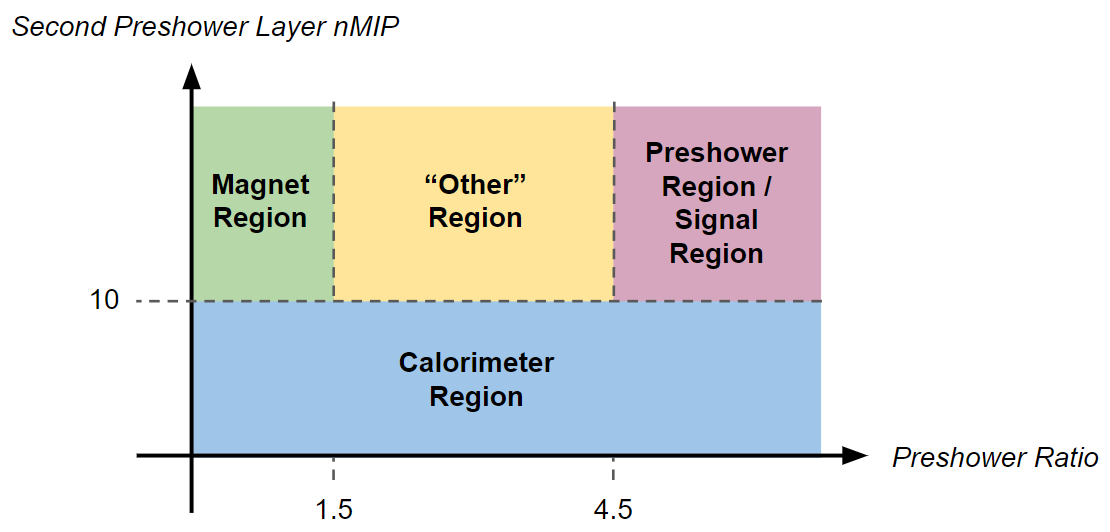}
\caption{Schematic of the regions used in the analysis. The signal region is a high-energy subset of the preshower region, with an additional energy requirement of at least 1.5 TeV.}
\label{fig:neut_regions}
\end{figure}

\cref{fig:neut_scatter_rz} shows the distribution of the location of the neutrino interaction vertex in the $(z,r)$ plane in simulation, as well as the decay vertex of a representative ALP-W signal model with $m_a = 120$~MeV and $g_{aWW} = 10^{-4}~\text{GeV}^{-1}$. The radius $r=\sqrt{x^2+y^2}$ is the distance from the central detector axis. Events with a minimum calorimeter energy deposit of 100 GeV are shown in the left column of \cref{fig:neut_scatter_rz}, and events with a larger than 1 TeV energy deposit in the right column. Neutrinos of both electron and muon flavour can be seen interacting throughout the detector volume, in particular in detector areas with larger material density such as the magnet, preshower scintillator system, and the calorimeter.
The ``calorimeter'' region is shown in the first row of \cref{fig:neut_scatter_rz}, clearly favouring neutrinos interacting within the calorimeter volume. In the second row of \cref{fig:neut_scatter_rz}, the background composition in the ``magnet'' region is given. The last row highlights the ``preshower'' and signal regions, dominated by neutrinos interacting in the preshower and, at higher energies, dominated by $a_W$ decays.

\begin{figure}[tbp]
\centering

\includegraphics[width=0.45\textwidth]{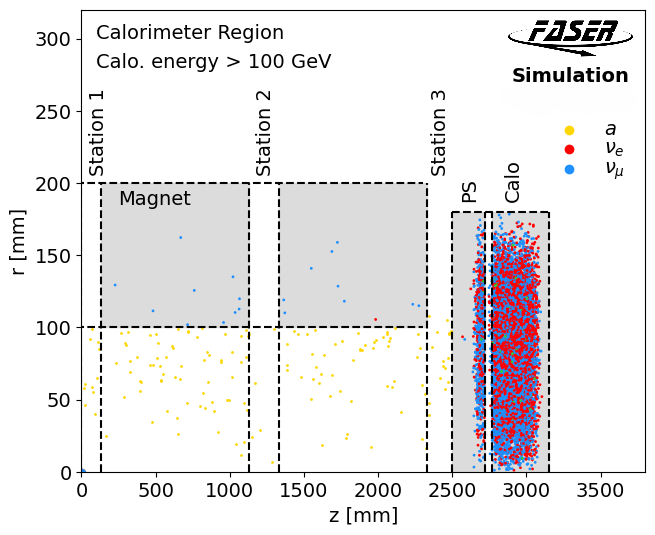}
\includegraphics[width=0.45\textwidth]{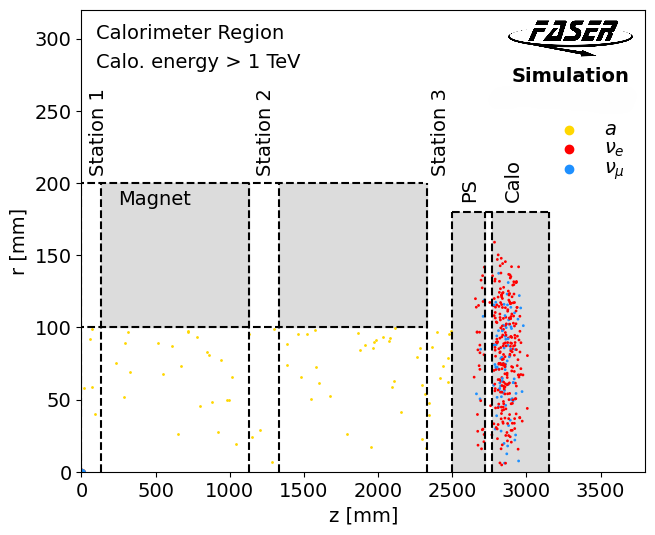}
\includegraphics[width=0.45\textwidth]{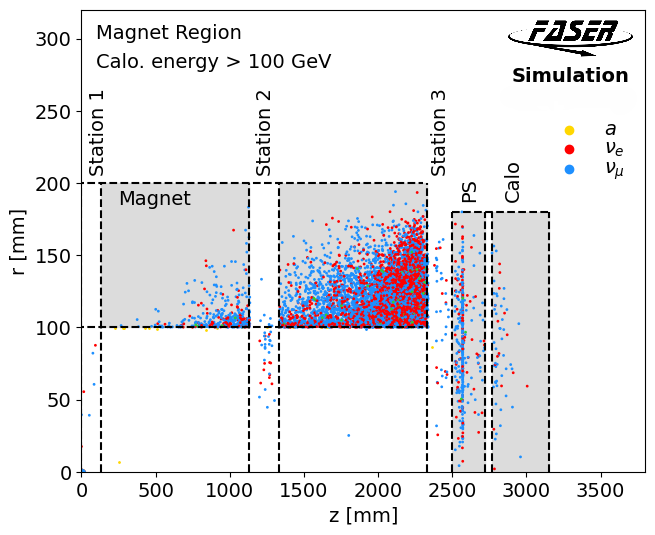}
\includegraphics[width=0.45\textwidth]{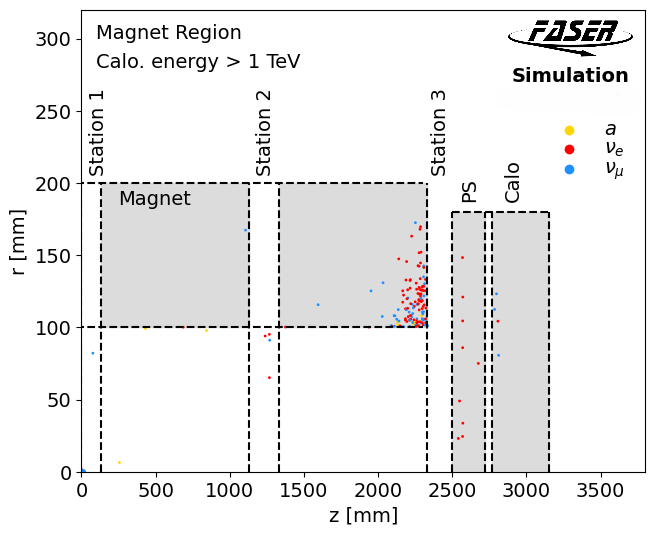}
\includegraphics[width=0.45\textwidth]{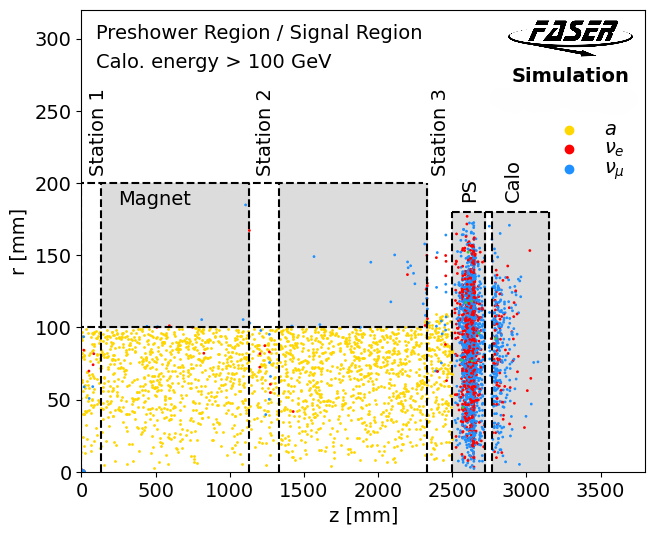}
\includegraphics[width=0.45\textwidth]{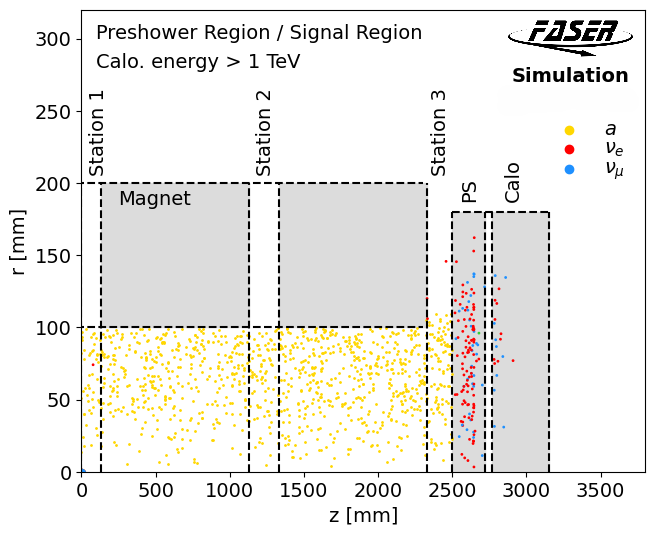}
 
\caption{Distributions in the $(z,r)$ plane of the neutrino (ALP-W, $a$) interaction (decay) vertex within the FASER detector for the different regions with different requirements on the calorimeter energy. Only the part of the detector downstream of the decay volume and timing scintillator station is shown.}
\label{fig:neut_scatter_rz}
\end{figure}

The ``magnet'' and ``calorimeter'' regions show high efficiency ($>80\%$) and purity among neutrino events ($>90\%$) for selecting true neutrinos interacting in the magnet and calorimeter, respectively. The efficiency is defined as the percentage of target neutrinos that are selected, and the purity is defined as the percentage of target neutrinos selected relative to all neutrinos selected. The ``preshower'' region's efficiency is $<40\%$, but this achieves a purity of around $80\%$ in neutrinos interacting in the preshower material. In combination with the ``calorimeter'' and ``Other'' region, this offers a good validation of all neutrino compositions.
\begin{figure}[htb]
\centering

\includegraphics[width=0.49\linewidth]{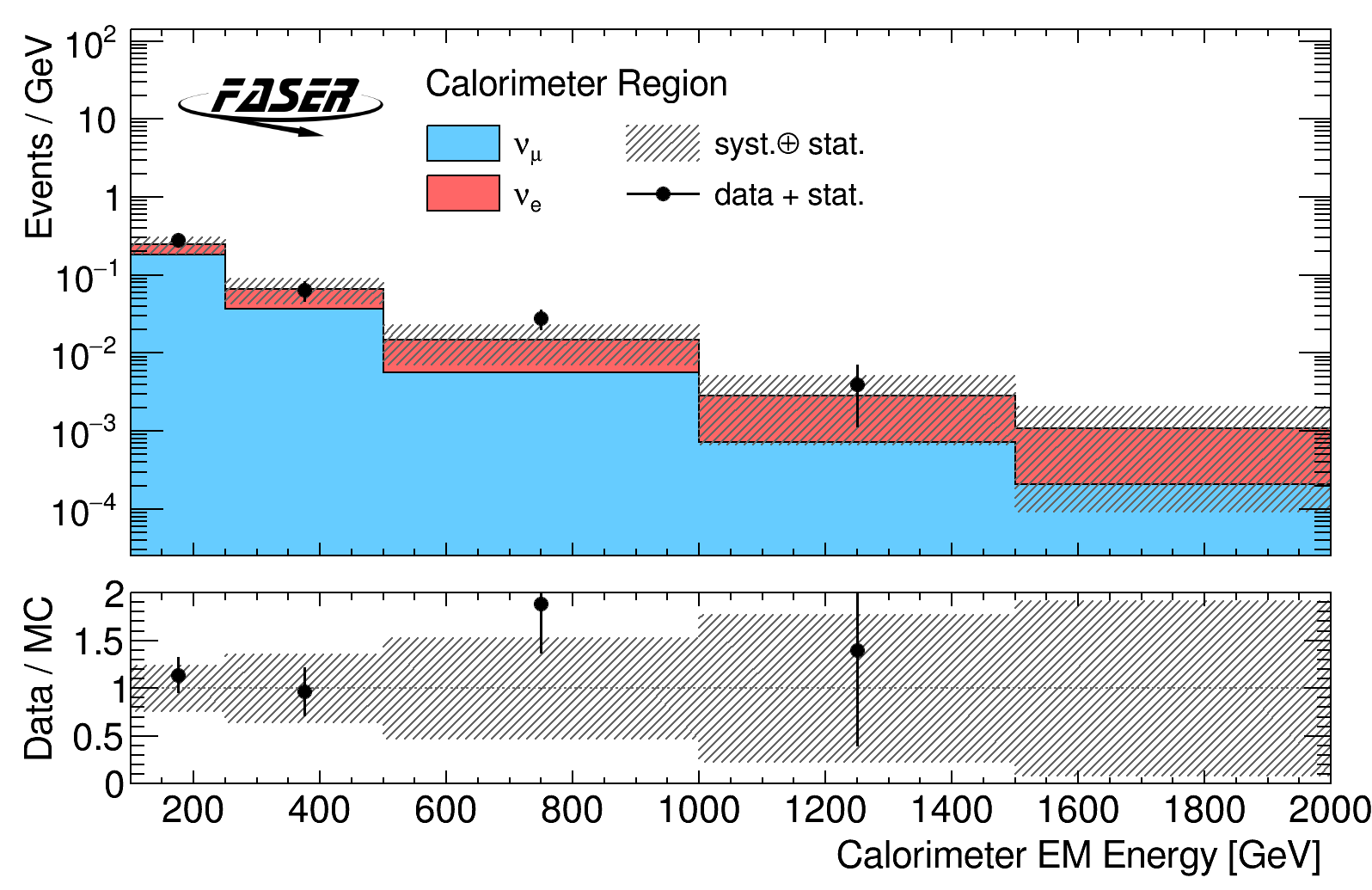}
\includegraphics[width=0.49\linewidth]{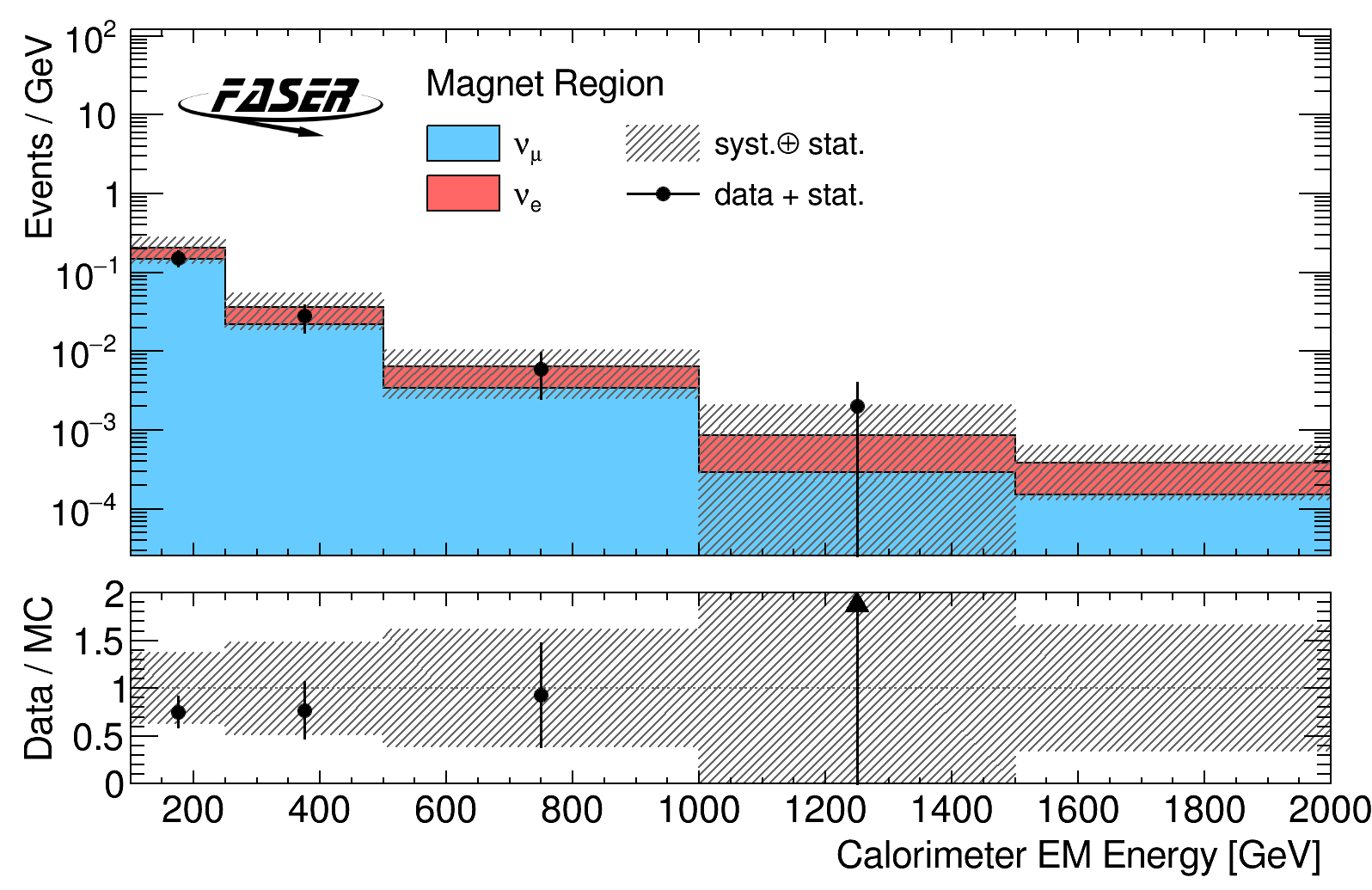}

\caption{Calorimeter energy distributions in the ``calorimeter'' (left) and ``magnet'' (right) region. The uncertainty band includes MC statistical uncertainties, experimental uncertainties, and uncertainties on the neutrino background flux. The last bin contains all events above 1.5 TeV.\label{fig:neutrinoVRs} Equivalent distributions split by neutrino production mechanism can be found in \cref{app:FlavourBreakdown}. \label{app:fig:neutrinoVRs:flav}
}
\end{figure}

\cref{tab:VRYieldsTable} shows the observed number of data events in each validation region, as well as the expected neutrino contribution, split in terms of $\nu_e$ and $\nu_\mu$, as predicted by MC. The MC describes the data well in all regions within statistical and systematic uncertainties, the latter dominated by the uncertainty on the neutrino flux modelling, which is around $30\%$ to $60\%$. A breakdown of the composition by neutrino production mechanism can be found in Appendix~\ref{app:FlavourBreakdown}.

\begin{table}[htp]
    \centering
    \begin{tabular}{|l|c|} \hline
\multicolumn{2}{|l|}{``Preshower'' region}\\ \hline
$\nu_{e}$ & \ \ 5.4 $\pm$ 2.7 (flux) $\pm$ 0.6 (exp.) $\pm$ 0.2 (stat.) \ \ \\
$\nu_{\mu}$ & 13.2 $\pm$ 2.4 (flux) $\pm$ 1.8 (exp.) $\pm$ 0.3 (stat.) \\
Total  & \textbf{18.7 $\pm$ 5.4 (29.1\%)} \\  
Data & \textbf{15}\\ \hline
\multicolumn{2}{|l|}{``Calorimeter'' region}\\ \hline
$\nu_{e}$ & 22.7 $\pm$ 12.8 (flux) $\pm$ 0.7 (exp.) $\pm$ 0.4 (stat.) \\
$\nu_{\mu}$ & 40.0 $\pm$ 6.9 (flux) $\pm$ 2.8 (exp.) $\pm$ 0.5 (stat.) \\
Total &  \textbf{62.9 $\pm$ 19.7 (31.4\%)}\\
Data & \textbf{74}\\ \hline
\multicolumn{2}{|l|}{``Magnet'' region}\\ \hline
$\nu_{e}$ & 13.9 $\pm$ 10.3 (flux) $\pm$ 1.4 (exp.) $\pm$ 0.3 (stat.) \\
$\nu_{\mu}$ & 29.5 $\pm$ 8.0 (flux) $\pm$ 3.8 (exp.) $\pm$ 0.4 (stat.) \\
Total &  \textbf{43.6 $\pm$ 18.3 (41.9\%)}  \\
Data & \textbf{34}\\ \hline
\multicolumn{2}{|l|}{``Other'' region}\\ \hline
$\nu_{e}$ & 6.5 $\pm$ 3.6 (flux) $\pm$ 0.8 (exp.) $\pm$ 0.2 (stat.) \\
$\nu_{\mu}$ & 15.3 $\pm$ 2.7 (flux) $\pm$ 2.2 (exp.) $\pm$ 0.3 (stat.) \\
Total &  \textbf{21.9 $\pm$ 7.0 (32.1\%)} \\
Data & \textbf{17}\\ \hline
    \end{tabular}
\caption{Breakdown of the neutrino composition and data yields in the ``magnet'', ``Other'', ``calorimeter'' and ``preshower'' regions (excluding events passing signal region selections). Listed are data and neutrino yields as predicted from MC in 57.7~$\ifb$, split by electron and muon neutrino components. Uncertainties on the flux as well as experimental uncertainties further discussed in~\cref{sec:Uncertainties} are also given. \label{tab:VRYieldsTable}}
\end{table}

\cref{fig:neutrinoVRs} shows the energy distribution in the ``calorimeter'' region and the ``magnet'' region, comparing data to MC predictions. The ``calorimeter'' region is dominated by muon neutrinos, particularly at lower energies. The ``magnet'' region is dominated by muon neutrinos produced in light hadron decays, showing good agreement between data and MC within uncertainties.
 The ``preshower'' region most closely emulates the neutrino event topologies in the signal region, serving as additional validation of the background modelling. The high-energy tail includes events from the signal region, and it is shown and further discussed in \cref{sec:Results}.

\subsection{Other Background Contributions}

\paragraph*{\bf Veto System Inefficiency}

One potential source of background events is muons traversing the detector volume. FASER’s veto system prevents significant amounts of muons from contaminating the regions of interest. Possible inefficiencies of each of the five veto scintillators are assessed using data by specifically choosing events with one well-reconstructed track passing through all the veto scintillators. For each individual plane, the inefficiency is determined by counting the proportion of events with signal corresponding to half of a MIP crossing the scintillator. This background is considered to be negligible based on the measured per-layer inefficiencies of the scintillators being below $10^{-5}$ in 2022 and 2023 data and the expected number of muons being of the order of $10^8$.  It has been assumed that the scintillator inefficiencies are uncorrelated since the scintillators and high voltage supplies are independent for the veto layers. 

\paragraph*{\bf Background from Large-Angle Muons}

Muons traversing the detector volume at a large angle might miss the veto system, interact, and deposit high-energy within the calorimeter. Unlike in the search for dark photons~\cite{FASER:2023tle}, there are no requirements for tracks associated to the muon that can be used to veto such background, hence several checks are performed on MC and data to ensure it is negligible. Dedicated MC samples have been produced (see \cref{sec:SamplesAndReco}) of muons traversing FASER using \texttt{FLUKA}. %
No such MC events pass the selections applied. 
A partially data-driven method was used to validate the MC results by selecting events with inverted veto scintillator charge requirements in bins of calorimeter energy.
To increase the number of events in the selected regions, inverted timing scintillator requirements were also  applied. Events were then split in two regions, depending on whether they passed the preshower ratio selection or not. In the former case, the selection closely replicates that of the signal region, but the number of events is low and the regions are largely contaminated by neutrino events. In the latter case, a larger sample of muons is selected, mimicking large-angle muon events albeit significantly differently from signal region events. By considering estimates from both regions, one providing an upper limit and the other leading to an overly conservative estimate, it was determined that the contribution of large-angle muons is 9 to 40 times smaller than that of neutrinos in the signal region. As this leads to an estimate of at most $\mathcal{O}(0.01)$ events, this background contribution is considered negligible in the analysis.

\paragraph*{\bf Neutral Hadron Background }

Background events from neutral hadrons generated in muon interactions within the rock in front of FASER may occur if these neutral hadrons pass through the veto system and subsequently interact or decay within the detector decay volume, resulting in signal in the preshower and a high calorimeter energy deposit. The likelihood of this background is significantly reduced by the requirements that the neutral hadron traverse the entire eight interaction lengths of the FASER$\nu$ detector and that the parent muon scatters in a way that avoids the veto scintillators. Additionally, the high calorimeter energy requirement suppresses neutral-hadron background events further. In the dark photon analysis~\cite{FASER:2023tle}, the estimate of the neutral hadron background passing the analysis selections was found to be negligible. In the present analysis, the veto scintillator requirements are the same, and the calorimeter energy requirement is 1.5~TeV, compared to 500~GeV in the dark photon analysis. Hence, such background is considered negligible.

\paragraph*{\bf Non-Collision Background}

For non-collision background studies, events were collected during periods without colliding bunches in IP1. Cosmic data worth 33 days of beam-free data-taking were examined, equivalent to the full 2022 and 2023 physics data-taking duration. No events with calorimeter energy deposits exceeding 100 GeV were observed, 
and only 9 events, irrespective of energy, pass the baseline selections outlined in \cref{tab:sum}. This indicates the negligible impact of cosmic-ray events when considering the other analysis requirements. \\
\indent Beam background from LHC Beam 1 (B1), the incoming beam to ATLAS in the FASER location, is the most relevant non-collision background for FASER. Potential detector activity arises from beam-gas interactions or beam tails interacting with the beampipe aperture, resulting in particles coming from a direction where FASER is much less shielded. Low-energy activity can be observed in FASER correlated with B1 bunches passing the back of the detector, 127 bunch-crossings before particles from the collisions of the same bunch would be recorded in FASER. This beam background is studied by checking the detector activity in events with bunch crossing identifiers corresponding to  proton bunches in LHC B1 passing the back of FASER, but which do not correspond to colliding bunches at IP1. Despite observing events with no signal in the veto scintillators and timing scintillators, beam background is suppressed to a negligible level when calorimeter timing selections are applied. Studies of collision timing and beam background show that the timing of such B1 events and collision events are well separated. The timing selection outlined in \cref{sec:EventSelection} effectively removes all B1 background.

As a result, the overall contribution from non-collision backgrounds, including cosmic rays and beam background, is deemed negligible.

\section{Systematic Uncertainties \label{sec:Uncertainties}}

Systematic uncertainties arise from various sources and apply to both expected signal yields and background estimates. 

The largest source of uncertainty that affects signal yields is the uncertainty associated with the modelling of the flux of SM particles produced in the forward direction of the LHC, from which signal events are produced \cite{HeavyHadronPrescription,Fieg:2023kld}. 
Uncertainties on the production of B-mesons are obtained from scale variations, while uncertainties on light hadron production are estimated using the spread of generators compared to the “central value” provided by the \verb|EPOS-LHC| prediction. The shift in the yields for each component is added in quadrature. An additional 20\% uncertainty, to account for uncertainties in the modeling of the $B\rightarrow X_{s}a$ branching ratio \cite{HeavyHadronPrescription}, is also included in the signal theoretical systematic uncertainties. As an example, the overall theoretical uncertainties on the $a_W$ signal yields are around 60\%.

The experimental uncertainties on the signal yields arise from the modelling of the detector response in the MC simulation. The uncertainty on the calorimeter energy scale calibration is evaluated following the procedure used in Ref.~\cite{FASER:2023tle}, that considers the difference in the calibration of the energy scale between data and MC simulation. It is quantified as around 6\% across the energy range of interest, and its effect on the signal yield uncertainty varies across signal models. 

Correction factors are evaluated for the simulated response of the preshower, both in terms of the second preshower layer charge deposit, measured in terms of energy deposit of a MIP or nMIP, and the preshower ratio. 
Photon conversion events from FASER data as well as single-particle electron events from testbeam data are used to obtain the correction factors. These are estimated through fitting the deposited energy distributions for each layer and extracting the most probable value for data and MC. The latter are then corrected to match the data. The correction factors are 1.20 and 1.13 for the second preshower layer nMIP and the PS ratio, respectively, and show no dependence on the energy. The systematic uncertainties on the correction factors for the second preshower layer MIP-equivalent charge deposit and PS ratio are evaluated considering the difference between the factors obtained with FASER and the testbeam data and are 20\% for the second preshower layer and 12\% for the PS ratio. The resulting effect on the yields is evaluated as a shift in the respective parameter values used in region definitions and is of the order of 1\% and 5\% for the second preshower layer and PS ratio, respectively. All above mentioned uncertainties were evaluated for each signal point in the ALP-W parameter space, as well as for other models considered for reinterpretation.
Overall, experimental uncertainties on the calorimeter energy scale and preshower-related quantities are $\mathcal {O}(20\%)$ on the signal yield.

The luminosity uncertainty for the 2022 and 2023 dataset was taken from ATLAS and is quoted to be $2.0 \%$ \cite{ATL-DAPR-PUB-2023-001,ATL-DAPR-PUB-2024-001, Avoni:2018iuv, DAPR-2021-01} and was applied to both data and MC. The statistical uncertainty from the number of MC simulated signal events is also included and ranges from 1 to 4\%.

The impact of changes in the beam half-crossing angle on the signal and background yields was also studied as a potential source of uncertainty. These small changes in the crossing angle caused a shift in the LOS position in FASER, leading to a potential $\sim$ 7\% variation in the expected signal and neutrino background yields. However, considering the luminosity fractions corresponding to different crossing angles, the overall effect on the full dataset is below 1\% for signal and background. Given that the theoretical uncertainty on the signal and neutrino background yields is above \(\mathcal{O}(50\%)\), this small effect is negligible and is not considered in the final analysis.

\cref{table:signal_unc} shows a break down of different theory and experimental  systematic uncertainties for three representative $a_W$ signal samples.

\begin{table}[tbp]
\centering
\begin{tabular}{|c||c|c|c|}   \hline
& \multicolumn{3}{c|}{Signal Samples} \\  
Uncertainties  & \makecell{$m_{a} = 140 \text{ MeV}$ \\ $g_{aWW} = 2 \times 10^{-4}$ GeV$^{-1}$} & \makecell{$m_{a} = 120 \text{ MeV}$ \\ $g_{aWW} = 1 \times 10^{-4}$ GeV$^{-1}$} & \makecell{$m_{a} = 300 \text{ MeV}$ \\ $g_{aWW} = 2 \times 10^{-5}$ GeV$^{-1}$ } \\ \hline
\multirow{3}{*}{Theo.} & \multirow{2}{*}{59.4\%} & \multirow{2}{*}{57.3\%} & \multirow{2}{*}{58.0\%}\\ 
 & \multicolumn{3}{c|}{} \\
 & \multicolumn{3}{c|}{theoretical uncertainties including flux and branching ratio} \\ \hline \hline
\multirow{3}{*}{Luminosity} &\multirow{2}{*}{2.0\% }  &\multirow{2}{*}{2.0\% } &  \multirow{2}{*}{2.0\% } \\ 
 & \multicolumn{3}{c|}{} \\
 & \multicolumn{3}{c|}{uncertainty on the luminosity estimate} \\ \hline 
\multirow{3}{*}{Calo E-scale} & \multirow{2}{*}{3.6\%} &\multirow{2}{*}{16.3\% } & \multirow{2}{*}{15.8\%} \\
& \multicolumn{3}{c|}{} \\
& \multicolumn{3}{c|}{uncertainty on calorimeter energy scale} \\ \hline 
\multirow{3}{*}{ PS ratio}  & \multirow{2}{*}{ 7.9\%} &\multirow{2}{*}{6.9\%} & \multirow{2}{*}{8.4\%} \\
& \multicolumn{3}{c|}{} \\
& \multicolumn{3}{c|}{uncertainty on the preshower ratio} \\ \hline 
\multirow{3}{*}{Second PS} & \multirow{2}{*}{0.6\%}& \multirow{2}{*}{0.6\%} &\multirow{2}{*}{0.6\%}  \\
& \multicolumn{3}{c|}{} \\
& \multicolumn{3}{c|}{uncertainty on the second preshower layer charge deposit} \\ \hline \hline
\multirow{3}{*}{ MC stat.} & \multirow{2}{*}{1.8\%} & \multirow{2}{*}{3.5\%} & \multirow{2}{*}{2.9\%} \\
& \multicolumn{3}{c|}{} \\
& \multicolumn{3}{c|}{statistical uncertainty of the MC sample} \\ \hline
\end{tabular}
\caption{The various sources of systematic uncertainties associated with the $a_W$ signal. Theory uncertainties include the uncertainty associated with the flux and branching ratio. Experimental uncertainties include the uncertainty on the luminosity, the calorimeter energy, the second preshower layer charge and the preshower ratio. These are shown for three $a_W$ MC signal points: $m_{a} = 140 \text{ MeV}, g_{aWW} = 2 \times 10^{-4}$ GeV$^{-1}$; $m_{a} = 120 \text{ MeV}, g_{aWW} = 10^{-4}$ GeV$^{-1}$; and $m_{a} = 300 \text{ MeV}, g_{aWW} = 2 \times 10^{-5}$ GeV$^{-1}$. The MC statistical uncertainty is also reported. \label{table:signal_unc}}
\end{table}

The main source of systematic uncertainty for the SM background arises from the theoretical modeling of the flux of neutrinos used to quantify the contributions due to their interactions. The composition of neutrinos arriving in FASER originates from both light and charm hadron decays as described in \cref{sec:BkgEstimation}. 
The hadron flux uncertainty is accounted for using the prescription of generator and scale variations detailed in Ref.~\cite{neutrinofluxPaper}. 
The overall impact on the background estimate for neutrino interactions is around 90\%. The same experimental uncertainties as detailed for the signal above are also applied to the neutrino background MC. Additionally, an uncertainty in modeling the preshower geometry, not relevant for the signal as non-interacting, is determined by estimating the amount of missing material in the MC simulation, which is approximately 6\%.

\section{Results} \label{sec:Results}

The total background expectation in the signal region is summarised in~\cref{tab:backgroundsummary}. Also shown are the expected yields for three benchmark ALP-W mass and coupling models representative of the parameter space targeted by this analysis. One event is observed in the signal region, consistent with the background-only hypothesis and within $0.6$ standard deviations of the SM expectation. This data event has a calorimeter energy of $1.6~\text{TeV}$, a preshower ratio of 9.0, and a large charge deposit in the second preshower layer. A visualisation of the event is given in ~\cref{app:eventdisplay}.

\begin{table}[tbh]
\centering
\begin{tabular}{|c|c|}
\hline
\bf{Source} & \bf{Event Rate}  \\
\hline
Neutrino Background &
$\begin{array}{rcl}
0.44&\pm& 0.35 \text{ (flux)} \\ 
     &\pm& 0.01 \text{ (Luminosity)}  \\
     &\pm& 0.15 \text{ (Calo E-scale)}  \\
     &\pm& 0.06 \text{ (PS ratio)} \\
     &\pm& 0.02 \text{ (second PS)} \\
     &\pm& 0.02 \text{ (PS geometry)} \\
     &\pm& 0.05 \text{ (MC stat.)} \\
 \textbf{Total: 0.44} &{\pm}& \textbf{0.39 (88.6\%)}
 \end{array}$
 \\
\hline
$a_W$ ($m_{a} = 140 \text{ MeV}, g_{aWW} = 2 \times 10^{-4}$ GeV$^{-1}$) &
$\begin{array}{rcl}
70.7 &\pm& 42.0 \text{ (theo.)} \\
     &\pm& 6.4 \text{ (exp.)} \\
     &\pm& 1.3 \text{ (MC stat.)} \\
 
\end{array}$ 
\\
\hline
$a_W$ ($m_{a} = 120 \text{ MeV}, g_{aWW} = 1 \times 10^{-4}$ GeV$^{-1}$) &
$\begin{array}{rcl}
91.1 &\pm& 52.2 \text{ (theo.)} \\
     &\pm& 16.2 \text{ (exp.)} \\
     &\pm& 3.2 \text{ (MC stat.)} \\

\end{array}$ 
\\
\hline
$a_W$ ($m_{a} = 300 \text{ MeV}, g_{aWW} = 2 \times 10^{-5}$ GeV$^{-1}$) &
$\begin{array}{rcl}
4.0 &\pm& 2.3 \text{ (theo.)} \\
    &\pm& 0.6 \text{ (exp.)} \\
    &\pm& 0.1 \text{ (MC stat.)} \\
 
\end{array}$ 
\\
\hline 
\bf{Data} & \bf{1} \\ \hline
\end{tabular}
\caption{Summary of the expected number of events for the neutrino background and three representative ALP-W models, along with the number of events observed in the experimental data. A breakdown of the different sources of experimental and theoretical systematic uncertainties is also provided.}
\label{tab:backgroundsummary}
\end{table}

\cref{fig:SRunblinded} shows the neutrino background expectation in the ``preshower'' and signal regions, with uncertainty bands including the experimental and theoretical systematic uncertainties. To illustrate the possible signal contributions, three representative $a_W$ signal predictions are overlaid. The signal region is dominated by electron neutrinos produced in light- and charm-hadron decays.

\begin{figure}[htp]
    \centering
\includegraphics[width=0.6\textwidth]{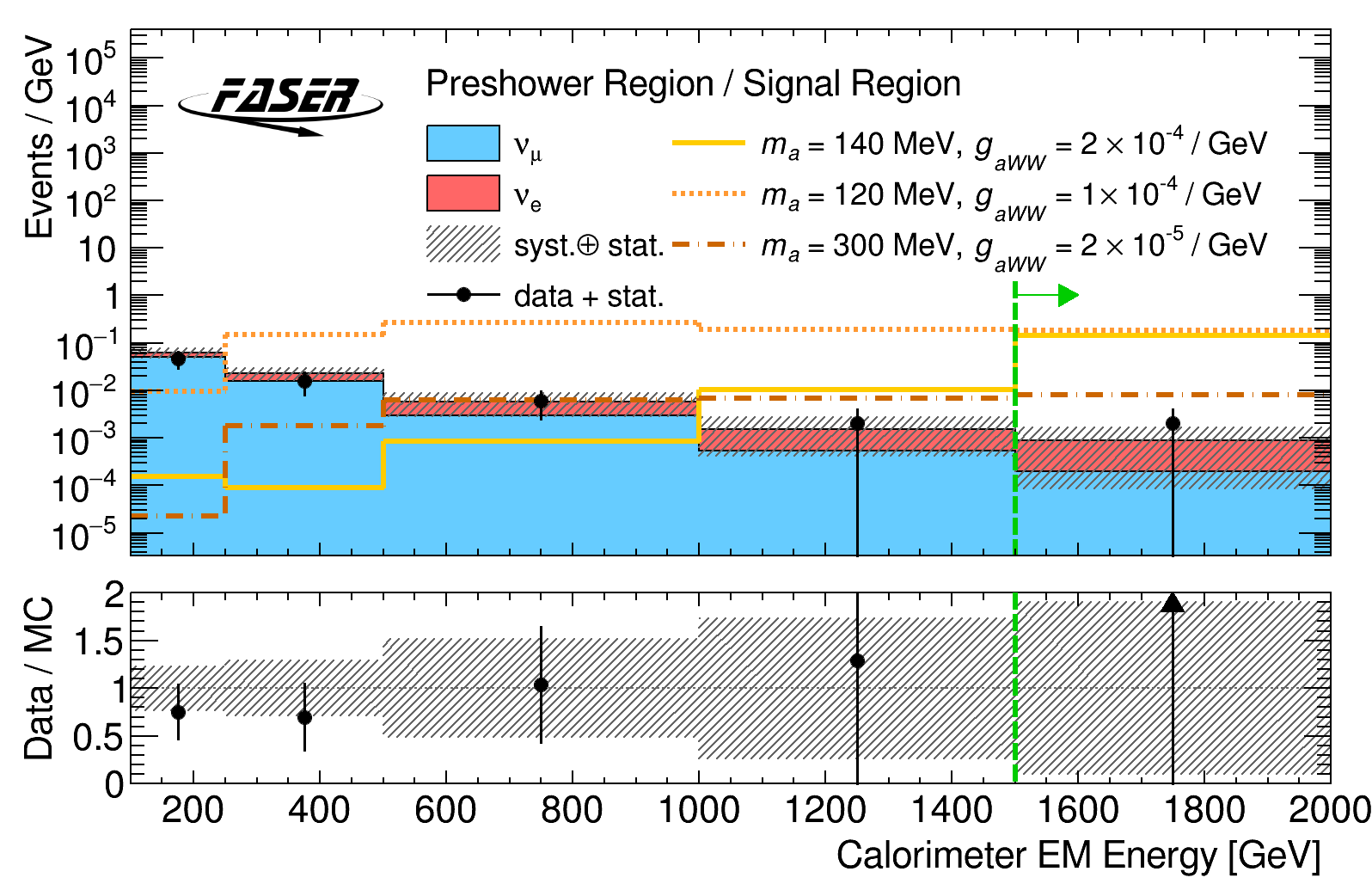}
    \caption{Calorimeter energy distribution in the ``preshower'' and signal regions, showing the neutrino background composition, separated according to neutrino flavour. The last high-energy bin above 1.5 TeV, highlighted with a green arrow, presents the signal region and includes the overflow. The neutrino background contributions, separated by neutrino production mechanism, are given in~\cref{app:FlavourBreakdown}.}
    \label{fig:SRunblinded}
\end{figure}

A statistical interpretation of the result has been performed following a profile likelihood estimation performed within the HistFitter statistical analysis framework~\cite{Baak:2014wma}. Following a convention of evaluating the $CL_s$~\cite{Read:2000ru} values at 90\% confidence level (C.L.), contours encompassing the excluded parameter space in the ALPs, up-philic, U(1)$_B$ and 2HDM models are produced. 

\cref{fig:ALPWLim} shows the excluded parameter space in the ALP-W coupling versus mass plane. Reported in grey are existing experimental exclusion limits from a wide range of experiments~\cite{izaguirre2016new, Gori:2020xvq, Kling_2020}. A detailed breakdown of the existing limits, shown here as joint excluded parameter space, can be found in ~\cref{app:existingconstraints}.
For ALPs coupled to weak gauge bosons, FASER is sensitive to previously unexplored parameter space with masses between 100 and 250 MeV and couplings ranging from $3 \times 10^{-5}$ to $5 \times 10^{-4}$~GeV$^{-1}$. Masses as heavy as 300 MeV can be excluded for a coupling of $2 \times 10^{-5}$ GeV$^{-1}$. This is complementary to searches at kaon factories which are sensitive to lower masses, less than 100 MeV.

\begin{figure}[tbp]
    \centering
    \includegraphics[width=0.7\textwidth]{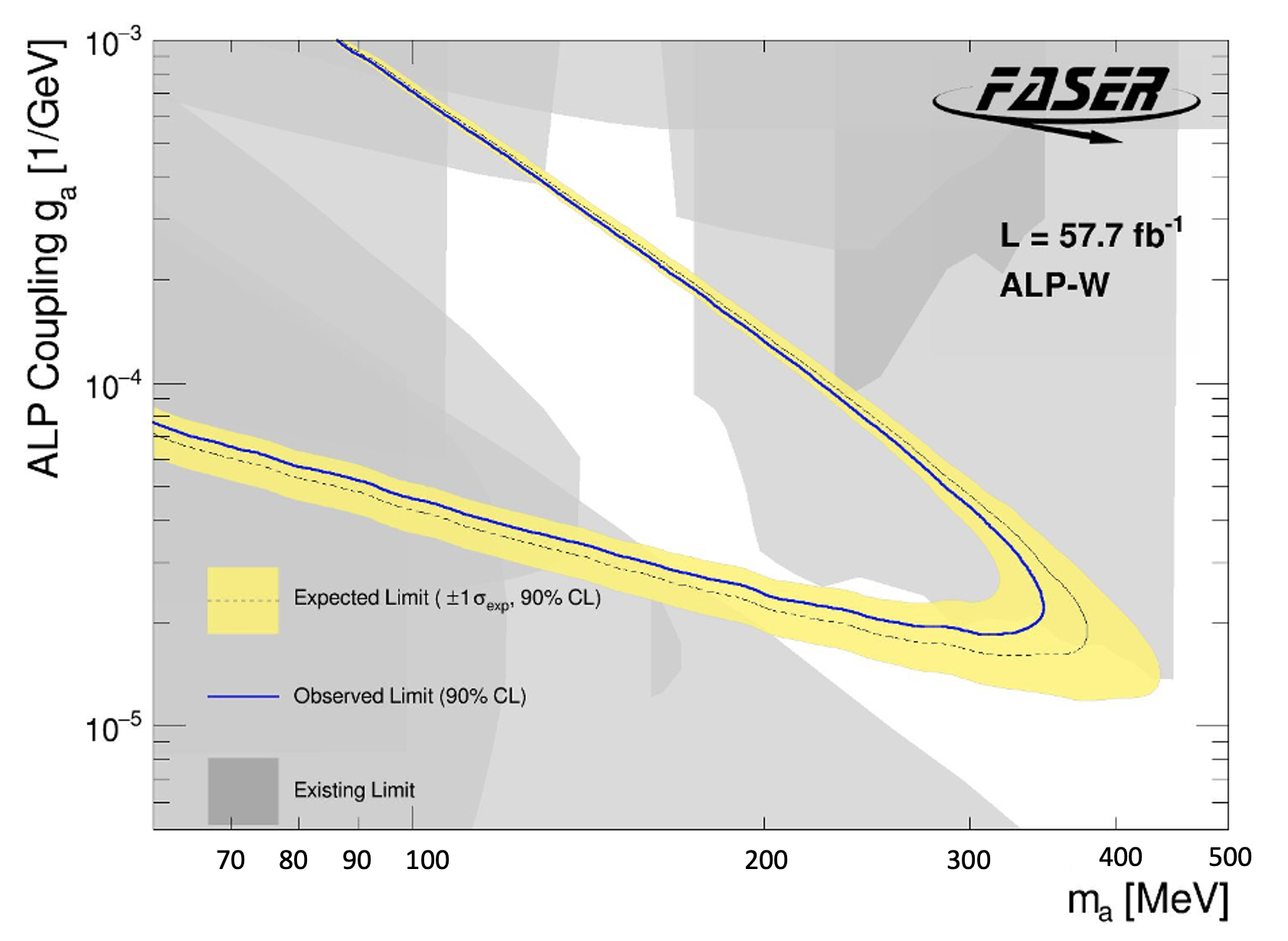}
    \caption{Interpretation of the signal region yield as ALP-W exclusion limits. Limits are provided at 90\% confidence level in the ALP~-W mass ($m_{a}$) and coupling ($g_{aWW}$) plane.}
    \label{fig:ALPWLim}
\end{figure}

Exclusion limits for the ALP-photon model are shown in the left panel of \cref{fig:ALPgluonPhoton}. FASER can probe ALP masses up to 80 MeV and couplings down to $10^{-4}$ GeV$^{-1}$, again demonstrating FASER’s complementary reach to existing constraints.

Results for the ALP-gluon model are shown in the right panel of \cref{fig:ALPgluonPhoton}. At low masses, ALP-gluons decay only to two photons, with hadronic decays such as $3\pi^0$ and $\pi^+ \pi^- \gamma$ occurring above approximately 0.4 GeV. FASER can probe some unconstrained regions near the $\pi^0$, $\eta$ mass where production rates are enhanced due to resonant mixing with these pseudo-scalars.

An interpretation of the signal region yield as limits in the parameter space of the U(1)$_B$ models is shown in the left panel of \cref{fig:U1BPlusUpPhiliv}, as a function of the new gauge boson mass and the coupling, $g_B$. At low masses, where \( m_{Z_B} < m_{\pi^0} \), the \( Z_B \) particle decays radiatively into electrons. As the mass increases, the decay channel \( Z_B \rightarrow \pi^0 \gamma \) becomes accessible, resulting in a final state with three photons. This decay mode remains dominant until \( m_{Z_B} \) reaches approximately 600 MeV, at which point the decay channel into \( \pi^0 \pi^+ \pi^- \) becomes possible. From 600 MeV onward, the small region of sensitivity is due to the \(\omega\) and \(\rho\) resonance, where the production rate of $Z_{B}$ is enhanced. The dashed lines represent exclusion limits from other experiments assuming enhanced rare $B$ and $K$ decays (e.g. $K\rightarrow \pi Z_B$ ). They are referred to as ``anomaly models"~\cite{Dror:2017ehi} because new heavier fields are introduced to resolve divergences such that meson decay rates increase, making the constraints qualitatively different from FASER constraints where such an assumption is not made.

The exclusion limit for the up-philic model is shown in the right panel of \cref{fig:U1BPlusUpPhiliv} as a function of the new scalar particle mass and its coupling, $g_u$. For $m_S > 2m_{\pi^0}$, higher LHC energies and boost factors enhance the sensitivity to shorter-lived up-philic scalars, particularly in the mass range $2 m_{\pi^0} < m_S < 2m_{\pi^+}$. This range allows probing $g_u$ values between the lower and upper constraints from previous searches. Previous searches sensitive to $S \to \pi^+ \pi^-$ decays did not constrain masses below this threshold, despite possible decays to neutral pions. Estimations for FASER take into account both charged and neutral pion decays.

Finally, \cref{fig:2HDMLim} shows the exclusion limits for the 2HDM models with a CP-even scalar Higgs decaying to two photons, $ H \rightarrow \gamma\gamma$. FASER can probe large values of tan $\beta \sim 1000 $ and masses of $H$ up to $B$-hadron masses of approximately 4~GeV, extending significantly into previously unconstrained regions. Larger luminosities can extend the reach to larger values of tan $\beta$ but the masses that can be probed are limited by the production mechanism, $B \rightarrow KH$.

This analysis highlights the complementarity of FASER with other experiments, probing regions of parameter space that were previously unexplored or poorly constrained. By searching for multi-photon final states, FASER is able to set stringent limits across multiple models.

\begin{figure}
    \centering
    \includegraphics[width=0.49\linewidth]{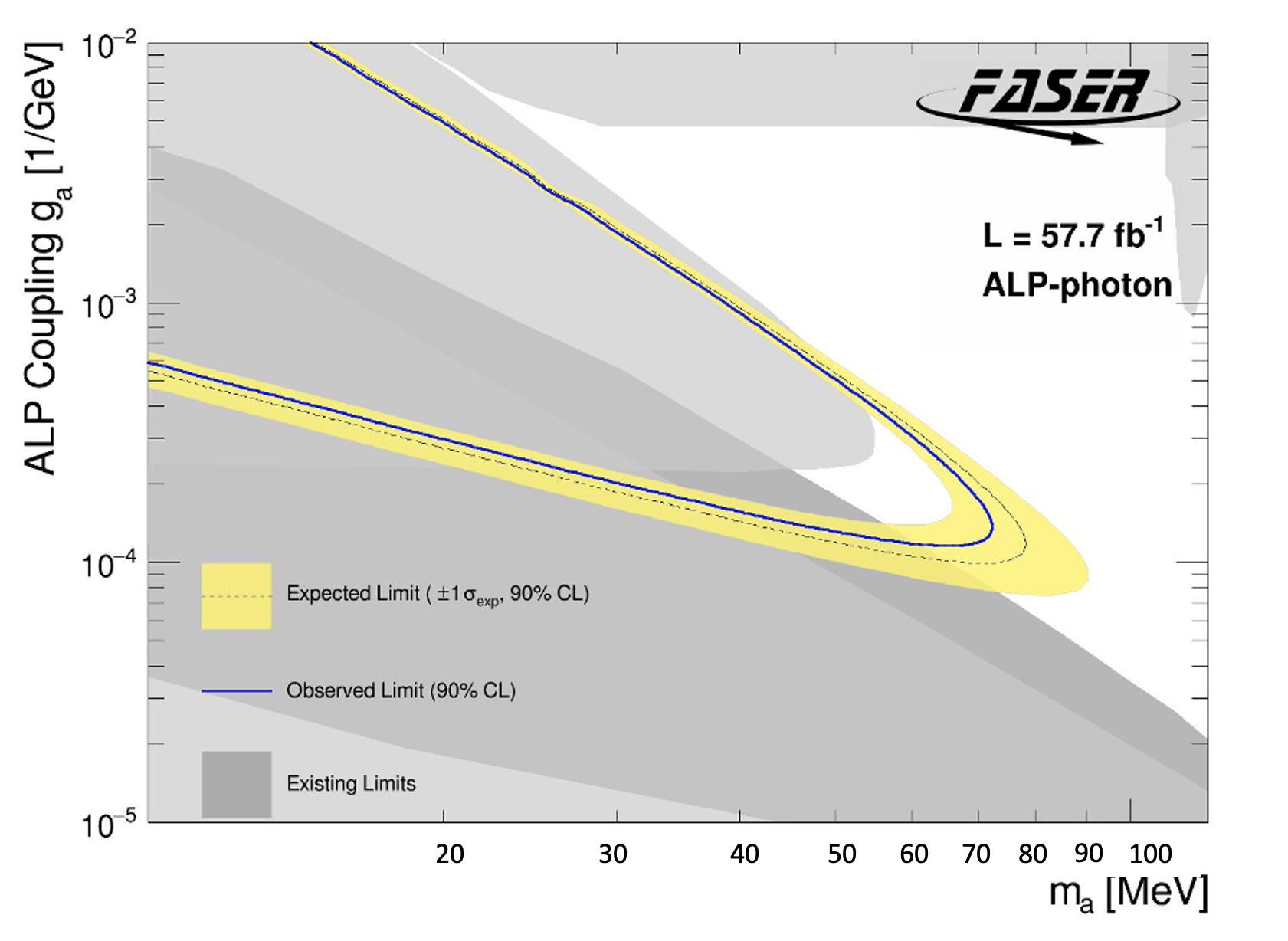}
    \includegraphics[width=0.49\linewidth]{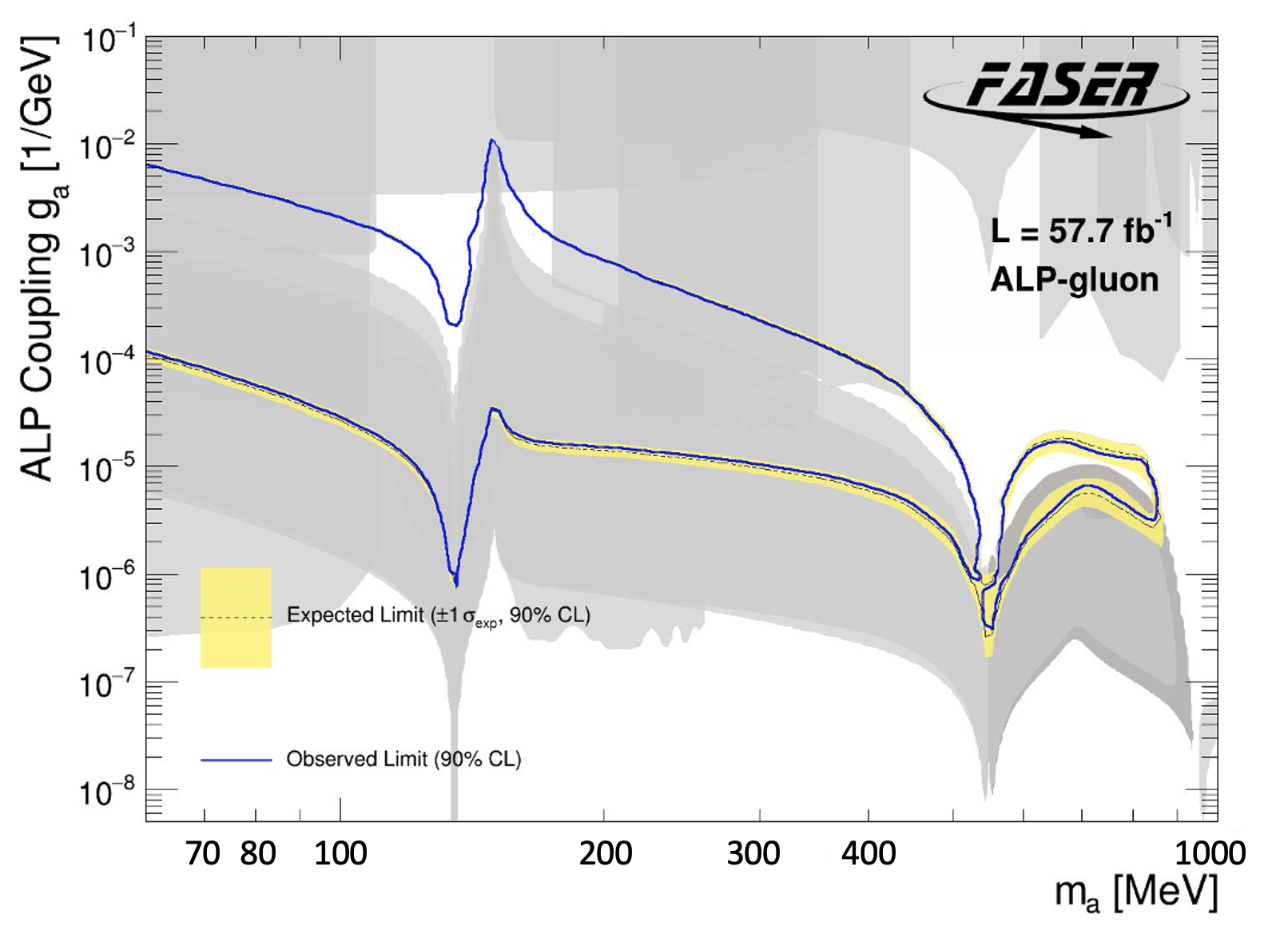}
    \caption{Interpretation of the signal region yield as ALP-photon (left) and ALP-gluon (right) exclusion limits at 90\% confidence level.}
    \label{fig:ALPgluonPhoton}
\end{figure}

\begin{figure}
    \centering
    \includegraphics[width=0.50\linewidth]{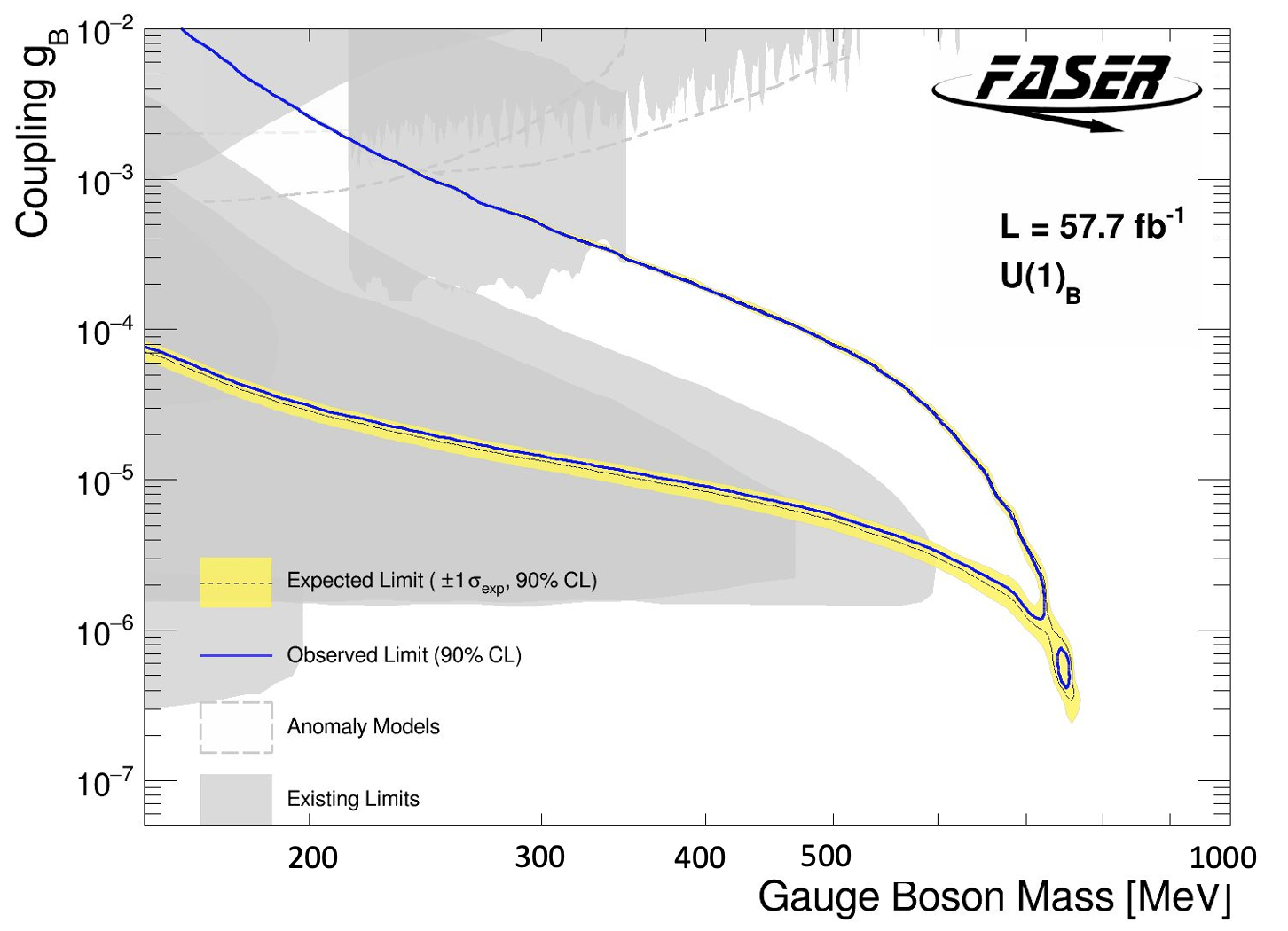}
    \includegraphics[width=0.49\linewidth]{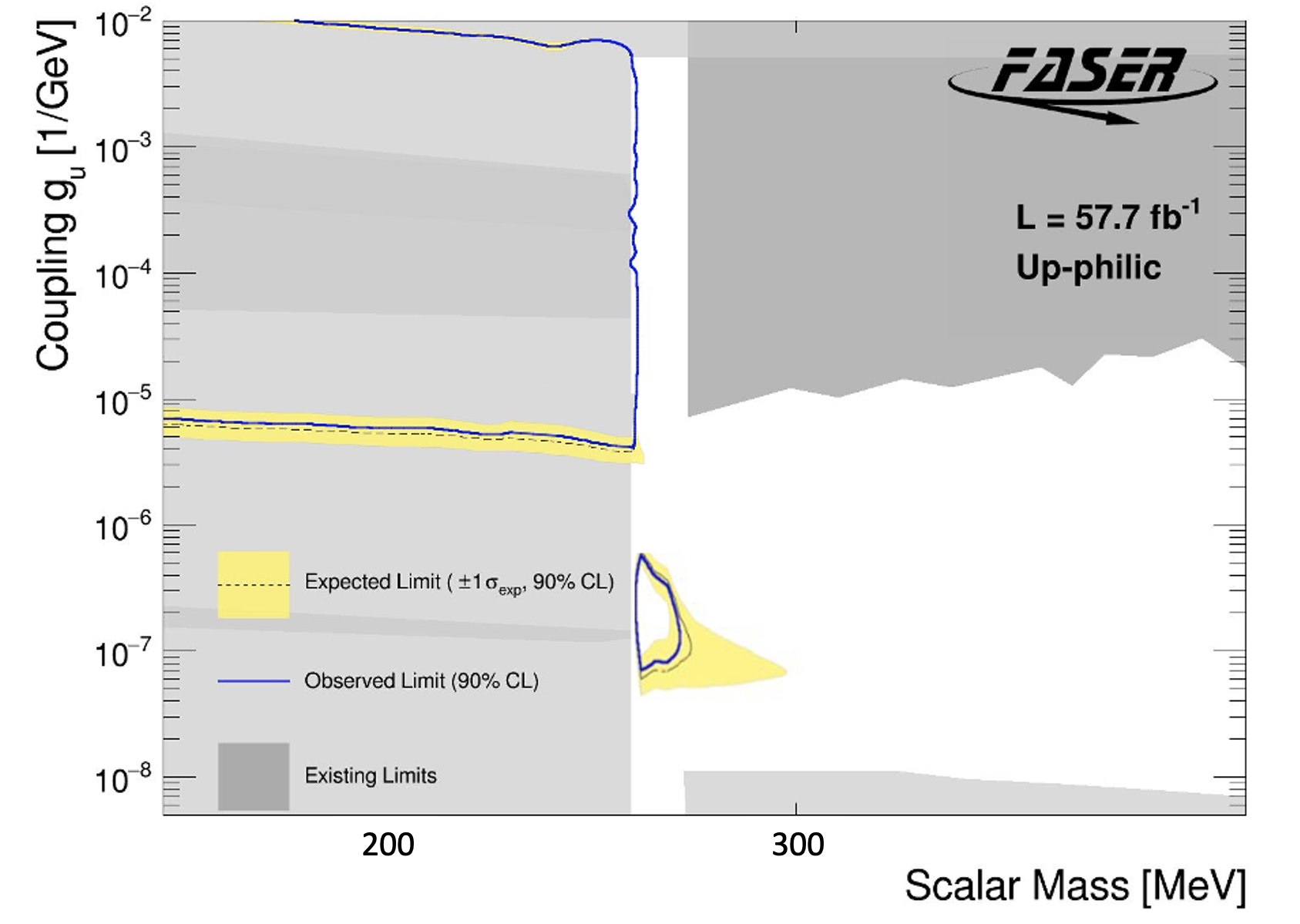}
    \caption{Interpretation of the signal region yield as U(1)$_B$ gauge boson (left) and up-philic (right) exclusion limits. Limits are provided at 90\% confidence level. For the definition of "anomaly models" for the U(1)$_B$ interpretation, see text. The yellow (expected limit) band is not visible in some regions, such as the upper part of some exclusion curves and the vertical segment of the up-philic exclusion limit, because it overlaps with the blue (observed limit) line, and its thickness is smaller than that of the observed limit curve.}
    \label{fig:U1BPlusUpPhiliv}
\end{figure}

\begin{figure}[tbp]
    \centering
    \includegraphics[width=0.7\textwidth]{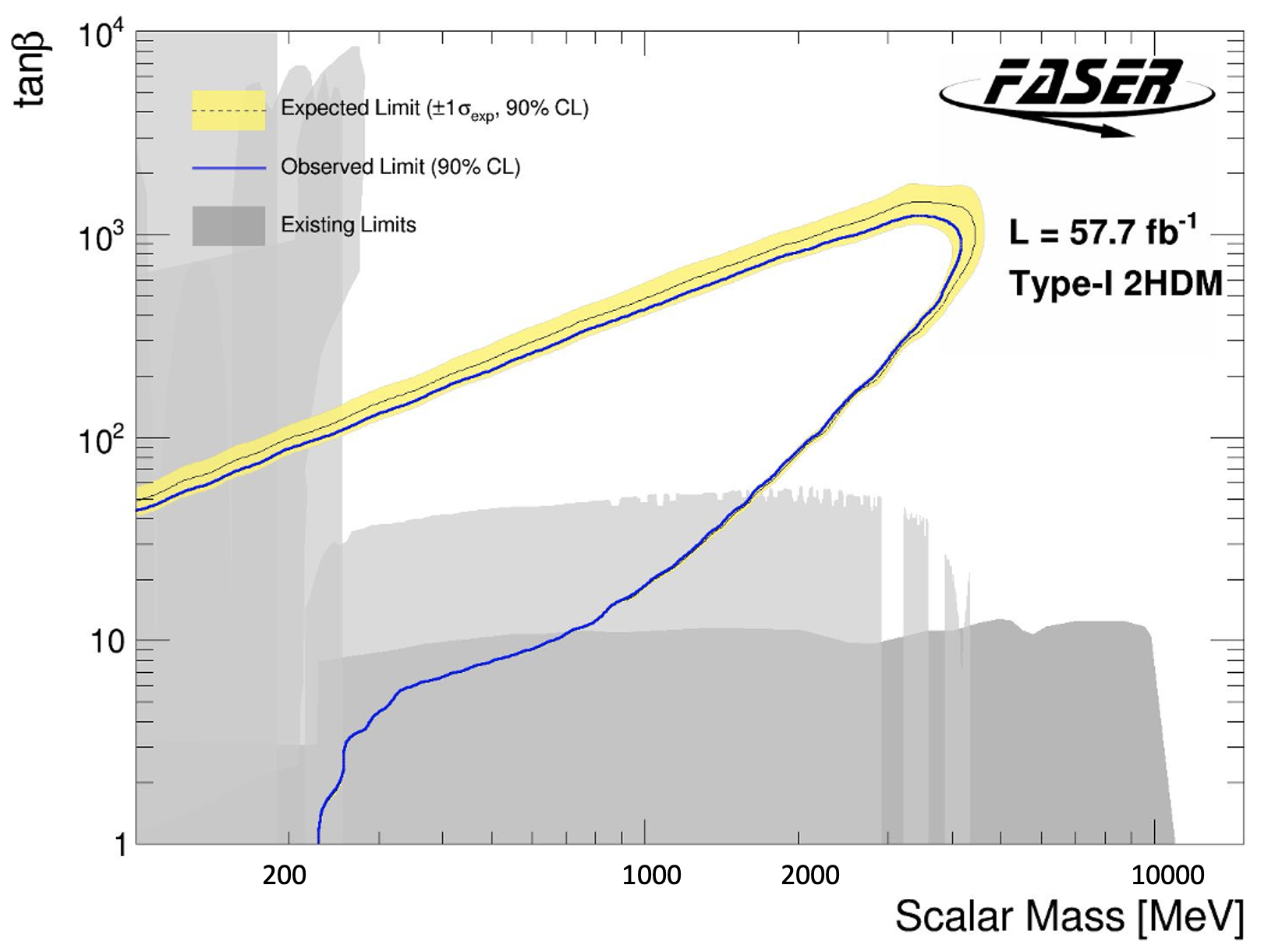}
    \caption{Interpretation of the signal region yield as 2HDM (type-I) exclusion limits at 90\% confidence level.}
    \label{fig:2HDMLim}
\end{figure}

\section{Conclusions \label{sec:conclusions}}

This paper presents FASER's first search for new particles decaying into photons, marking its first exploration of BSM physics predominantly produced in heavy-flavour decays. Data collected by FASER in 2022 and 2023 from proton-proton collisions at the LHC with a center-of-mass-energy of 13.6 TeV have been studied. 
Multiple ALP models and other multiphoton signature models have been considered. SM background sources that can mimic a similar detector signal as ALPs have been studied. The dominant background stems from neutrinos crossing the FASER detector volume and interacting with its material. Other backgrounds such as cosmic muons and beam-gas interactions have been studied and can be considered negligible in the context of this analysis. 
One data event was observed in the signal region, with a background expectation of $0.44 \pm 0.39$.  Coupling strengths of the ALP to weak gauge bosons between $3 \times 10^{-5}$ to $5 \times 10^{-4}$~GeV$^{-1}$ were excluded in previously-unprobed parameter space with ALP masses between 100 and 250 MeV. ALPs as heavy as 300 MeV were excluded for a coupling strength of $2 \times 10^{-5}$ GeV$^{-1}$. In addition, multi-photon models such as the type-I 2HDM, U(1)$_B$ vector boson, and up-philic scalar were investigated, excluding previously unprobed parameter space.

\FloatBarrier
\section*{Acknowledgments}
\label{sec:Acknowledgments}
\input{acknowledgments}

\appendix
\section{Additional Reinterpretation: Dark Photon Model}
\label{app:darkPhoton}
\FloatBarrier

Dark photons, $A'$, with masses between $2m_e$ and $2m_\mu$ are expected to decay predominantly into electron pairs, with other decay channels being negligible. A reinterpretation of the results of this paper in terms of dark photon models where electron-positron pairs from $A'$ decays are identified as EM deposits with no tracking requirements, is therefore possible. In the case of the $A'$, it must decay after the timing scintillator and before the PS, giving an effective decay volume of approximately 2.6m. This differs from ALPs, which can also decay before the timing scintillator, as their decay does not involve charged particles. This is shown in \cref{fig:DPLim}, where the results are also compared to those previously published by FASER using  $27\,\text{fb}^{-1}$ of data in a dedicated analysis requiring reconstructed tracks associated to the charged particles. The region excluded in the $m_{A'}$ vs. kinetic mixing parameter space is extended up to approximately 100~MeV in dark photon mass. The larger signal acceptance of the presented analysis is extending the sensitivity at large kinetic mixing values, whereas at lower kinetic mixings the higher energy requirement is reducing the sensitivity. This shows how an approach utilising a combination of selections with and without tracking requirements can significantly boost the sensitivity to dark photon models. 
\begin{figure}[htbp]
    \centering
    \includegraphics[width=0.7\textwidth]{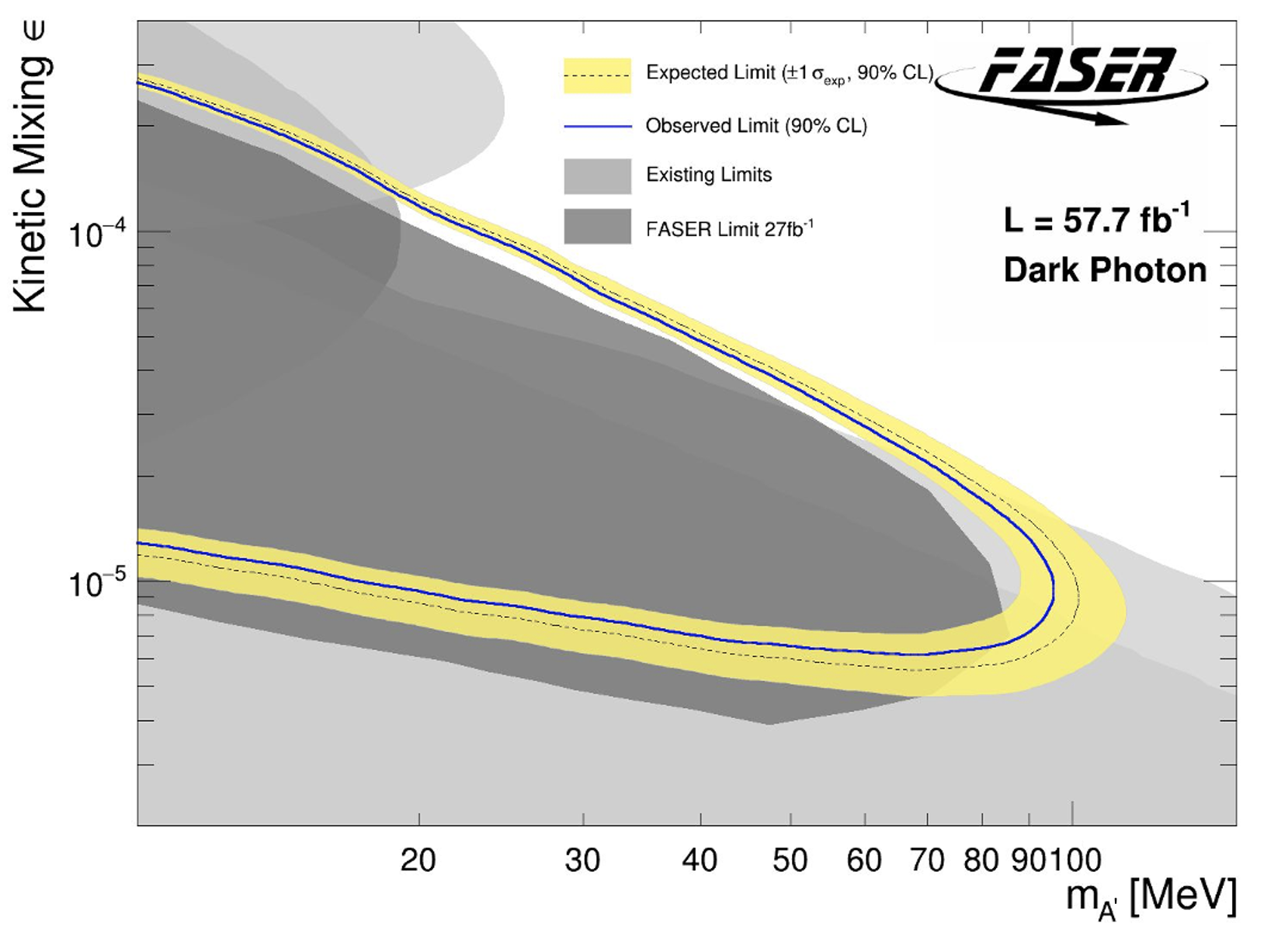}
    \caption{Interpretation of the signal region yield as dark photon exclusion limits at 90\% confidence level in the $(m_{A’}, \epsilon)$ plane, where $m_{A’}$ is the dark photon mass and $\epsilon$ is the kinetic mixing parameter. }
    \label{fig:DPLim}
\end{figure}

\FloatBarrier
\section{Overview of Signal Model Production and Decays \label{app:SigProdAndDecay}}
In \cref{fig:app:ProdDecMerged}, the production mechanisms and branching ratios for the signal models discussed in \cref{sec:SignalModels} are given. Mass ranges relevant for FASER sensitivity are selected. Central production rate predictions are shown as solid lines and uncertainty bands include flux uncertainties. The given predictions include the expectation within $\theta < 0.2~\text{mrad}$ of the LOS and a minimum energy of the long-lived particles of 100 GeV. The central prediction is taken from \verb|EPOS-LHC| and the variation is calculated from the other generators discussed in~\cref{sec:SamplesAndReco}. %

\begin{figure}
    \centering
    \subfigure[ALP-W]{\includegraphics[width=0.41\textwidth]{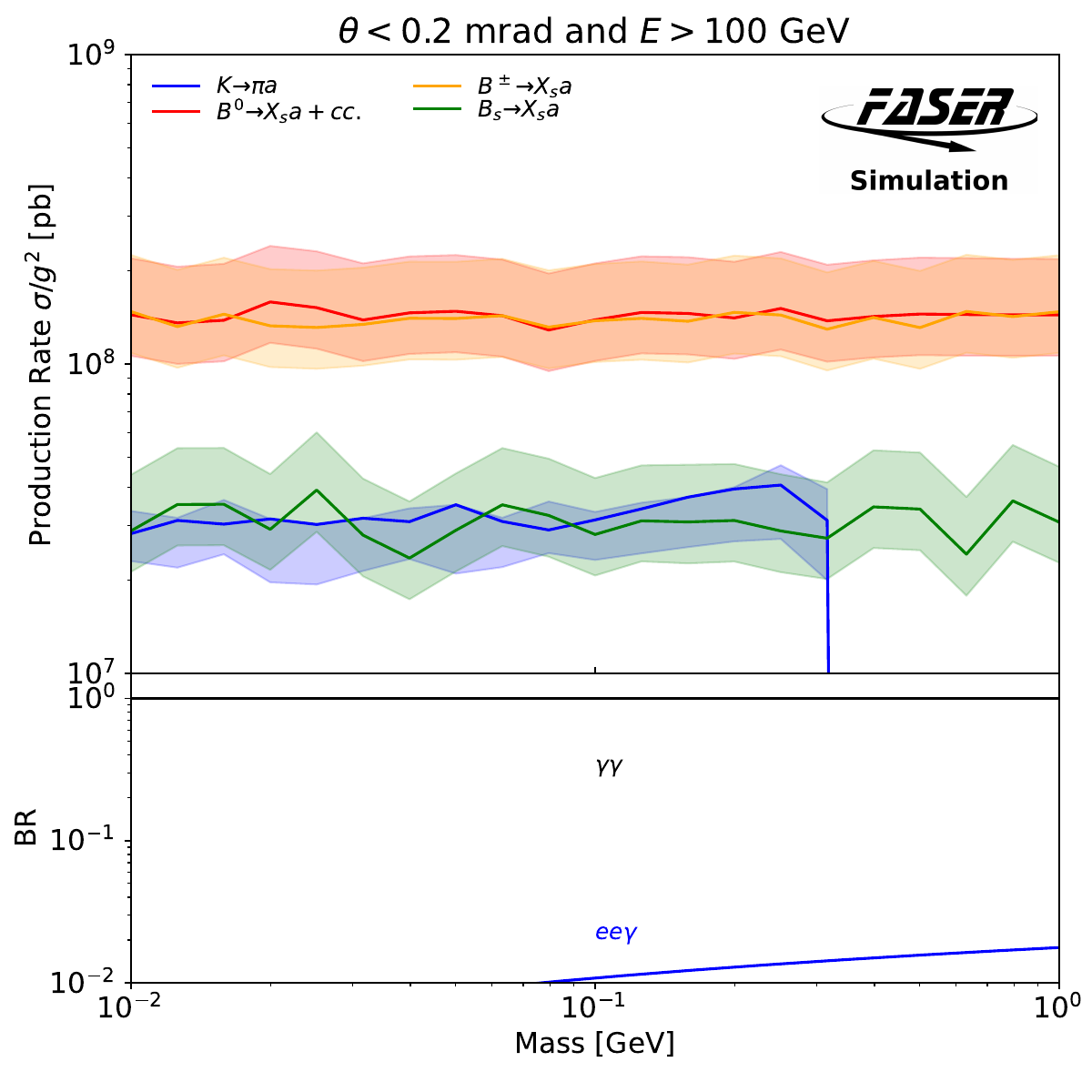}}
    \subfigure[ALP-photon]{    \includegraphics[width=0.41\textwidth]{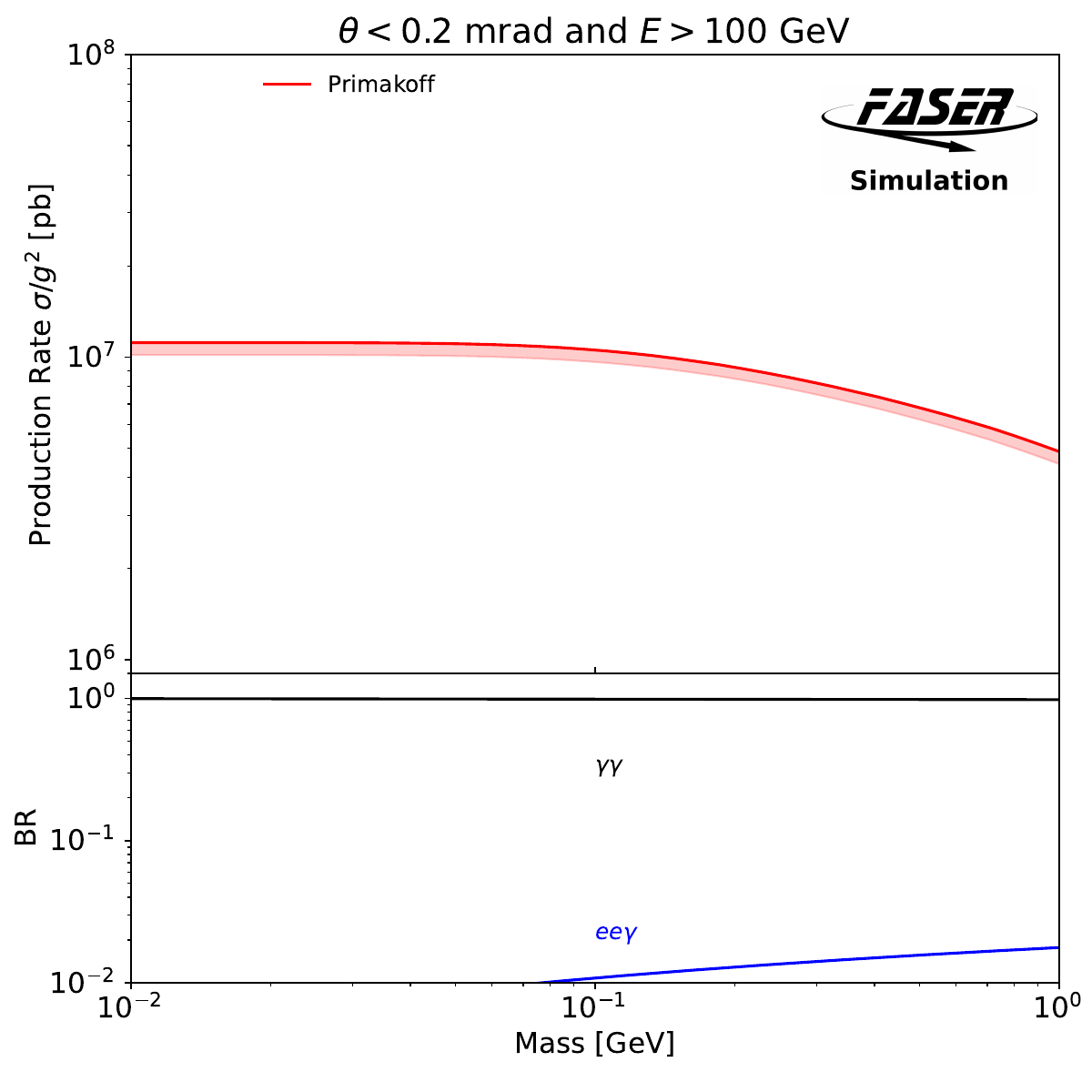}}\\
    \subfigure[ALP-gluon]{   \includegraphics[width=0.41\textwidth]{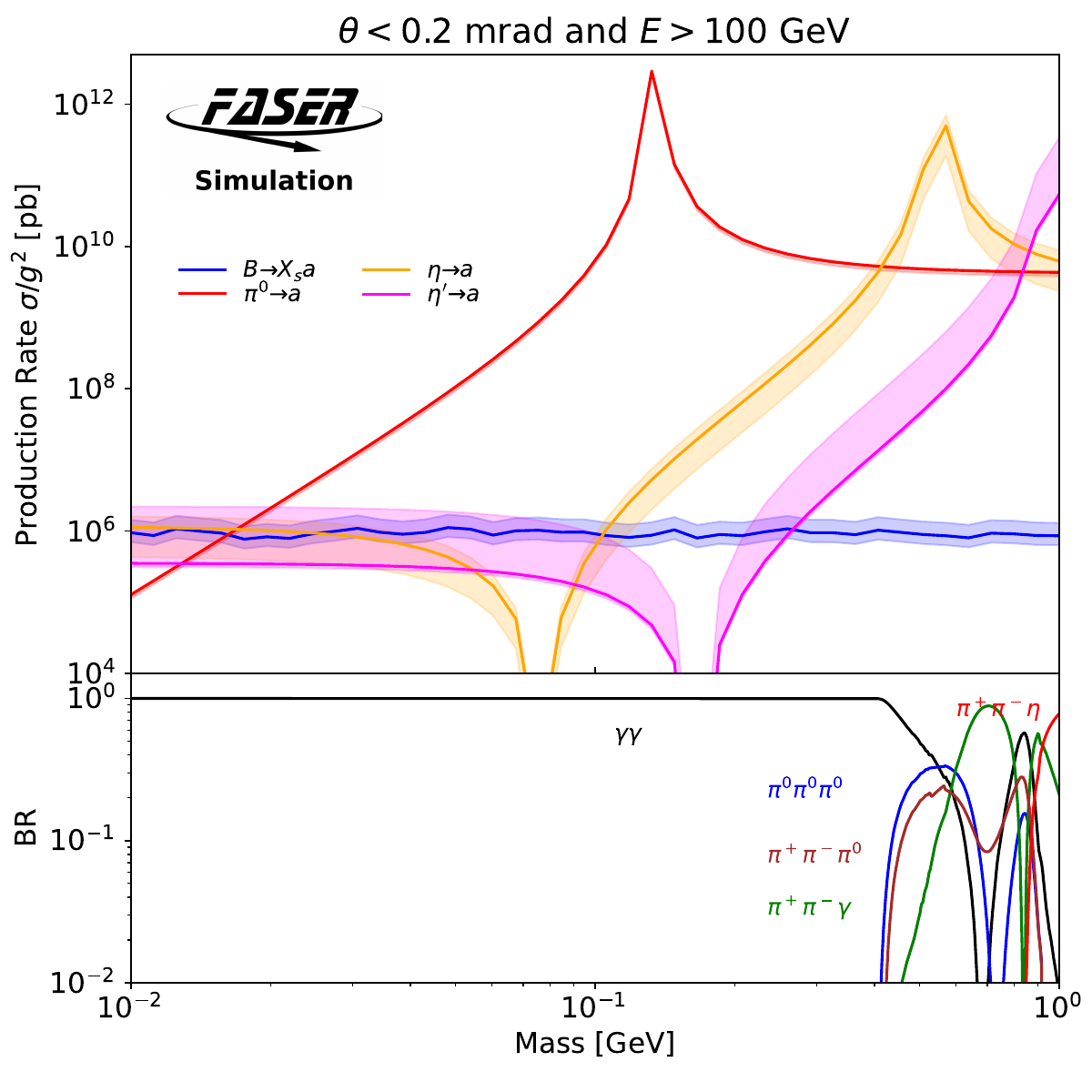}}
    \subfigure[2HDM (type-I)]{   \includegraphics[width=0.41\textwidth, height =7.1cm]{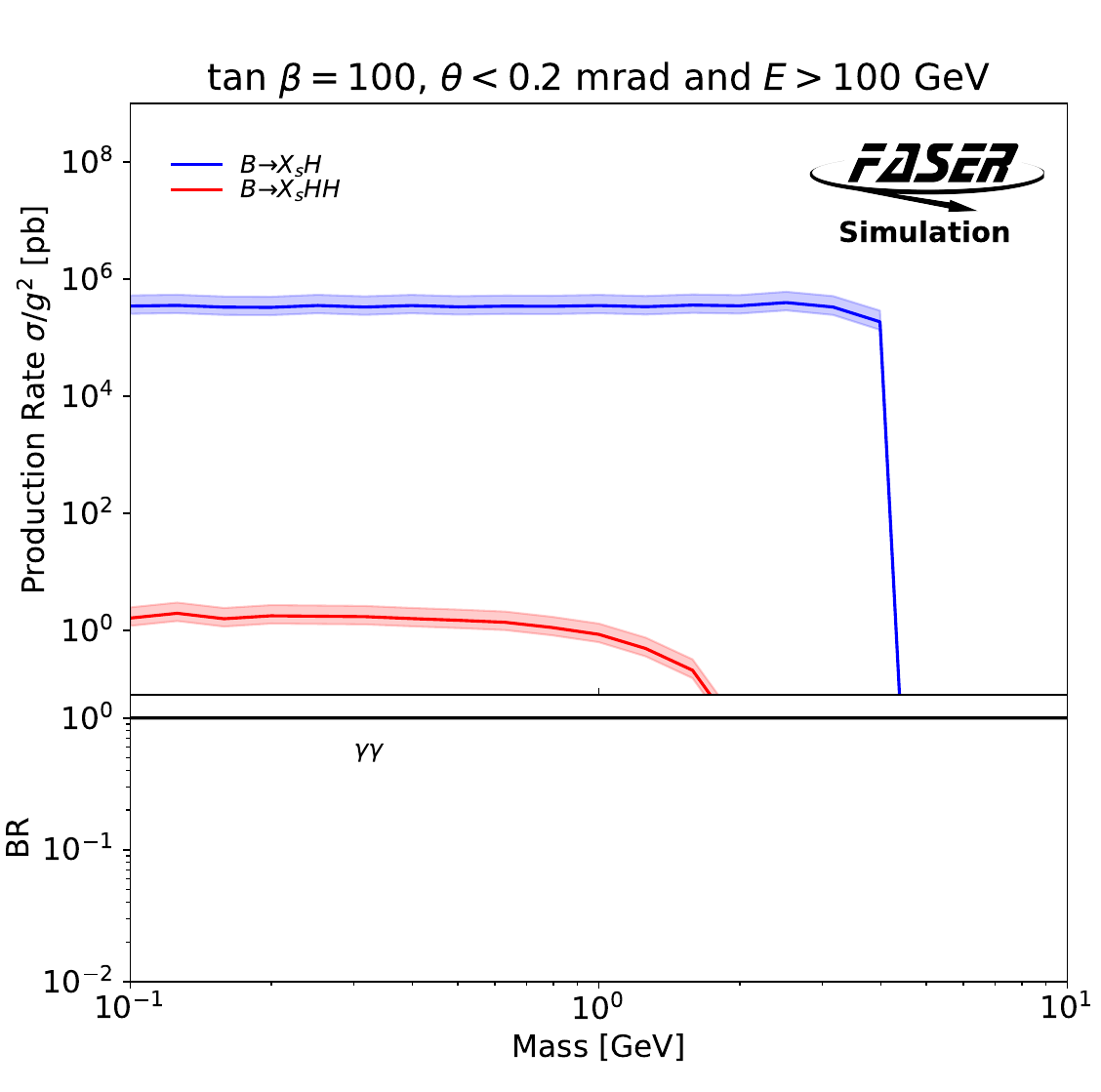}}\\
    \subfigure[U(1)$_B$]{\includegraphics[width=0.41\textwidth]{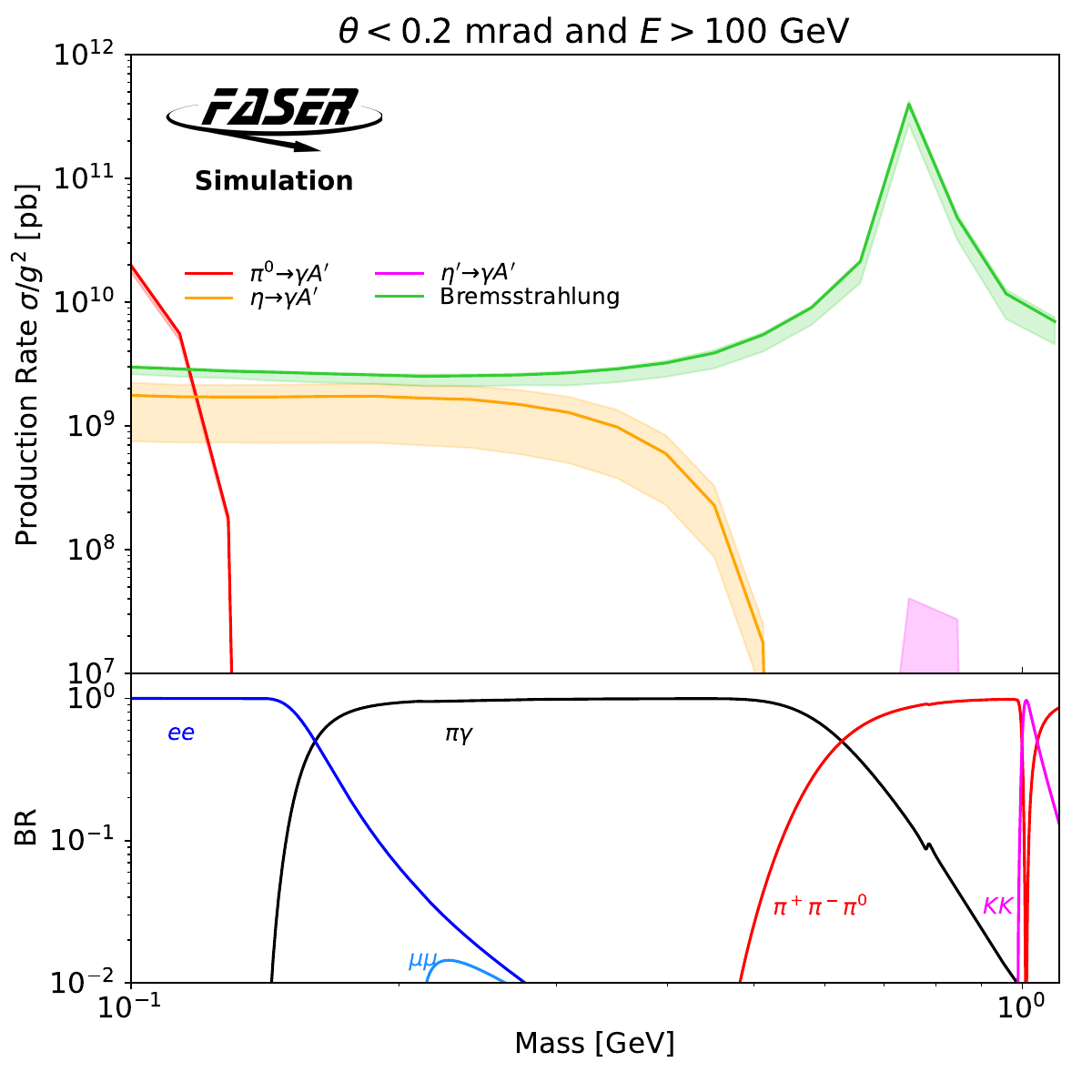}}
    \subfigure[Up-philic scalar]{   \includegraphics[width=0.41\textwidth]{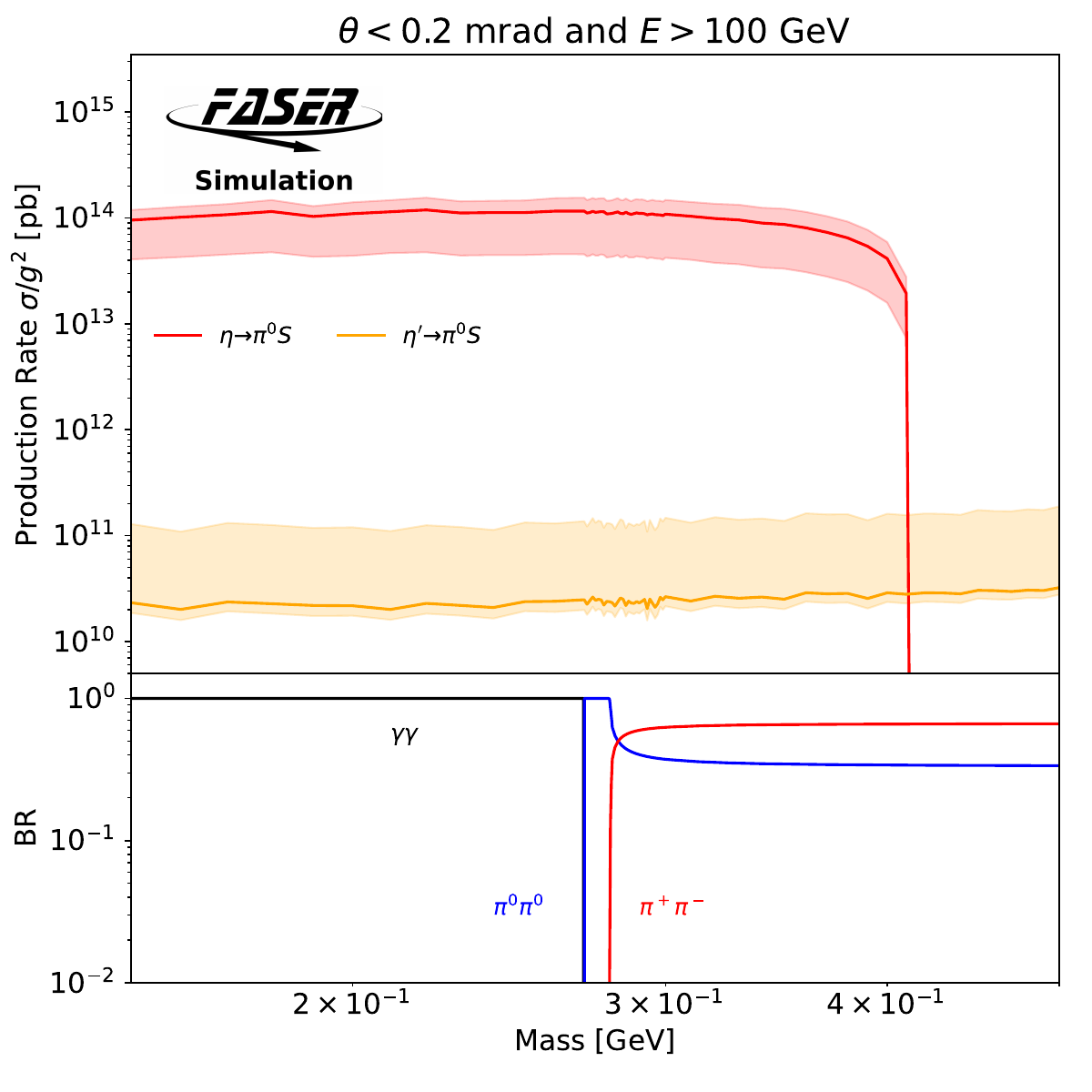}}\\
    
    \caption{Production rates and branching ratio overview for signal models described in  \cref{sec:SignalModels}. The central prediction as solid lines and uncertainty bands including flux uncertainties are shown. In this case, differently to~\cref{footnote1}, $\theta$ is the polar angle with respect to the LOS. \label{fig:app:ProdDecMerged}}
\end{figure}

\FloatBarrier
\section{Parametrised Signal Efficiencies}
\label{app:parametrisedEfficiencies}

To evaluate parametrised signal efficiencies, the following decay final states have been considered: $\gamma\gamma$, $\pi^0 \gamma$, $ \pi^0 \pi^0$, $\pi^0\pi^0\pi^0$, $ee$, $\pi^+\pi^-\gamma$ and $\pi^+\pi^-\pi^0$. Dedicated high-statistics simulation samples with a flat energy distribution were used to calculate the efficiency of a long-lived particle (LLP) to pass all analysis selections as a function of the LLP energy. The result is shown in  \cref{fig:parametrisedEff}.

For the $\gamma\gamma$, $\pi^0 \gamma$, $ \pi^0 \pi^0$, $\pi^0\pi^0\pi^0$ and $ee$ final states, the efficiency is well described by a \textit{tanh}-like function: it turns on sharply at an LLP energy of 1.5~TeV, corresponding to the calorimeter energy cut, and approaches a constant value at high energies, whose value is set by the preshower selections. Low photon multiplicity final states (like $\gamma\gamma$) have a mildly higher efficiency at high energies than the high photon multiplicity final states (like $3\pi^0 \to 6\gamma$) since the large number of photons is more likely to saturate the preshower and hence reduce the preshower ratio below the selection value of 4.5. The di-electron channel has a lower efficiency, since only events after the timing station pass the analysis selections. For the two charged pion channels, the efficiency increases much more slowly with energy. This is because the measured calorimeter energy is typically much lower than the LLP energy. In practice, the sum of two \textit{tanh}-like functions provide a good fit.

\begin{figure}[htp]
    \centering
    \subfigure[\label{fig:parametrisedEff:gammas}]{\includegraphics[width=0.65\linewidth]{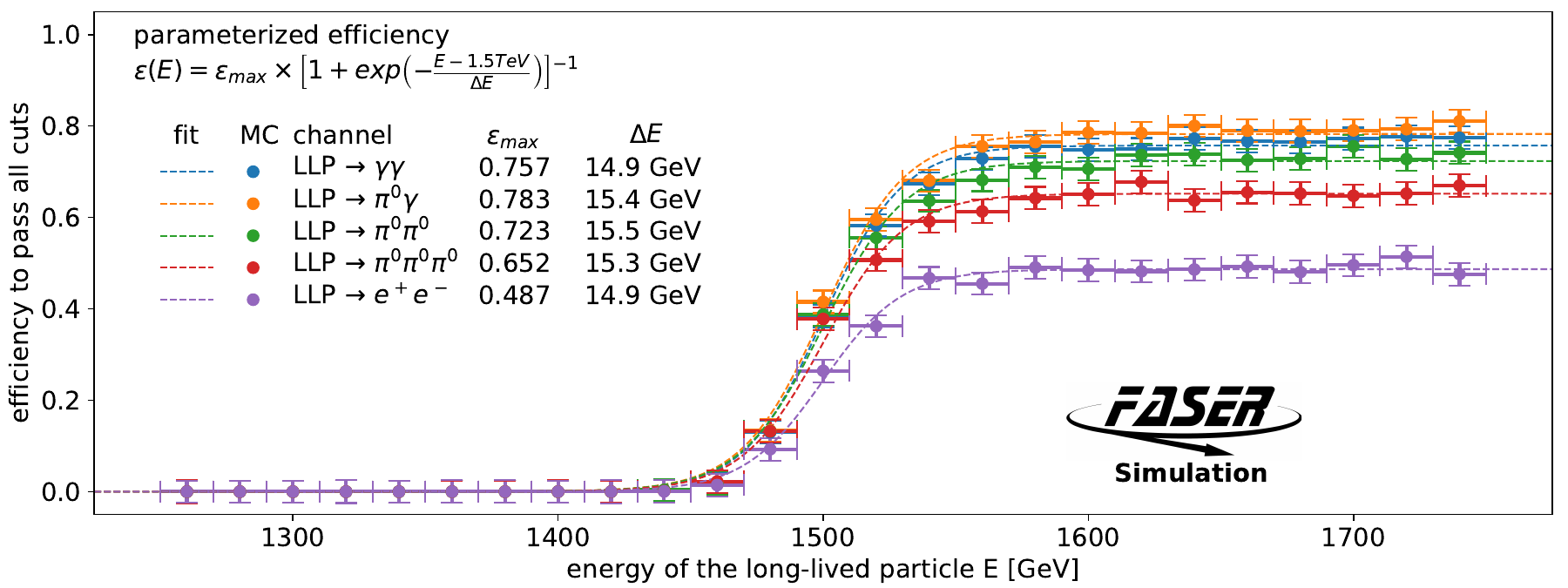}}
    \subfigure[\label{fig:parametrisedEff:pigammas}]{\includegraphics[width=0.65\linewidth]{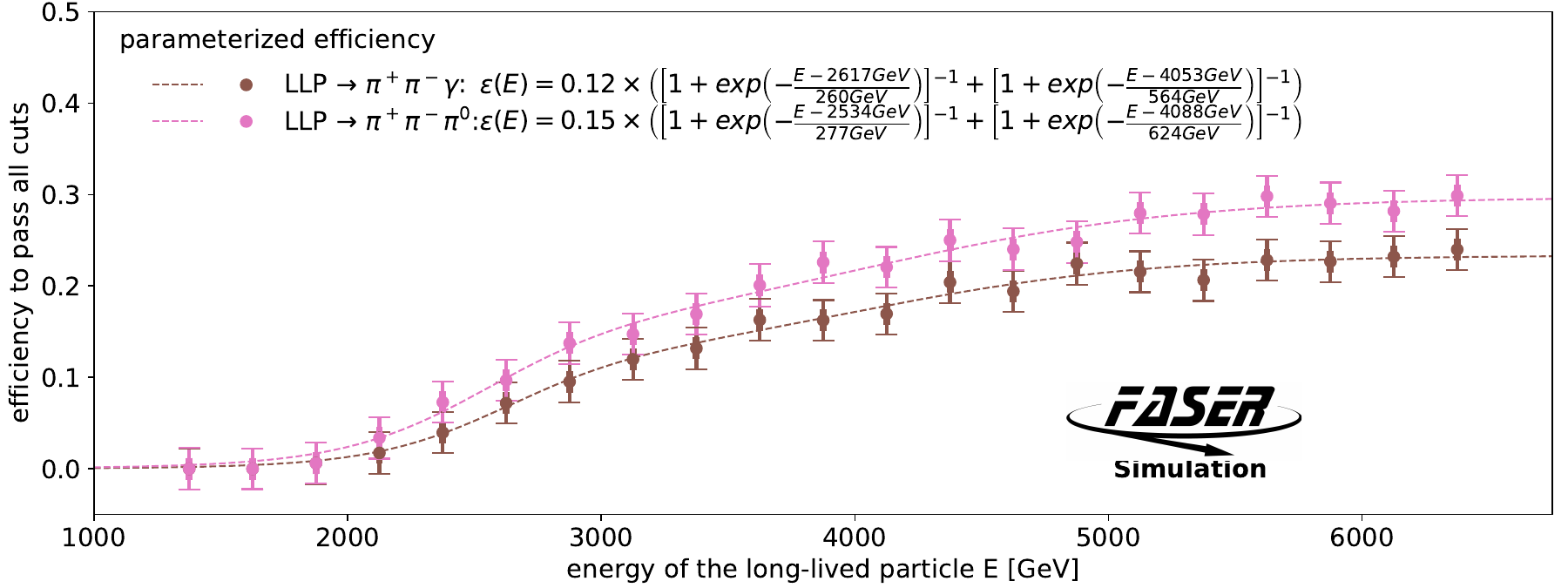}}
    \caption{Parametrised signal efficiencies for long-lived particles decaying inside of the FASER decay volume. }
    \label{fig:parametrisedEff}
\end{figure}

\FloatBarrier
\section{FASER Limits in Relation to Previous Experiments}
\label{app:existingconstraints}
\FloatBarrier
\cref{fig:Limits_Colour} shows the observed ALP-W limit. Existing constraints are also shown from BaBar~\cite{BaBar:2021ich}, E137~\cite{Bjorken:1988as}, LEP~\cite{L3:1994shn,Anashkin:412002,Jaeckel:2015jla}, E949~\cite{Artamonov_2005}, KOTO~\cite{KOTO:2018dsc}, KTeV~\cite{Abouzaid_2008} and CDF~\cite{CDF:2013lma,Bauer:2017ris}.

\cref{fig:Limits_ColourALP-photon} shows the observed ALP-photon limit. Existing constraints from LEP~\cite{OPAL:2002vhf}, PrimEx~\cite{PrimEx:2010fvg,Aloni:2019ruo}, NA64~\cite{NA64:2020qwq}, NuCal~\cite{Dobrich:2019dxc,Blumlein:1990ay} and E137~\cite{Bjorken:1988as} are shown.

The ALP-gluon limits in comparison with results from NA62~\cite{NA62:2023olg,NA62:2021zjw,NA62:2020pwi}, BaBar~\cite{BaBar:2021ich}, flavour constraints~\cite{Aloni:2018vki}, NuCal~\cite{Blumlein:1991xh,Aloni:2018vki}, CHARM~\cite{CHARM:1985anb}, NA48~\cite{NA48:2002xke,Goudzovski:2022vbt}, E949~\cite{E949:2005qiy,BNL-E949:2009dza,Jerhot:2022chi} and from the total decay width of the kaon~\cite{Goudzovski:2022vbt} are shown in \cref{fig:Limits_ColourALPGluon}.

\begin{figure}[t!]
    \centering
        \subfigure[\label{fig:Limits_Colour}]{\includegraphics[width=0.45\textwidth]{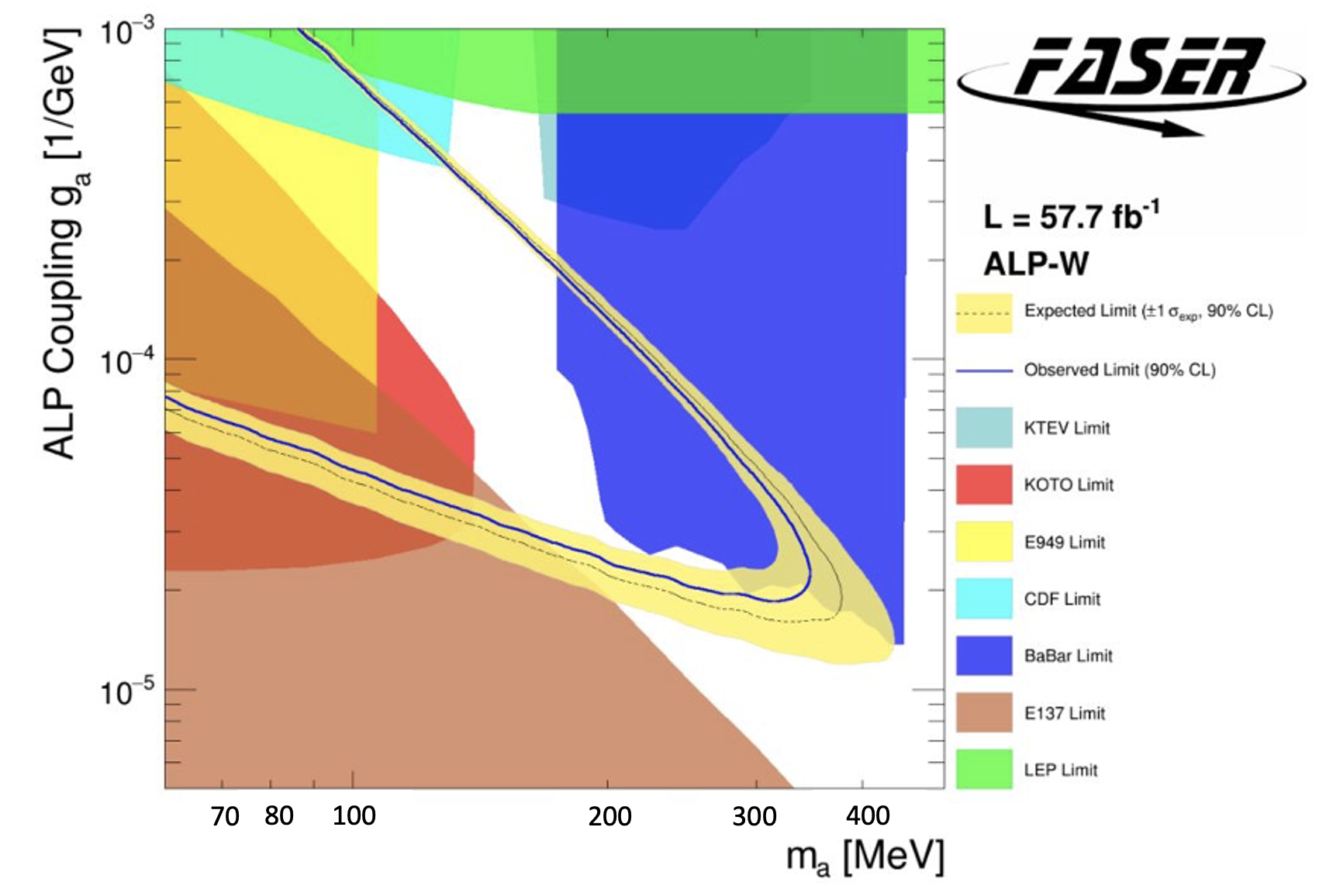}}
    \hfill
        \subfigure[\label{fig:Limits_ColourALP-photon}]{\includegraphics[width=0.45\textwidth]{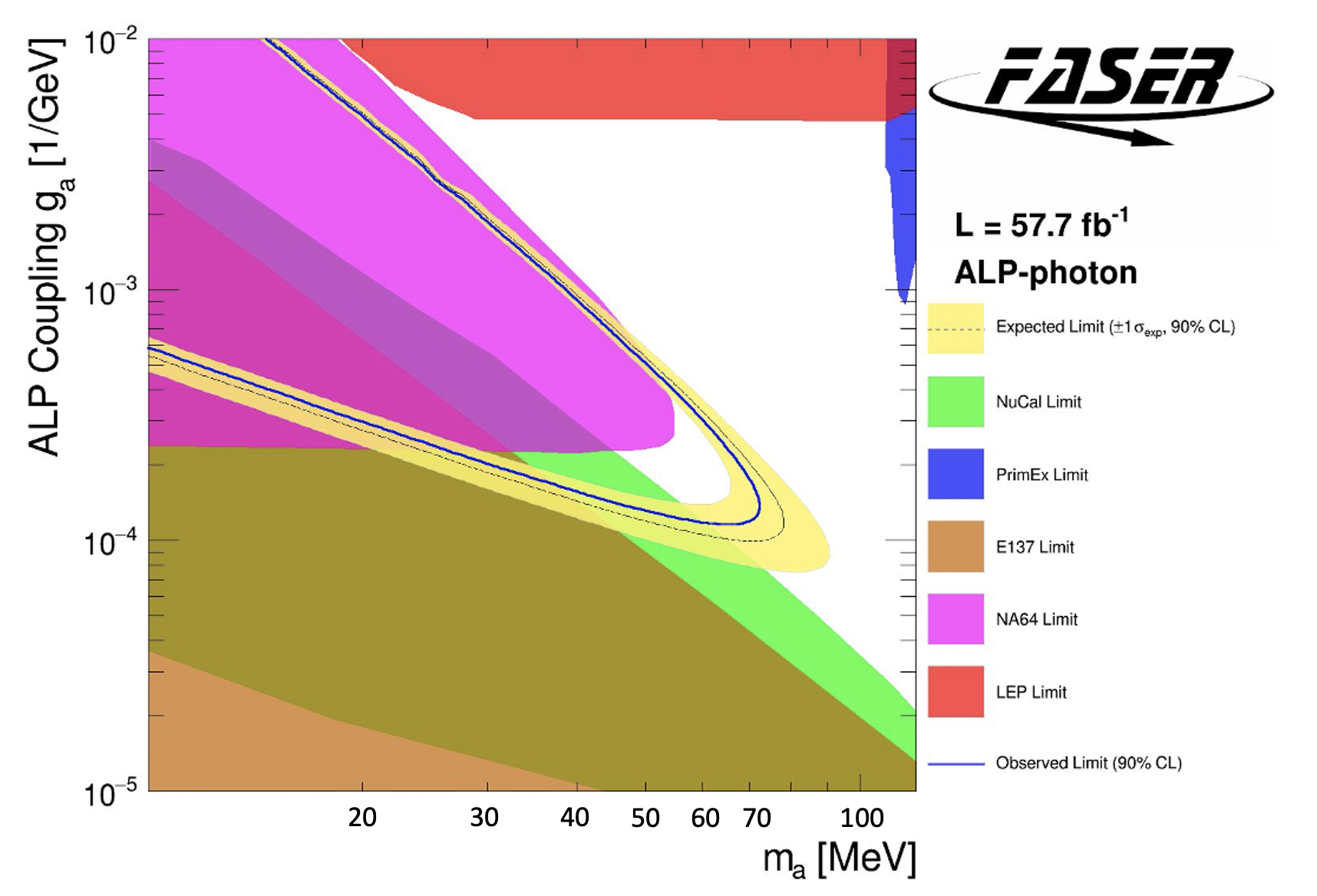}}
    \hfill\\
        \subfigure[\label{fig:Limits_ColourALPGluon}]{\includegraphics[width=0.45\textwidth]{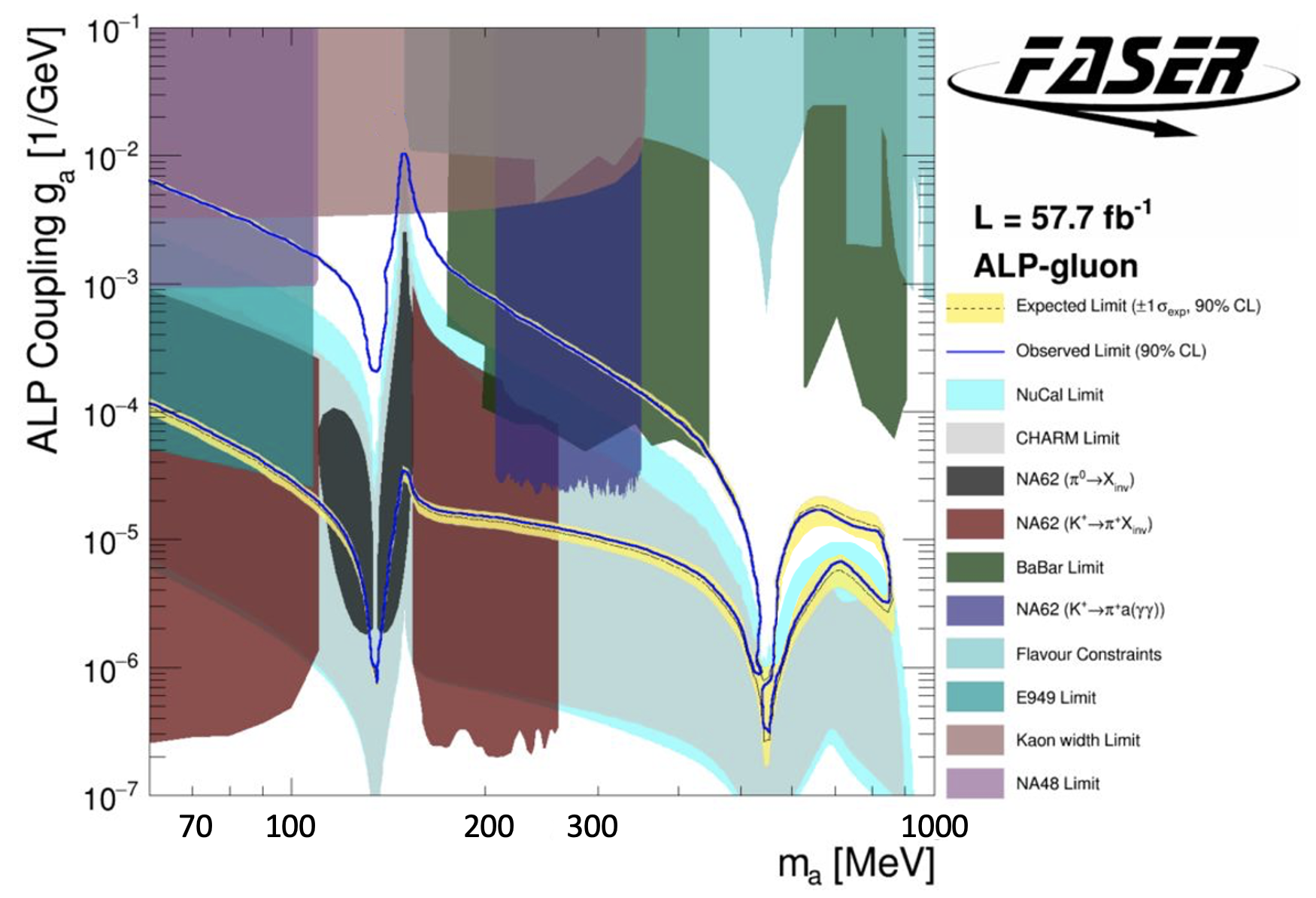} }
    \caption{Interpretation of the signal region yields for different exclusion limits}
\end{figure}

\begin{figure}[t!]
    \centering
        \subfigure[\label{fig:Limits_ColourU1B}]{\includegraphics[width=0.45\textwidth]{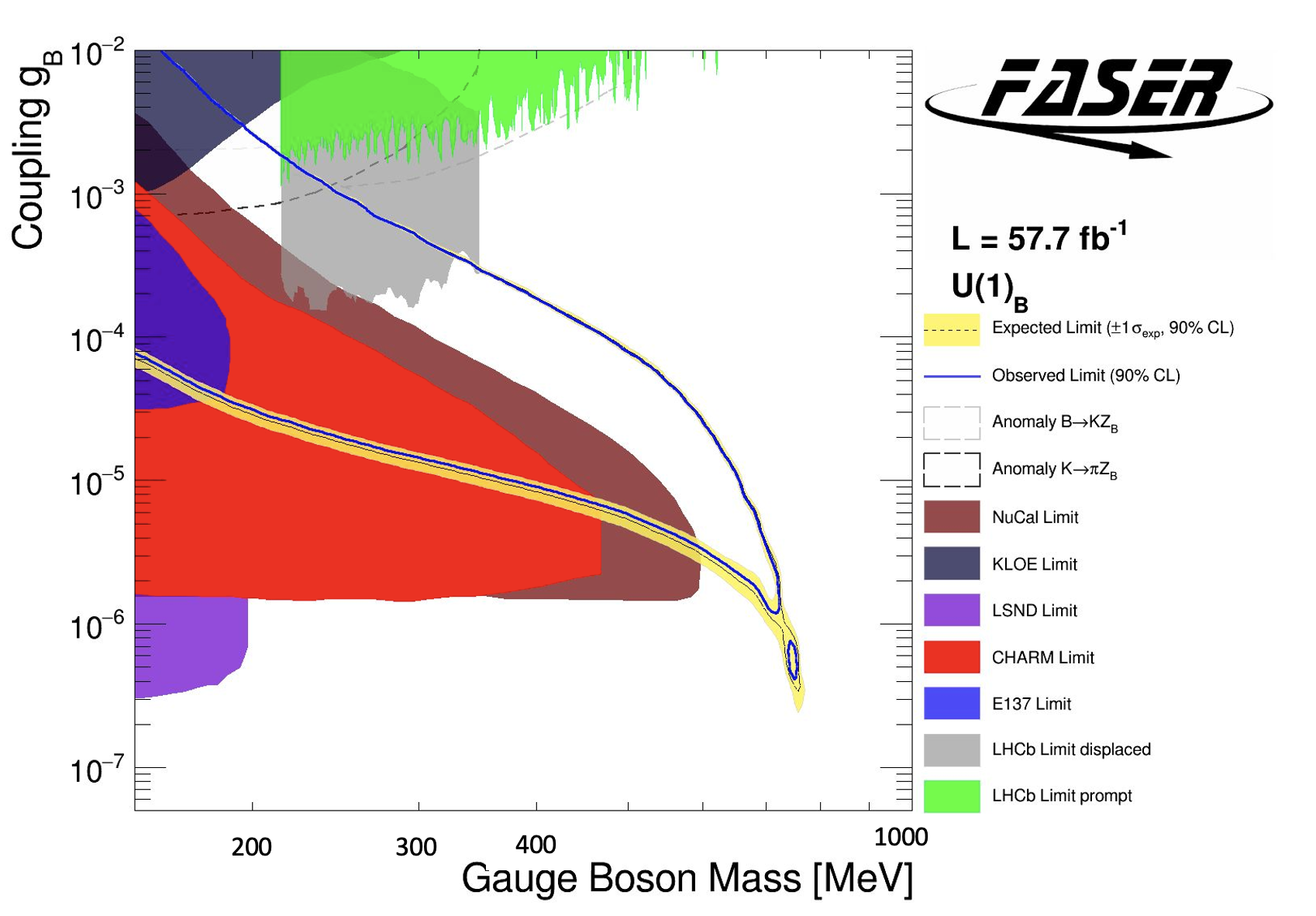}}
        \subfigure[\label{fig:Limits_ColourUP}]{\includegraphics[width=0.45\textwidth]{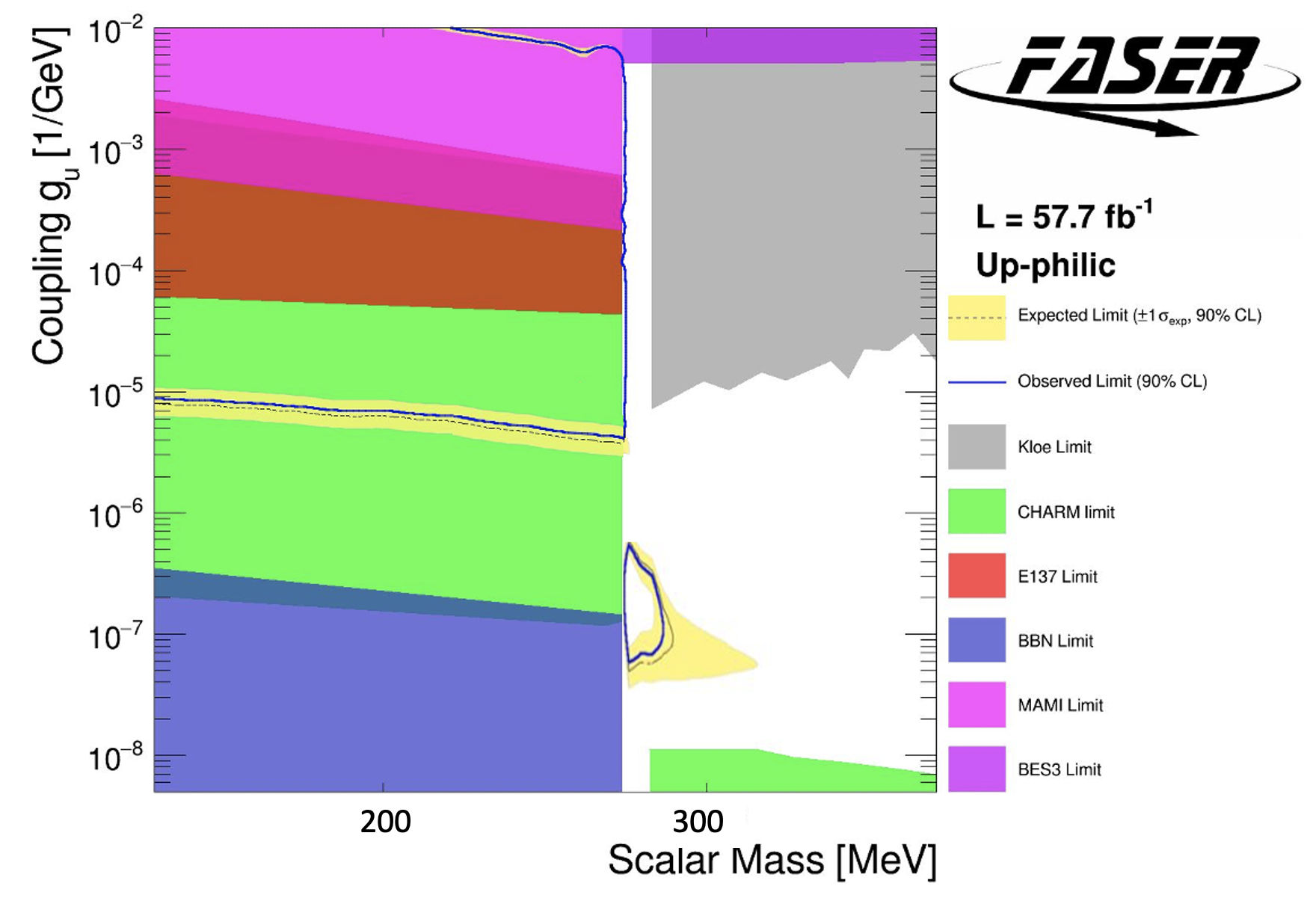}}\\
        \subfigure[\label{fig:Limits_Colour2HDM}]{\includegraphics[width=0.45\textwidth]{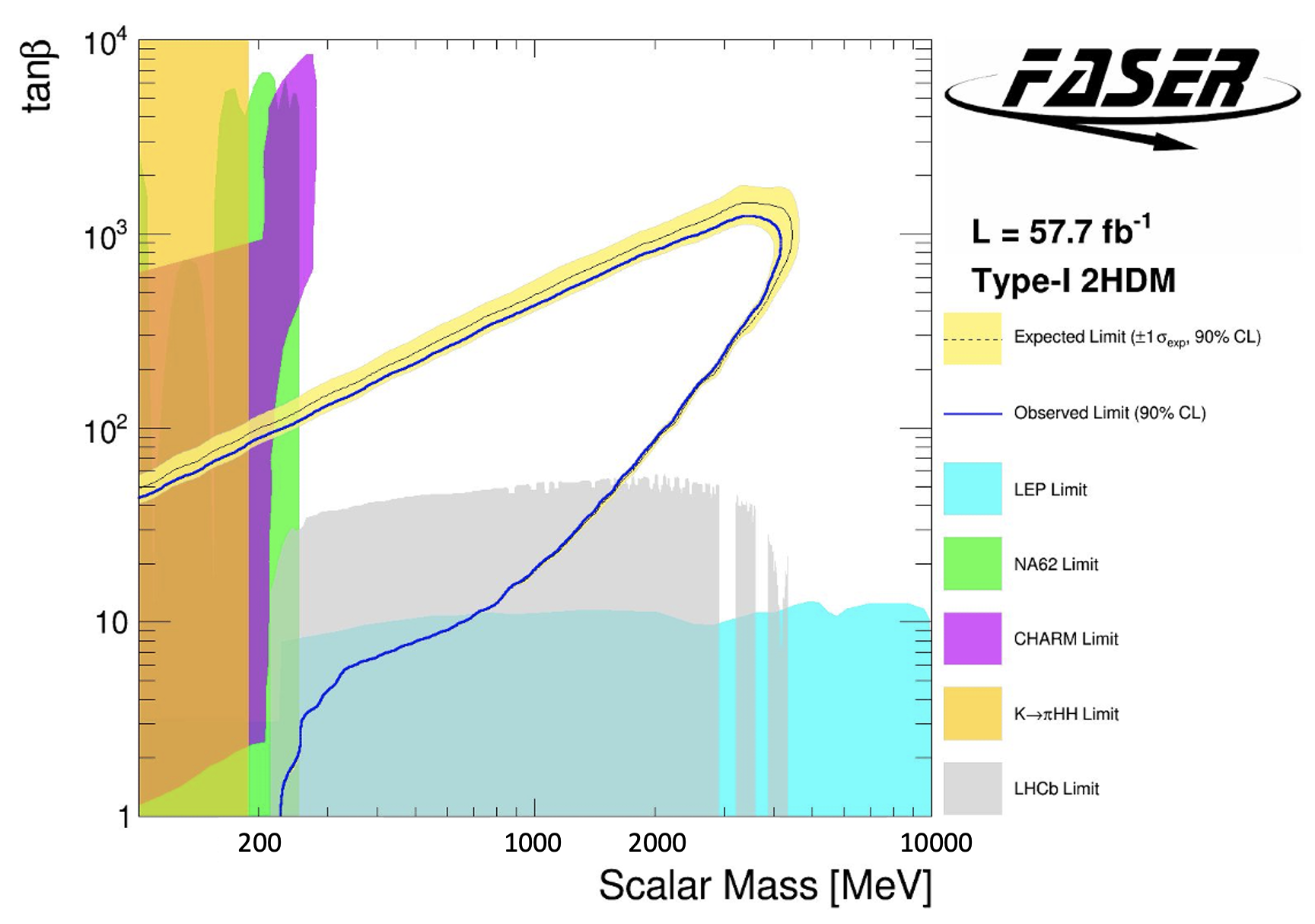} }
    \caption{Interpretation of the signal region yields for different exclusion limits}
\end{figure}

\cref{fig:Limits_ColourU1B} displays the existing constraints for the U(1)$_B$ model in comparison to the FASER limit presented. Here limits from NuCal~\cite{Blumlein:1990ay, Blumlein:1991xh}, CHARM~\cite{CHARM:1985anb}, KLOE~\cite{Anastasi:2015qla, KLOE-2:2012lii}, E137~\cite{Bjorken:1988as}, LHCb~\cite{LHCb:2019vmc} and LSND~\cite{LSND:2001aii,Bauer:2018onh} are considered. Additionally, limits from anomalous currents are included~\cite{Dror:2017ehi}.

The existing constraints for the up-philic model are shown in \cref{fig:Limits_ColourUP}. Here considered are MAMI~\cite{A2atMAMI:2014zdf}, BES3~\cite{BESIII:2016tdb}, E137~\cite{Bjorken:1988as}, KLOE~\cite{KLOE-2:2016zfv}, CHARM~\cite{CHARM:1985anb}, and from the number of effective degrees of freedom that could affect the Big Bang  Nucleosynthesis (BBN)~\cite{Cooke:2013cba,Batell:2018fqo, Millea:2015qra}.

\begin{figure}[tbh]
\centering
\includegraphics[width=0.75\textwidth]{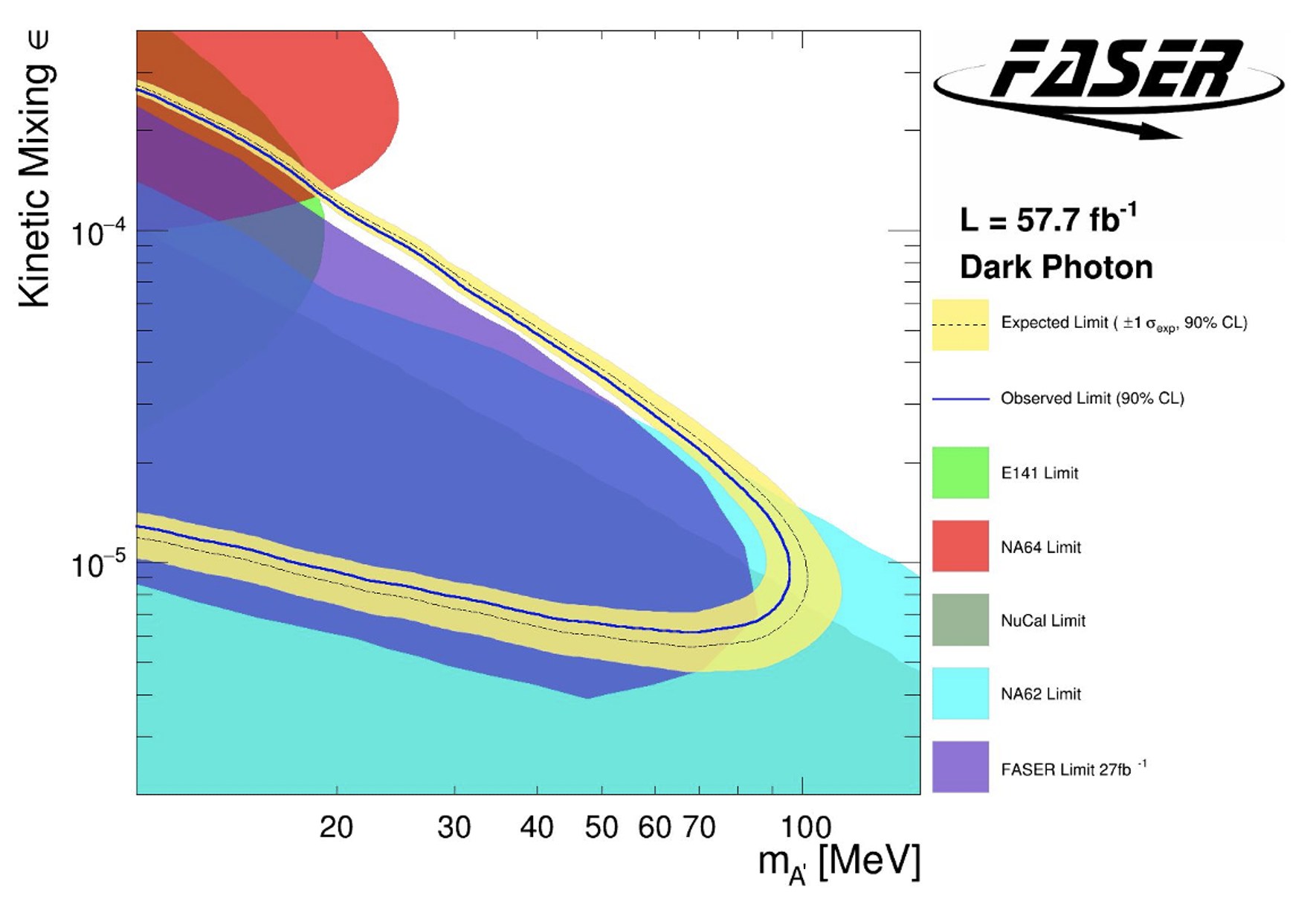}
\caption{Interpretation of the signal region yields as dark photon exclusion limit.}
\label{fig:Limits_ColourDP}
\end{figure}

The FASER limit interpreted within the 2HDM in comparison with results from LEP~\cite{OPAL:2008tdn,ALEPH:2013htx,Arbey:2018}, NA62~\cite{NA62:2021zjw}, CHARM~\cite{CHARM:1985anb, Krnjaic:2015mbs}, LHCb~\cite{LHCb:2015nkv} and $K\rightarrow~\pi HH$~\cite{NA62:2021zjw} are shown in \cref{fig:Limits_Colour2HDM}.

An interpretation of the results in a dark photon model are given in~\cref{fig:Limits_ColourDP}. This includes the initial sensitivity of FASER with a track-based analysis~\cite{FASER:2023tle} as discussed in~\cref{app:darkPhoton}. Additionally, search results from NA62~\cite{NA62:2023nhs}, NA64~\cite{NA64:2019auh},  NuCal~\cite{Blumlein:1991xh,Blumlein:2013cua} and E141~\cite{Riordan:1987aw,Andreas:2012mt} are shown.

\FloatBarrier
\section{Event Display}
\FloatBarrier
The data event, here named the \textit{ALPtrino} event, observed in the signal region is visualised through its detector signature in \cref{app:fig:evtdisplay}.

\label{app:eventdisplay}
\begin{figure}
    \centering
    \includegraphics[width=\linewidth]{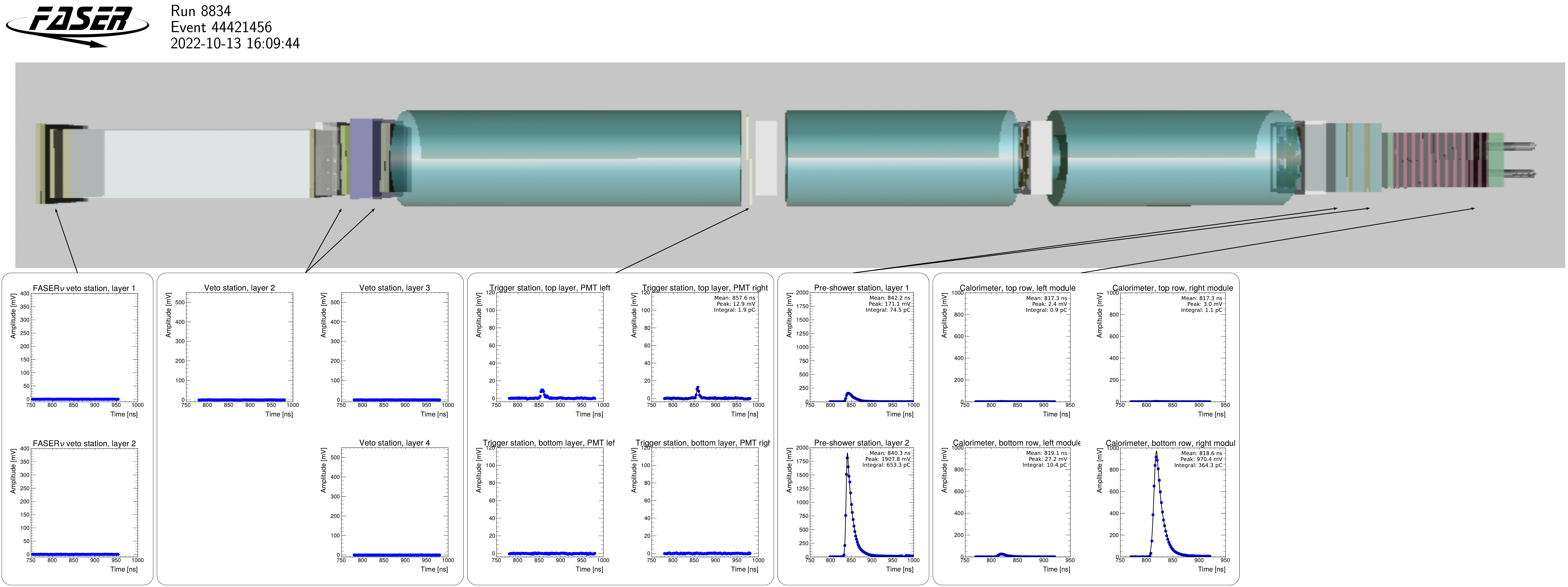}
    \caption{Event display of the ALPtrino event recorded by FASER on 13 October 2022 with 13.6 TeV stable beams. The waveforms for signals in the scintillators and calorimeter modules are shown in blue, fitted to a Crystal Ball function. The timing and veto scintillator charge deposits are below the thresholds considered for the signal region definition. A clear charge deposit in the second preshower layer can be seen. The event has been triggered by the calorimeter modules, with an overall reconstructed energy of 1.6~TeV. The ATLAS interaction point is 480~m to the left of the detector shown. In the title of the waveform plots, left and right is defined facing the downstream direction.} 

    \label{app:fig:evtdisplay}
\end{figure}

\FloatBarrier
\section{Neutrino Compositions by Production Mechanism}
\label{app:FlavourBreakdown}

A breakdown by neutrino production mechanism composition of the neutrino background in the signal region, as well as in the ``calorimeter'', ``magnet'', ``Other'' and ``preshower'' regions, is given in ~\cref{tab:app:VRYieldsTable} and~\cref{tab:app:neut_comp}. \cref{app:fig:neutrinoVRs:prod} shows the energy distribution in the ``calorimeter'' and ``magnet'' region. \cref{app:fig:SRunblinded} shows the energy distribution in the ``preshower'' and signal regions.

\begin{figure}[htp]
    \centering
\includegraphics[width=0.6\textwidth]{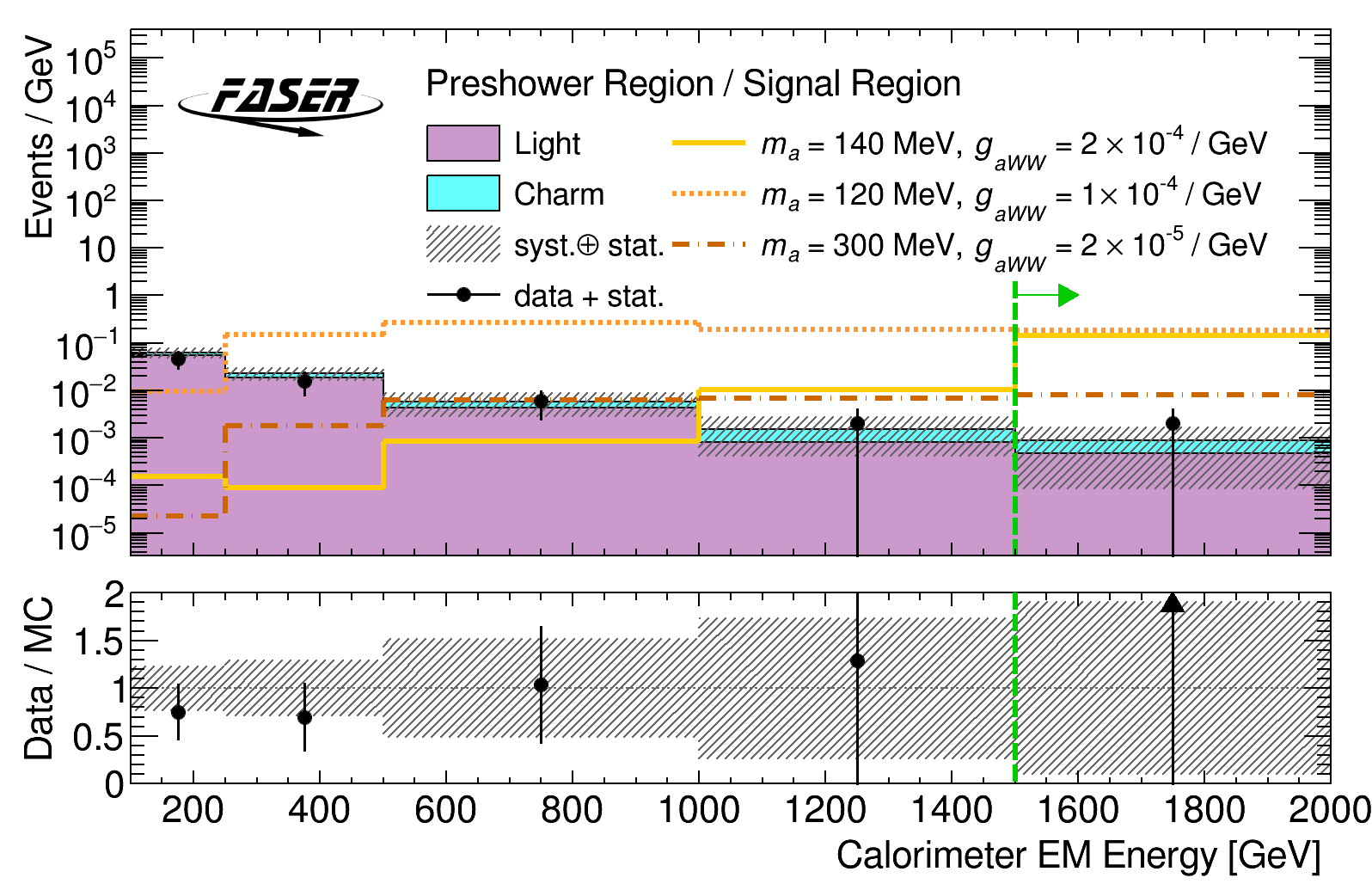}

\caption{Calorimeter energy distribution in the ``preshower'' and signal regions, showing the neutrino background composition separated according to neutrino production mechanism. The last high-energy bin above 1.5 TeV, highlighted with a green arrow, presents the signal region and includes overflows.}
    \label{app:fig:SRunblinded}
\end{figure}

\begin{figure}
\centering
\includegraphics[width=0.49\linewidth]{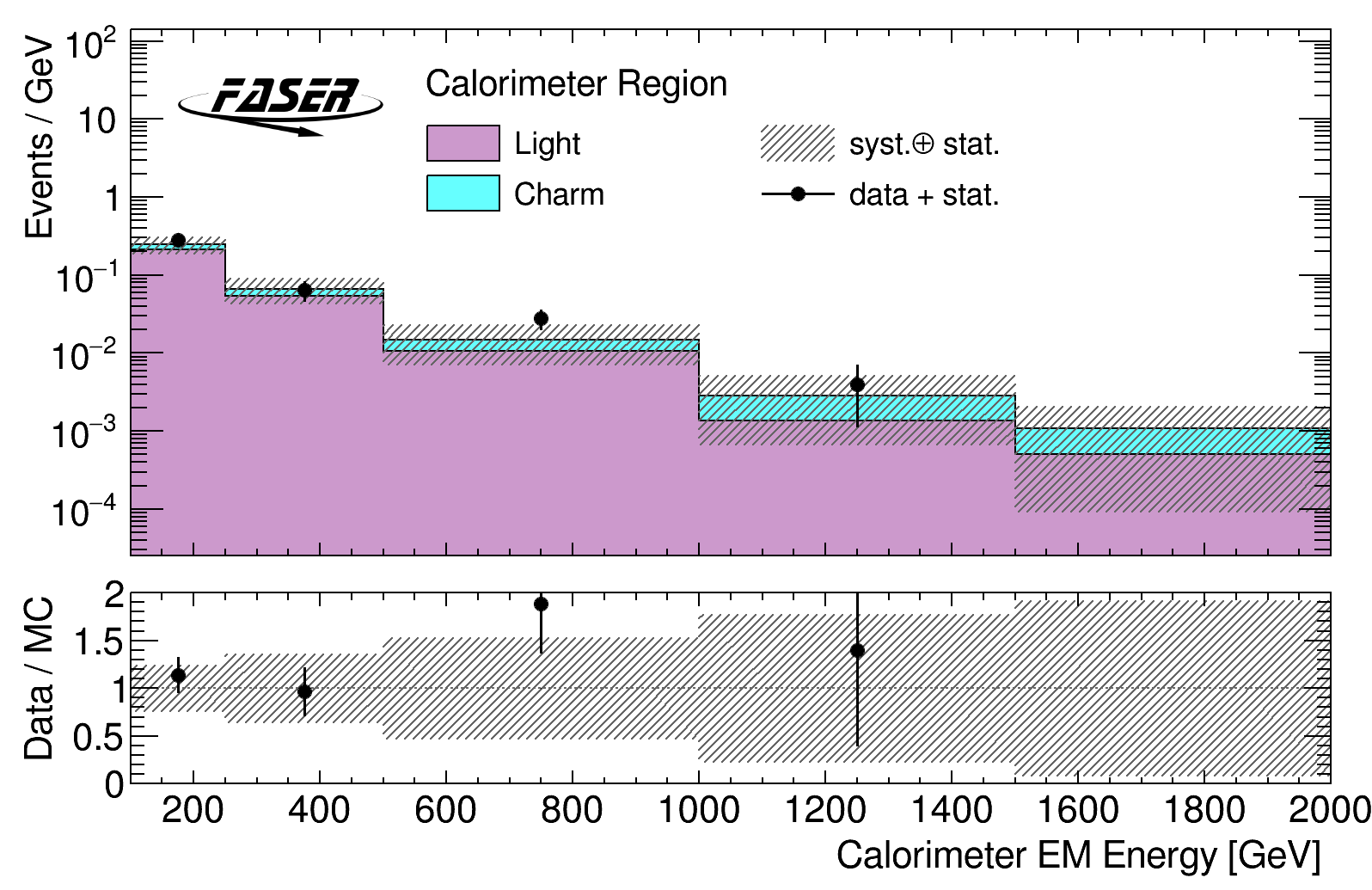}
\includegraphics[width=0.49\linewidth]{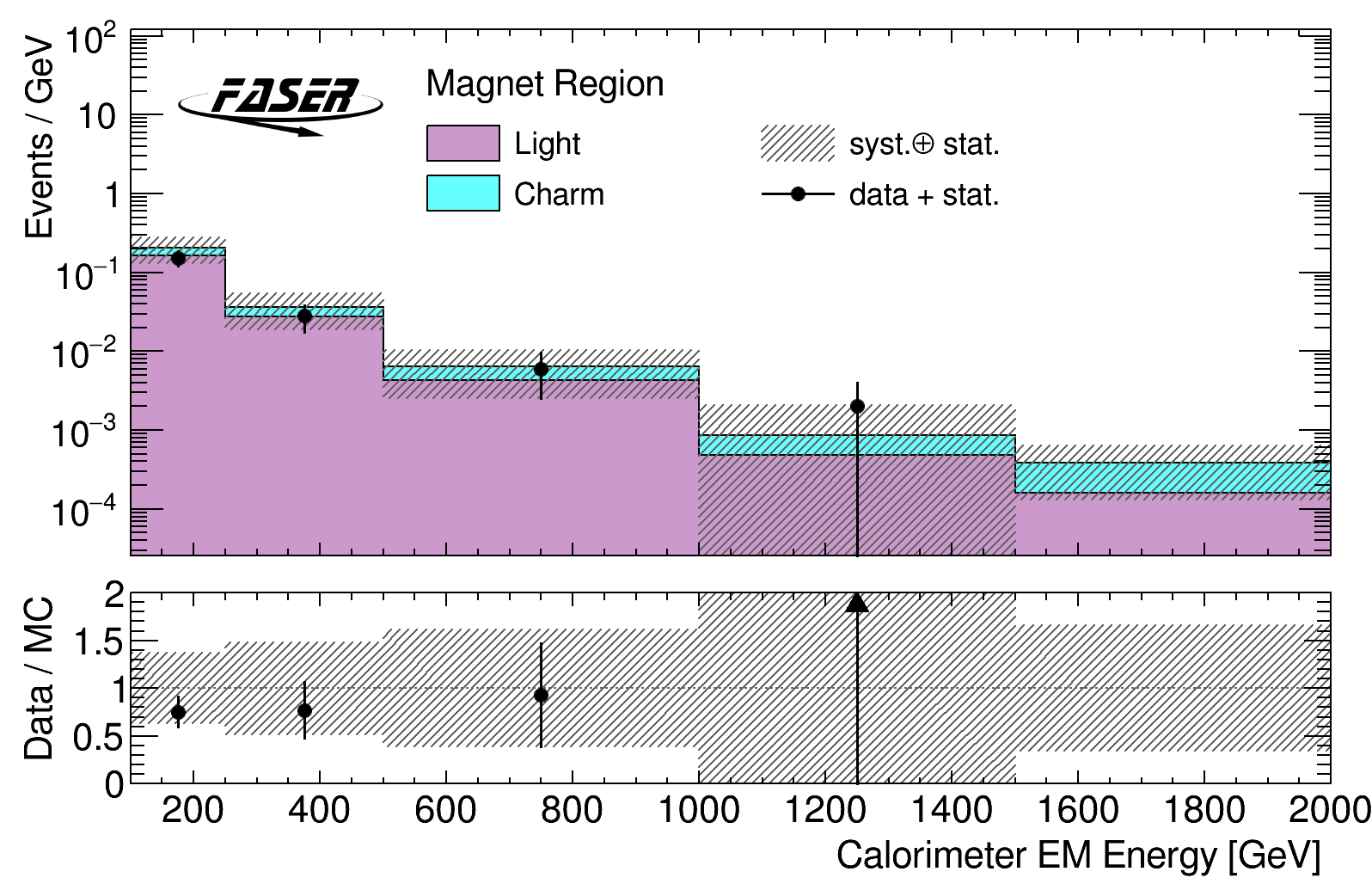}
\caption{Calorimeter energy distributions in the ``calorimeter'' (left) and ``magnet'' (right) region broken by production modes. The uncertainty band includes MC statistical uncertainties, experimental uncertainties, and uncertainties on the neutrino background flux. The last bin contains all events above 1.5 TeV.\label{app:fig:neutrinoVRs:prod}}
\end{figure}

\begin{table}[]
    \centering
    \begin{tabular}{|l|c|} \hline
     \multicolumn{2}{|l|}{``Magnet'' region} \\ \hline
Light & 33.7$^{+\mathrm{6.7}}_{-\mathrm{3.4}}$ (flux) $\pm$ 4.3 (exp.) $\pm$ 0.4 (stat.) \\
Charm & 9.9$^{+\mathrm{16.2}}_{-\mathrm{4.6}}$ (flux) $\pm$ 0.9 (exp.) $\pm$ 0.2 (stat.) \\

Total &  \textbf{43.6 $\pm$ 18.3 (41.9\%)}  \\
Data & \textbf{34}\\ \hline
\multicolumn{2}{|l|}{``Other'' region} \\ \hline
Light & 17.9$^{+\mathrm{1.3}}_{-\mathrm{0.8}}$ (flux) $\pm$ 2.5 (exp.) $\pm$ 0.3 (stat.) \\
Charm & 4.0$^{+\mathrm{6.2}}_{-\mathrm{1.9}}$ (flux) $\pm$ 0.5 (exp.) $\pm$ 0.2 (stat.) \\

Total &  \textbf{21.9 $\pm$ 7.0(32.1\%)} \\
Data & \textbf{17}\\ \hline
\multicolumn{2}{|l|}{``Calorimeter'' region} \\ \hline

Light & 51.8$^{+\mathrm{2.0}}_{-\mathrm{3.4}}$ (flux) $\pm$ 3.1 (exp.) $\pm$ 0.5 (stat.) \\
Charm & 11.1$^{+\mathrm{19.1}}_{-\mathrm{5.1}}$ (flux) $\pm$ 0.5 (exp.) $\pm$ 0.3 (stat.) \\
Total &  \textbf{62.9 $\pm$ 19.7 (31.4\%)}\\
Data & \textbf{74}\\ \hline
\multicolumn{2}{|l|}{``Preshower'' region} \\ \hline
Light & 15.5$^{+\mathrm{0.9}}_{-\mathrm{1.2}}$ (flux) $\pm$ 2.0 (exp.) $\pm$ 0.3 (stat.) \\
Charm & 3.1$^{+\mathrm{4.7}}_{-\mathrm{1.5}}$ (flux) $\pm$ 0.4 (exp.) $\pm$ 0.1 (stat.) \\
Total  & \textbf{18.7 $\pm$ 5.4 (29.1\%)} \\ 
Data & \textbf{15}\\ \hline
    \end{tabular}
\caption{Breakdown of the neutrino composition and data yields in the ``magnet'', ``other'', ``calorimeter'' and ``preshower'' regions (excluding events passing signal region selections). Listed are data and neutrino yields as predicted from MC in 57.7~$\ifb$, also split in light and charm production components. Uncertainties on the flux as well as experimental uncertainties further discussed in \cref{sec:Uncertainties} are also given. \label{tab:app:VRYieldsTable}}
\end{table}

\begin{table}[tbh]
    \centering
    \begin{tabular}{|c|c|}
    \hline
    \multicolumn{2}{|c|}{$>$ 1.5 TeV signal region} \\ \hline
\ \ Light & 0.23$^{+\mathrm{0.01}}_{-\mathrm{0.11}}$ (flux) $\pm$ 0.11 (exp.) $\pm$ 0.04 (stat.) \\
\ \ Charm \ \ & \ \ 0.20$^{+\mathrm{0.34}}_{-\mathrm{0.09}}$ (flux) $\pm$ 0.06 (exp.) $\pm$ 0.03 (stat.) \\
\textbf{Total} & \textbf{0.44 $\pm$ 0.39 (88.6\%)} \\ \hline

    \end{tabular}
\caption{Summary of the MC estimate of the neutrino background broken by production mechanism in the signal region. Uncertainties on the flux, as well as experimental uncertainties further discussed in \cref{sec:Uncertainties}, are also given. The MC events are normalised to 57.7 fb$^{-1}$ and MC statistical uncertainties are given.}
\label{tab:app:neut_comp}
\end{table}

\clearpage

\FloatBarrier
\bibliographystyle{utphys}
\bibliography{references}

\end{document}

%% file: header.tex
\usepackage{amsmath,amssymb,amsthm,amsfonts}
\usepackage{graphicx,tabularx}
\usepackage{color}
\usepackage{multirow}
\usepackage{comment}
\usepackage{enumitem}
\usepackage{subfigure}
\usepackage{orcidlink}
\usepackage{cleveref}
\usepackage{slashed}
\usepackage[normalem]{ulem}
\usepackage{scalerel}
\usepackage{wrapfig}
\usepackage{cancel}
\usepackage{xcolor}
\usepackage{multirow}
\usepackage{setspace}

\newcommand{\PRE}[1]{{#1}} 

\newcommand{\be}{\begin{equation} \begin{aligned}}
\newcommand{\ee}{\end{aligned} \end{equation}}
\newcommand{\beqa}{\begin{eqnarray}}
\newcommand{\eeqa}{\end{eqnarray}}

\def\figureautorefname~#1\null{Fig.\,#1\null}
\def\tableautorefname~#1\null{Tab.\,#1\null}
\def\equationautorefname~#1\null{Eq.\,(#1)\null}

\newcommand{\mev}{\text{MeV}}

\newcommand{\tev}{\text{TeV}}

\newcommand{\ifb}{\text{fb}^{-1}}

\RequirePackage[normalem]{ulem}

\crefname{section}{Sec.}{Secs.}
\crefname{figure}{Fig.}{Figs.}
\crefname{equation}{Eq.}{Eqs.}
\crefname{table}{Table}{Tables}
\crefname{appendix}{Appendix}{Appendices}

\usepackage{placeins}

%% file: authorlist240903.tex
\author{Roshan Mammen Abraham\,\orcidlink{0000-0003-4678-3808}}
\affiliation{Department of Physics and Astronomy, University of California, Irvine, CA 92697-4575, USA}

\author{Xiaocong Ai\,\orcidlink{0000-0003-3856-2415}}
\affiliation{School of Physics, Zhengzhou University, Zhengzhou 450001, China}

\author{John Anders\,\orcidlink{0000-0002-1846-0262}}
\affiliation{CERN, CH-1211 Geneva 23, Switzerland}

\author{Claire Antel\,\orcidlink{0000-0001-9683-0890}}
\affiliation{D\'epartement de Physique Nucl\'eaire et Corpusculaire, University of Geneva, CH-1211 Geneva 4, Switzerland}

\author{Akitaka Ariga\,\orcidlink{0000-0002-6832-2466}}
\affiliation{Albert Einstein Center for Fundamental Physics, Laboratory for High Energy Physics, University of Bern, Sidlerstrasse 5, CH-3012 Bern, Switzerland}
\affiliation{Department of Physics, Chiba University, 1-33 Yayoi-cho Inage-ku, 263-8522 Chiba, Japan}

\author{Tomoko Ariga\,\orcidlink{0000-0001-9880-3562}}
\affiliation{Kyushu University, Nishi-ku, 819-0395 Fukuoka, Japan}

\author{Jeremy Atkinson\,\orcidlink{0009-0003-3287-2196}}
\affiliation{Albert Einstein Center for Fundamental Physics, Laboratory for High Energy Physics, University of Bern, Sidlerstrasse 5, CH-3012 Bern, Switzerland}

\author{Florian~U.~Bernlochner\,\orcidlink{0000-0001-8153-2719}}
\affiliation{Universit\"at Bonn, Regina-Pacis-Weg 3, D-53113 Bonn, Germany}

\author{Emma Bianchi\,\orcidlink{0009-0001-2381-0650}}
\affiliation{D\'epartement de Physique Nucl\'eaire et Corpusculaire, University of Geneva, CH-1211 Geneva 4, Switzerland}

\author{Tobias Boeckh\,\orcidlink{0009-0000-7721-2114}}
\affiliation{Universit\"at Bonn, Regina-Pacis-Weg 3, D-53113 Bonn, Germany}

% ...

%% file: acknowledgments.tex
We thank CERN for the excellent performance of the LHC and the technical and administrative staff members at all FASER institutions for their contributions to the success of the FASER experiment. We thank the ATLAS Collaboration for providing us with accurate luminosity estimates for the Run 3 LHC $pp$ collision data. We thank the LHCb Collaboration for the loan of their calorimeter modules which were crucial for this analysis. We also thank the CERN STI group for providing detailed FLUKA simulations of the muon fluence along the LOS. 

This work was supported in part by Heising-Simons Foundation Grant Nos.~2018-1135, 2019-1179, and 2020-1840, Simons Foundation Grant No.~623683, U.S. National Science Foundation Grant Nos.~PHY-2111427, PHY-2110929, and PHY-2110648, JSPS KAKENHI Grant Nos.~19H01909, 22H01233, 20K23373, 23H00103, 20H01919, and 21H00082, the joint research program of the Institute of Materials and Systems for Sustainability, ERC Consolidator Grant No.~101002690, BMBF Grant No.~05H20PDRC1, DFG EXC 2121 Quantum Universe Grant No.~390833306, Royal Society Grant No.~URF$\backslash$R1$\backslash$201519, UK Science and Technology Funding Councils Grant No.~ST/ T505870/1, the National Natural Science Foundation of China, Tsinghua University Initiative Scientific Research Program, and the Swiss National Science Foundation.